\documentclass[12pt,letterpaper]{umd-thesis}    % for either Masters or PhD
\usepackage[dvips]{graphics}
\usepackage{psfig}
\usepackage{rotating}
\usepackage{amsthm}
\usepackage{amstex}
\usepackage{subfigure}
\spacing{1}
\title{
SPURIOUS LYAPUNOV EXPONENTS\\ 
COMPUTED USING THE\\
ECKMANN-RUELLE PROCEDURE
}
\author{Joshua A. Tempkin}
\department{Department of Mathematics}
\advisor{Professor James A.\ Yorke}
\chairtitle{Chairman/Advisor}
\committee      {       Professor Celso Grebogi\\
			Professor Edward Ott \\
			Professor J.\ Robert Dorfman \\
                        Professor Denny Gulick 
                }
\dedication{\centering 
To Margaret, for waiting,\\
and\\
To Andrea, for not waiting
}    
\setboolean{hasfigures}{true}
\setboolean{hastables}{true}
\setboolean{hascopyright}{true}
\abstractfile{abstract.tex}
\acknowledgements{I would like to thank my advisor, Dr. James Yorke, for his support and
guidance throughout this thesis project.  He has taught me a lot about
mathematics and how to communicate ideas.  His insistence that proofs be
unencumbered by notation helped me numerous times to clarify my ideas.  
His intuition is astonishing and his nurturing patience seems limitless.  
I have grown tremendously since we began working together.

I also would like to thank my thesis committee, Dr. Celso Grebogi, Dr.
Edward Ott, Dr. Robert Dorfman, and Dr. Denny Gulick, for being
available and helpful whenever I had questions.  In addition, I had
several interesting discussions with Dr. Brian Hunt and with Dr. Tim Sauer
of George Mason University.  Dr. Sauer was especially helpful with the
work in Appendix B.

During my first few years of graduate study, Dr. Robert Warner and Dr.
Frances Gulick took me under their wings and encouraged me.  Millie
Stengel and the secretaries in the Math Department have helped in many
many ways, big and small, over the years.

The students and postdocs in the Chaos Group have been very supportive.  
They are a great bunch of people, one and all, and several deserve special
mention:  Linda Moniz, Dj Patil, Myong-Hee Sung, Mike Roberts, Mitrajit
Dutta, and Guocheng Yuan.

I also want thank to my parents.  Their unwavering faith in me and their
support, encouragement, and love have made a big difference in my life.

Finally, special thanks, heartfelt gratitude, and warm appreciation go to
my dear and loving wife, Margaret, for {\it absolutely everything}.  She
has put up with more from me than I will admit to, and still she
comes away smiling.  No more late nights in the basement, honey, I
promise!  Now I can get a ``real'' job!
}
\date{1999}
\begin{document}
%
% make the title page and so forth
%
\makefrontmatter
%
% 
%
% Thesis goes here
%
% SPURIOUS LYAPUNOV EXPONENTS COMPUTED USING THE ECKMANN-RUELLE PROCEDURE
%     Ph.D. Thesis      by Joshua A. Tempkin
% 
%  THESISMASTER.TEX -- July 24, 1999
%

% Macro file
\newcommand{\half}{{1 \over 2}}
\newcommand{\qedbox}{\vbox{\hrule\hbox{\vrule\kern3pt\vbox{\kern6pt}\kern3pt\vrule}
\hrule}}
\newcommand{\QED}{\unskip\nobreak\hfil\penalty50\hskip.75em\null\nobreak\hfil
\qedbox{\parfillskip=0pt \finalhyphendemerits=0 \par} \bigskip}
\newcommand{\THM}[1]{\bigskip\noindent{\bf {#1}.} \ }
\newcommand{\PROOF}{\THM {Proof}}
\newcommand{\SKETCH}{\THM {Sketch of Proof}}
\newcommand{\CASE}[1]{\smallskip\noindent{CASE {#1}:} }
\newcommand{\NOTE}{\THM {Note}}
\newcommand{\SECTION}[1]{\centerline{\underbar{\bf {#1}}}\medskip}
\newcommand{\reals}{I\!\!R}
\newcommand{\REF}[1]{\cite{#1}}
\newcommand{\henon}{H\'enon }

\newcommand{\harr}[1]{\smash{\mathop{\hbox to .5in{\rightarrowfill}}
\limits^{\scriptstyle#1}}}
\newcommand{\varr}[1]{\llap{$\scriptstyle #1$}\left\downarrow \vcenter to
.5in{}\right.}
\newcommand{\diagram}[1]{{\normallineskip=8pt \normalbaselineskip=0pt
\matrix{#1}}}

\newcommand{\dataset}{\{y_i\}_{i=0}^\infty}
\newcommand{\dt}{\Delta t}
\newcommand{\del}{\Delta \!}

\newcommand{\Ctay}{C_{Taylor}}

\newtheorem{thm}{Theorem}[chapter]
\newtheorem{lemma}[thm]{Lemma}
\newtheorem{prop}[thm]{Proposition}
\newtheorem{defn}[thm]{Definition}

\newtheorem{thmspec}{Theorem}[chapter]
\newcounter{timchap}
\newcounter{timthm}

% Chapters
% CHAPTER 1  --  chap1.tex

%\input{erdmacro.tex}
\chapter{Introduction}

In the analysis of observed physical systems, it has become standard
practice to study an auxiliary system reconstructed from a time series of
measured data.  Successful reconstruction of the original system's
attractor is the basis of the method for Lyapunov exponent calculation
proposed by Eckmann and Ruelle \REF{ER, EKRC} and by Sano and Sawada
\REF{SS}.  However, since the reconstructed attractor often lies in a
larger dimensional space than the original system, the calculations
produce too many exponents.  This leads to an important question:  how do
we distinguish the true Lyapunov exponents of the underlying system from
the extra ``spurious'' ones present for the reconstructed system?  We
answer this question for two specific cases: (a) when a one-dimensional
system is reconstructed in $m$ dimensions and (b) when a two-dimensional
system is reconstructed in five dimensions.

The attractor reconstruction process begins by choosing a number $m$ and
observing the present state $p$ of the underlying system with that number
$m$ of independent measurements $\pi_i(p)$, $i=1,\dots,m$.  For each point
$p$ in the phase space, there is an $m$-dimensional vector $\pi(p) =
(\pi_1(p), \dots, \pi_m(p))$.  This produces a {\bf measurement function}
$\pi$ that associates points in $\reals^m$ with points in the phase space
of the underlying dynamical system.  In practice, the measurement function
$\pi$ often consists of time-delayed versions of a scalar measurement (see
Takens and others \REF{Takens,PCFS}).  Under certain genericity conditions
(see, for example, \REF{SYC}), the original dynamical phase space
attractor $A$ will be topologically equivalent to its reconstructed image
$\pi(A)$ in $m$-dimensional Euclidean space.  See Figure 1.1.

\begin{figure}
\begin{center}
\subfigure[]{\scalebox{.5}
{\begin{turn}{-90}\includegraphics{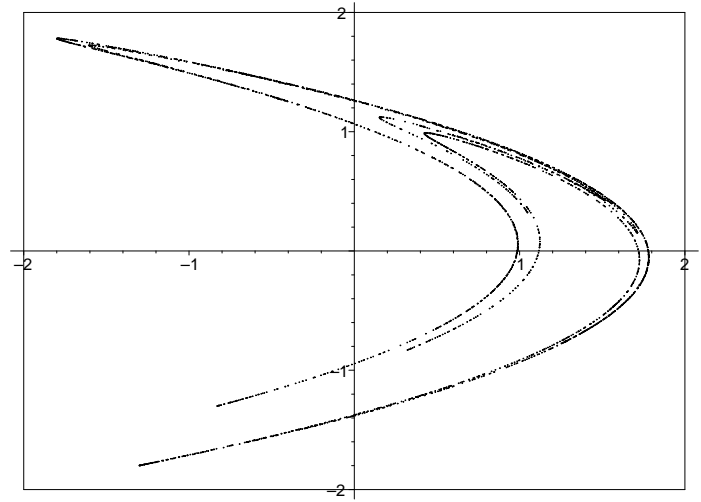}\end{turn}}}
\subfigure[]{\scalebox{.5}
{\begin{turn}{-90}\includegraphics{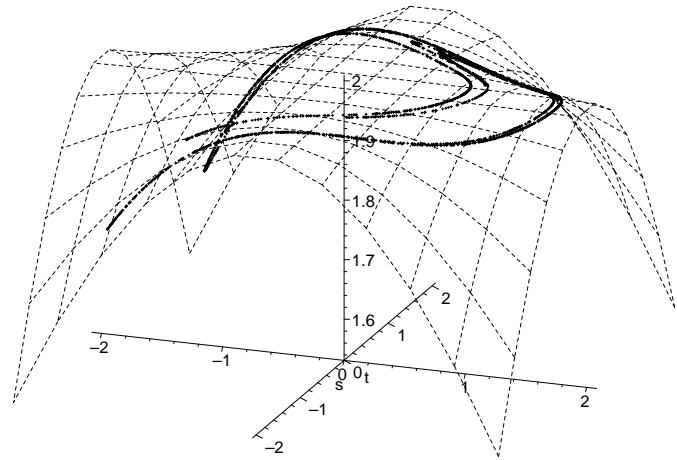}\end{turn}}}
\end{center}
\caption[A measurement function reconstructs the attractor.]{The
measurement function $\pi$ reconstructs the true attractor $A$ of the
underlying dynamics (shown in (a)) as the set $\pi(A)$ in some $\reals^m$
(shown in (b)).  When the reconstruction dimension $m$ is large enough,
the generic (smooth) measurement function $\pi$ will be a one-to-one
correspondence between $A$ and $\pi(A)$.}
\end{figure}

The set of measurement vectors $\pi(p)$ in the reconstructed attractor
$\pi(A)$ can be studied for geometrical and dynamical properties.  
Looking to the reconstructed attractor for dynamical properties of the
original attractor was suggested in 1985 by Eckmann and Ruelle, et. al.
\REF{ER,EKRC} and also by Sano and Sawada \REF{SS}.  It is usually
necessary that $m$ be chosen large enough that there is a one-to-one
correspondence between points of the original attractor and points of the
reconstructed attractor \REF{SYC}.  This requirement often forces the
reconstruction space to have larger dimension than the original system.  
In these cases, Lyapunov exponent calculations in the reconstruction space
produce $m$ real numbers, not all of which can be Lyapunov exponents of
the original dynamical system.  For example, if a two-dimensional
dynamical system is reconstructed in five dimensions, current methods
compute five ``exponents'' in the reconstruction space.  At most two of
these can be Lyapunov exponents of the original system; the other numbers
are ``spurious'' exponents.

How, then, do we tell the true exponents from the spurious exponents?  
Parlitz \REF{Parlitz} proposed that recomputing the exponents using the
reversed time series would make the true exponents switch sign.  Bryant,
Brown, and Abarbanel looked at the local ``thickness'' of the data set to
identify spurious exponents \REF{BBA}.  In addition, some authors have
proposed removing the extra dimensions by projecting the reconstructed
dynamics to its tangent plane (see \cite[pp.\ 336--339]{HBCJM}
and \cite[pp.\ 2156-2157]{SzS}. In the present paper, we study
the original Eckmann-Ruelle algorithm in order to clarify its output.

The algorithm presented in \REF{EKRC} for computing Lyapunov exponents has
three major steps.  First, one reconstructs the attractor in some
$m$-dimensional Euclidean space of measurements as indicated above.  On
the reconstructed attractor, there is a time-$\tau$ map $F$ which takes
the $m$-vector $P_t$ at time $t$ to the $m$-vector $P_{t+\tau}$ at time
$t+\tau$.  This map $F$ represents the reconstructed dynamics, and
$F(P_t)$ is defined to be $P_{t+\tau}$.  In the second step, one computes
a local linearization matrix for $F$ at each point $P$ of the
reconstructed attractor by finding the $m \times m$ matrix $M$ (depending
on $P$) which satisfies as closely as possible
$$
P_{t+\tau}-F(P) \approx M (P_t-P)  \qquad \hbox{ for all $t$ for which
$P_t$ is close to $P$}.
$$
We call $M=M(P)$ the Eckmann-Ruelle linearization at $P$.  In the last
step (which will be discussed in more detail in Chapter 3), one computes
the Lyapunov exponents of $F$ from a matrix product of these local
linearizations.  That is, the Lyapunov exponents will be the various
values achieved by
$$
\lim_{n \to \infty} {1 \over n} \ln \left\| M_{n-1} M_{n-2} \cdots M_1
M_0\nu \right\|
$$
for various unit vectors $\nu \in \reals^m$, where $M_i = M(P_i)$ is the
best local linearization matrix (i.e., the Eckmann-Ruelle linearization)
at the point $P_i=\pi(p_i)$ in the trajectory.

To understand the output from this algorithm, it is crucial to determine
the Eckmann-Ruelle linearizations.  At first glance, one might think these
linearizations should produce derivative matrices, $DF_P$.  This, however,
will not happen in general.  Suppose that an attractor reconstructed in
$m$-dimensional Euclidean space lies within a lower-dimensional surface in
$\reals^m$.  The reconstructed dynamics are well-defined on this surface,
but they are not defined off of it, and so the classical $m \times m$
derivative matrix $DF_P$ will not exist.  See Figure 1.1b. Therefore, the
local linearizations cannot be derivative matrices.

In this paper, we study the local linearization matrices.  For the two
cases mentioned above, we will show that these matrices have several
important properties.  First and foremost, the linearizations of the
reconstructed dynamics are surprisingly good in the following sense (see
Theorem \ref{JJ1} in Chapter 2).  For any matrix $L$, if we ``linearize''
the reconstructed dynamics $F$ about the point $P$, we obtain for some
integer $k$
\begin{equation}\label{EQN1.1}
F(P_t) - F(P) - L\left(P_t - P\right) = O\!\left(\left\|P_t-P\right\|^k\right)
\quad \hbox{ as $P_t \to P$}.
\end{equation}
We write $g(x) = O(|x|^k)$ as $x \to 0$ to mean that there exists some
constant $C$ such that $\|g(x)\| \leq C |x|^k$ for all $x$ in some
neighborhood of $0$.  For most choices of $L$, we expect $k=1$ in
(\ref{EQN1.1}).  For the traditional derivative $L=DF_P$, we expect $k=2$
(when it exists).  However, the Eckmann-Ruelle linearization $L=M(P)$ in
our specific cases has $k$ larger than this.  In the case of a
one-dimensional system reconstructed in $m$ dimensions, we find $k=m+1$,
and in the case of a two-dimensional system reconstructed in 5 dimensions,
we find $k=3$.  The second important property of these matrices is that
there are natural coordinate systems with respect to which the
linearization matrices have a special upper-triangular matrix
representation.  This upper triangular form allows us to easily compute
the Lyapunov exponents (which depend on the diagonal entries when the
matrices are upper triangular).  Moreover, for generic measurement
functions, these exponents will be completely independent of the specific
measurement function $\pi$ used in the reconstruction, depending only on
the dynamics of the original system.  This key property allows us to
derive explicit formulas for the Lyapunov exponents produced by the
Eckmann-Ruelle procedure in the low noise limit.  We will demonstrate this
in Chapter 3.

Our paper is structured in the following manner.  In Chapter 2, we
determine the linear map that provides the best local linearization to the
reconstructed dynamics.  In Chapter 3, we give the matrix representation
alluded to above and derive formulas for the Lyapunov exponents produced
by the algorithm.  The goal of Chapter 4 is to prove that numerical
determinations of the local linearization matrix converge to the
Eckmann-Ruelle matrix as the radius of the neighborhood shrinks to zero.  
Chapter 5 presents numerical work illustrating our theoretical results.  
Finally, appendices are included which contain technical lemmas used in
the text.

% CHAPTER 2  --  chap2.tex
%\documentclass{report}
%\begin{document}

%\input{erdmacro.tex}
\chapter{Local Linearizations}\label{ch:loclin}

\newcommand{\Mt}{\widetilde M}
\newcommand{\dpidx}{{{\partial \pi} \over {\partial x}}(p)}
\newcommand{\dpidy}{{{\partial \pi} \over {\partial y}}(p)}
\newcommand{\ddpidxx}{{{\partial^2 \pi} \over {\partial x^2}}(p)}
\newcommand{\ddpidxy}{{{\partial^2 \pi} \over {\partial x \partial y}}(p)}
\newcommand{\ddpidyy}{{{\partial^2 \pi} \over {\partial^2 y}}(p)}
\newcommand{\dpifdx}{{{\partial (\pi \circ f)} \over {\partial x}}(p)}
\newcommand{\dpifdy}{{{\partial (\pi \circ f)} \over {\partial y}}(p)}
\newcommand{\ddpifdxx}{{{\partial^2 (\pi \circ f)} \over {\partial x^2}}(p)}
\newcommand{\ddpifdxy}{{{\partial^2 (\pi \circ f)} \over {\partial x \partial
y}}(p)}
\newcommand{\ddpifdyy}{{{\partial^2 (\pi \circ f)} \over {\partial y^2}}(p)}

A simple example where $f(p)=2p(mod \ 2\pi)$ on the interval $[0,2\pi]$
allows us to describe the general problem.  We reconstruct this interval
in $\reals^2$ via the measurement function $\pi(p) = \left(\cos(p),
\sin(p)\right)$.  The observed dynamics $F=\pi f \pi^{-1}$ maps points on
the unit circle: $\pi(p)$ to $\pi(f(p))$.  The following question is
central to our discussion:  for any given point $P=\pi(p)$ on the unit
circle $\pi([0,2\pi])$, which $2 \times 2$ matrix provides the best local
linearization of $F$ around $P$ (in the sense of (\ref{EQN1.1}))?  Since
$F$ is not defined off the unit circle, the traditional derivative of $F$
does not exist.  On the other hand, the linearizations of $F$ are $2
\times 2$ matrices while the derivatives of the original system $f$ are $1
\times 1$.  It follows that no linearization of $F$ can be a derivative
matrix.  What, then, is the best local linearization of $F$ near $P$ (if
one even exists)?

In this chapter, we elucidate the nature of the local linearization
matrices for the reconstructed dynamics.  In their papers \REF{ER, EKRC},
Eckmann and Ruelle discussed the best local linearization of the
reconstructed dynamics.  We will define the ``Eckmann-Ruelle
linearization'' to be the unique linear map with certain properties.  In
Theorems \ref{JJ1} and \ref{JJ3}, we will show that our definition
provides the best
local linearization.

In the simple example above, we specified the measurement function $\pi$
explicitly.  In practice, however, one rarely knows the measurement
function that arises from the attractor reconstruction process. Time-delay
embeddings are commonly used, but even then, one cannot know the
measurement function completely without knowledge of how the scalars in
the time-series relate to the states of the system.  This knowledge is
often unavailable in experimental situations.

For this reason, we focus on measurement functions with generic properties
(generic in the sense of prevalence).  A property is {\bf generic} in the
sense of prevalence if whenever a function lacks the property, arbitrarily
small perturbations of that function have the property with probability 1
\REF{HSY}.  In the situations that interest us, the generic measurement
function can be taken to be a smooth diffeomorphism from the underlying
phase space into $m$-dimensional Euclidean space $\reals^m$.  The
genericity conditions that guarantee the (topological) equivalence of the
underlying and reconstructed phase space attractors \REF{SYC} typically
force the reconstruction space to have larger dimension than the
underlying attractor.  In fact, the dimension $m$ of the reconstruction
space can be at least twice that of the underlying system.

We adopt the convention that lower-case letters refer to the underlying
system while upper-case letters refer to the reconstructed system.  For
example, $P=\pi(p)$ is the point in the reconstructed phase space
$\reals^m$ corresponding to the point $p$ in the underlying phase space.  
The convention will also extend to sets in the various spaces:  $u$ could
be a neighborhood of $p$ in the underlying phase space while $U$ could be
a neighborhood of $P$ in the reconstructed phase space.  We represent the
dynamical flow on the underlying phase space by the time-$\tau$ map $f$
for some $\tau$.  The measurement function $\pi$ maps the underlying phase
space into $\reals^m$ for some $m$.  In $\reals^m$, there is an induced
time-$\tau$ flow map $F=\pi f \pi^{-1}$ that maps each $\pi(p)$ to
$\pi(f(p))$, and we refer to $F$ as the reconstructed dynamics.  Assuming
that $\pi$ is a one-to-one correspondence on the underlying phase space
(i.e., that $p \not = q$ implies $\pi(p) \not = \pi(q)$), the map $F$ is
well-defined on the reconstructed phase space.

We examine the special case when the underlying map $f$ is one-dimensional
and the following specific assumptions about $f$, $\pi$, and a point
$P=\pi(p)$ hold.

\begin{description}
\item[(A1)] {\it The underlying dynamical system $f$ maps the unit
interval $[0,1]$ into itself.  In addition, we assume that $f$ has $m+1$
continuous derivatives, i.e., $f$ is $C^{m+1}$.}
\item[(A2)] {\it The measurement function $\pi$ maps $[0,1]$ into
$\reals^m$, and $\pi$ is $C^{m+1}$.}
\item[(A3)] {\it For this $p \in [0,1]$, $P=\pi(p) \in \reals^m$, and the
set $\pi^{-1}(P)$ contains only one point, namely $p$.}
\item[(A4)] {\it The first $m$ derivative vectors for $\pi$ at $p$, i.e.,
$\pi^{(n)}(p):= {{d^n \pi} \over {dp^n}}(p)$ for $n=1,2,\dots,m$, are
linearly independent in $\reals^m$.}
\end{description}

\noindent Properties (A3) and (A4) are generic properties in the space of
$C^{m+1}$ functions from $[0,1]$ into $\reals^m$.  By (A4), the derivative
vector $\pi'(p)$ is nonzero.  Together with (A3) and the Inverse Function
Theorem, this implies

\begin{description}
\item[(A3')] {\it There are neighborhoods $u \subseteq [0,1]$ of $p$ and
$U \subseteq \pi([0,1]) \subset \reals^m$ of $P$ such that $\pi(x) \in U$
if and only if $x \in u$, and $\pi|_u$ is a diffeomorphism.}
\end{description}

\noindent This statement has a useful consequence.  Since $\pi|_u$ is a
diffeomorphism, there is a constant $C_\pi > 1$ such that if
$P_1,P_2 \in U$ with $P_i = \pi(p_i)$, $p_i \in u$, $i=1,2$, then
\begin{equation}\label{EQN2.1}
{1 \over {C_\pi}} \|P_1-P_2\| \leq |p_1-p_2| \leq C_\pi \|P_1-P_2\|.
\end{equation}
This inequality will be useful in translating statements back and forth
between the underlying space and the reconstruction space.

We begin our analysis of this case by examining the Taylor expansion of
$\pi$ about a point $P=\pi(p)$ in $m$-space for which (A3) and (A4) hold.  
Since $\pi$ is one-to-one on the neighborhood $u$, any point $P+\del P$ in
$\pi(u)$ near $P$ can be written in the form
$$
P+\del P = \pi(p+h) = P + \sum_{n=1}^m { {\pi^{(n)}(p)} \over {n!} } h^n +
Rem(h)
$$
where $Rem(h)$ is the Taylor remainder term (an $m$-vector here).  There
is a constant $\Ctay > 0$ such that
$$
\left\|Rem(h)\right\| \leq \Ctay |h|^{m+1}.
$$
By hypothesis (A4), the vectors $\pi^{(1)}(p), \dots, \pi^{(m)}(p)$ are
linearly independent and form a basis for $\reals^m$ which we call the
{\bf canonical embedding basis} at $P$.  (See Figure 2.1.)
\begin{figure}
\begin{center}
\scalebox{.5}{\begin{turn}{-90}\includegraphics{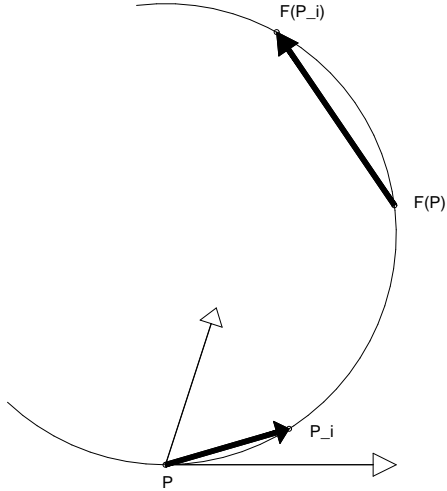}\end{turn}}
\end{center}
\caption[The canonical embedding basis in $\reals^2$.]{The derivative
vectors $\pi^{(1)}(p)$ and $\pi^{(2)}(p)$ (thin arrows) form the canonical
embedding basis for $\reals^2$ at $P=\pi(p)$. Also, the small tangent
vector $\del P_i = P_i - P$ maps to its image vector
$F(P_i) - F(P)$ near $F(P)$ (thick arrows).}
\end{figure}
The little vector $\del P$ at $P$ can be written conveniently in this basis:
\begin{equation}\label{EQN2.2}
\del P = \left(h, \half h^2, \dots, {1 \over {m!} }h^m\right)_P + Rem(h).
\end{equation}
We look at the image $F(P)=\pi(f(p))$ in essentially the same way.  The
Taylor expansion of $F(P+\del P) = F(\pi(p+h)) = \pi(f(p+h))$ is given by
$$
F(P+\del P) = F(P) + \sum_{n=1}^m { {(\pi \circ f)^{(n)}(p)} \over {n!} }
h^n + Rem_f(h).
$$
Here, $Rem_f$ is the Taylor remainder vector associated with $\pi \circ
f$.  Without loss of generality, we can choose the constant $\Ctay$ so
that we also have
$$
\left\|Rem_f(h)\right\| \leq \Ctay |h|^{m+1}.
$$
At this point, we are ready to define the Eckmann-Ruelle linearization.  
The definition will be justified by the properties stated in Theorem
\ref{JJ1}.

\begin{defn}
Assuming (A1) -- (A4), we define the {\bf Eckmann-Ruelle linearization}
$M=M(P)$ to be the unique linear map from $\reals^m$ to $\reals^m$
satisfying
\begin{equation}\label{EQN2.3}
M \ \pi^{(n)}(p) = (\pi \circ f)^{(n)}(p) 
\qquad \hbox{for each $n=1,\dots, m$}.
\end{equation}
\end{defn}

$M$ is well-defined and unique because the set of vectors $\{\pi^{(1)}(p),
\dots, \pi^{(m)}(p)\}$ on the left-hand side of (\ref{EQN2.3}) forms a
basis for $\reals^m$ by assumption (A4).  While it may be convenient to
think of $M$ as a matrix, no coordinate system has yet been specified.  
We will give a matrix representation for $M$ in the next chapter.  Now, we
prove that $M(P)$ is in fact the best linear approximation to the
reconstructed dynamics $F$ near $P$.

We say that $g(x) = O\!(|x|^k)$ as $x \to 0$ if there exists a constant
$C$ such that $\|g(x)\| \leq C |x|^k$ for all $x$ in some neighborhood of
0.  Note that the error term in (\ref{EQN2.4}) of Theorem \ref{JJ1} can be
far smaller than that for the usual Jacobian matrix, which would be
$O\!(\left\|\del P\right\|^2)$.

\begin{thm}[Local Linearization, $\reals^1\!\to\!\reals^m$]\label{JJ1}
Assume (A1) and (A2).  Let $P=\pi(p)$ be a point of $\pi([0,1])$ for which
(A3) and (A4) hold.  If $P$ is in the closure of a trajectory of $F$,
$P_0, P_1, \dots \in \reals^m$, then the Eckmann-Ruelle linearization
$M=M(P)$ defined by (\ref{EQN2.3}) is the unique linear map such that
\begin{eqnarray}\label{EQN2.4}
F(P + \del P) - F(P) - M \del P & = &O\!\left(\left\|\del P\right\|^{m+1}
\right) \\
\hbox{as $\left\|\del P\right\| \to 0$},& & \hbox{where $P+\del P \in
\pi([0,1])$.} \nonumber
\end{eqnarray}
\end{thm}

\PROOF
For small $\del P$ with $P+\del P \in \pi(u)$, we can write $P+\del
P = \pi(p+h)$.  Note that $F(P)=\pi(f(p))$ and $F(P+\del P) =
\pi(f(p+h))$.  First, we prove that the Eckmann-Ruelle linearization
$M=M(P)$ satisfies (\ref{EQN2.4}).  Using the definition of $M$ in
equation (\ref{EQN2.3}), we cancel terms from the two Taylor series:
\begin{multline*}
\lefteqn{F(P + \del P) -F(P) - M \del P} \\
\begin{split}
&= \pi(f(p+h)) - \pi(f(p)) - M \left(\pi(p+h)-\pi(p)\right) \\
&= \left(\sum_{n=1}^m (\pi\!\circ\! f)^{(n)}(p) {{h^n} \over {n!}}
+ Rem_f(h)\right) - M \left(\sum_{n=1}^m \pi^{(n)}(p) {{h^n} \over {n!}} +
Rem(h) \right) \\
&= Rem_f(h) - M Rem(h). 
\end{split}
\end{multline*}
Since $M$ is a fixed map and $|h| \leq C_\pi \left\|\del P\right\|$ by
(\ref{EQN2.1}), we have
\begin{eqnarray*}
\left\| F(P + \del P) - F(P) - M \del P  \right\|
& \leq & \left\|Rem_f(h)\right\| + \left\|M Rem(h)\right\| \\
& \leq & \Ctay (1 + \|M\|) |h|^{m+1} \\
& \leq & \Ctay (1 + \|M\|) C_\pi^{m+1} \|\del P\|^{m+1}
\end{eqnarray*}
for all $\left\|\del P\right\|$ sufficiently small, establishing
(\ref{EQN2.4}).

It remains to show that $M$ is the only linear map that satisfies
(\ref{EQN2.4}).  Suppose to the contrary there is another linear map $\Mt$
such
that
$$
\left\| F(P + \del P) - F(P) - \Mt \del P \right\| \leq C_1 \|\del P\|^{m+1} 
\qquad \hbox{as $\del P \to 0$, $P+\del P \in \pi(u)$}
$$
for some constant $C_1$.  Set $A:=\Mt - M$.  For small enough $\del P$
with $P+\del P \in \pi(u)$,
\begin{eqnarray*}
\left\|A \del P\right\|
&\leq& \left\| F(P\!+\!\del P) -F(P) -M\!\del P \right\| + \left\|
F(P\!+\!\del P) - F(P) - \Mt\!\del P \right\| \\
&\leq& C_2 \|\del P\|^{m+1}
\end{eqnarray*}
where $C_2 := C_1 + \Ctay (1 + \|M\|) C_\pi^{m+1}$.  With respect to the
canonical embedding basis at $P$, the vector $\del P$ has the form in
(\ref{EQN2.2}), and so,
\begin{eqnarray*}
\left\| A \left(h, \half h^2, \dots, {1 \over {m!} }h^m\right)_P \right\|
& \leq& \left\| A \del P \right\| + \left\|A Rem(h) \right\| \\
& \leq& C_2 \|\del P\|^{m+1} + \|A\| \ \Ctay |h|^{m+1} \\
& \leq& C_2 C_\pi^{m+1} |h|^{m+1} + \|A\| \ \Ctay |h|^{m+1} \\
& :=& C_3 |h|^{m+1} 
\end{eqnarray*}
Therefore, for $h \in u$ sufficiently small
$$
\left\| A \left(h, \half h^2, \dots, {1 \over {m!} }h^m\right)_P
\right\|^2 \leq C_3^2 |h|^{2m+2}.
$$
If we represent the matrix $A$ with respect to the canonical embedding
basis at $P$, then the left-hand side is a polynomial in $h$, call it
$p(h)$, with degree at most $2m$.  Since $P$ is a limit point of the
trajectory in $\reals^m$, there are infinitely many $h_i \in u$, $P+\del
P_i = \pi(p+h_i) \in U$, $h_i \to 0$ for which this polynomial satisfies
$|p(h_i)| \leq C_3^2 |h_i|^{2m+2}$. By Proposition \ref{PP1} of Appendix
A, $p(h)$ must be the zero polynomial.  This, in turn, forces all the
elements of the matrix $A$ to be zero.  Thus, $A=0$ and $\Mt = M$, proving
that $M=M(P)$ is indeed the unique linear map satisfying (\ref{EQN2.4}).
\QED

To make this method explicit (and for later use), we compute the
Eckmann-Ruelle linearization for our doubling map example.  Let $f(p) = 2p
(mod \ 2\pi)$ on $[0,2\pi]$ and $\pi(p) = \left(\cos(p), \sin(p)\right)$.  
Let $0 \leq p_0 < 2\pi$ and set $\pi(p_0) = (x_0,y_0)$
where $x_0 = \cos(p_0)$ and $y_0 = \sin(p_0)$.  Then,
$$
\pi^{(1)}(p_0) = \begin{pmatrix}-y_0 \\ x_0 \end{pmatrix}
\quad \hbox{and} \quad 
\pi^{(2)}(p_0) = \begin{pmatrix}-x_0 \\ -y_0  \end{pmatrix}.
$$
Also, we have $(\pi \circ f)(p) = \left(\cos(2p), \sin(2p)\right)$ and
$$
(\pi \circ f)^{(1)}(p_0) = \begin{pmatrix} -2\left(2 x_0 y_0\right) \\
2\left(x_0^2-y_0^2\right) \end{pmatrix}
\quad \hbox{and} \quad 
(\pi \circ f)^{(2)}(p_0) = \begin{pmatrix} -4\left(x_0^2-y_0^2\right) \\
-4\left(2 x_0 y_0\right) \end{pmatrix}.
$$
Since $M \pi^{(1)}(p_0) = (\pi \circ f)^{(1)}(p_0)$ and $M \pi^{(2)}(p_0)
= (\pi \circ f)^{(2)}(p_0)$ by (\ref{EQN2.3}), we are led to the matrix
equation
$$
M \begin{pmatrix} -y_0 & -x_0 \\ x_0 & -y_0 \end{pmatrix} = 
\begin{pmatrix} -2\left(2 x_0 y_0\right) & -4\left(x_0^2-y_0^2\right) \\
2\left(x_0^2-y_0^2\right) & -4\left(2 x_0 y_0\right) \end{pmatrix}
$$
which we can solve for $M$:
$$
M(x_0,y_0) = \begin{pmatrix}4x_0^3 & -4y_0^3 \\ 2y_0\left(2x_0^2+1\right)
& 2x_0\left(2y_0^2+1\right) \end{pmatrix}.
$$
In Chapter 5, we will describe numerical experiments for this example
showing that the linearization matrix calculated by the computer is close
to this $M(x_0,y_0)$.

We next examine the special case when a two-dimensional underlying system
$f$ is reconstructed into five-dimensional space.  We make the following
specific assumptions about $f$, $\pi$, and a point $P=\pi(p)$ in
$\reals^5$.

\begin{description}
\item[(B1)] {\it $f$ maps the unit square $[0,1] \times [0,1]$ into
itself, and $f$ is $C^3$.}
\item[(B2)] {\it $\pi$ maps $[0,1] \times [0,1]$ into $\reals^5$, and
$\pi$ is $C^3$.}
\item[(B3)] {\it For this $p \in [0,1]\times[0,1]$, $P=\pi(p) \in
\reals^5$, and the set $\pi^{-1}(P)$ contains only one point, namely $p$.}
\item[(B4)] {\it The first and second order partial derivative vectors for
$\pi$ at $p$, namely
$$
\displaylines{
\pi_x(p) := \dpidx, \hskip .65in \pi_y(p) := \dpidy \cr
\pi_{xx}(p) := \ddpidxx, \qquad \pi_{xy}(p) := \ddpidxy, \qquad 
\pi_{yy}(p) := \ddpidyy \cr
}$$
are linearly independent in $\reals^5$.}
\end{description}

\noindent Properties (B3) and (B4) are generic properties in the space of
$C^3$ functions from $[0,1] \times [0,1]$ into $\reals^5$.  As in the
previous case, we obtain a seemingly stronger statement as a consequence
of (B3), (B4), and the Inverse Function Theorem:

\begin{description}
\item[(B3')] {\it There are neighborhoods $u \subseteq [0,1] \times [0,1]$
of $p$ and $U \subseteq \pi([0,1] \times [0,1])$ of $P$ such that $\pi(x)
\in U$ if and only if $x \in u$, and $\pi|_u$ is a diffeomorphism.}
\end{description}

\noindent This statement implies the obvious analog of (\ref{EQN2.1}) for
$p_i \in \reals^2$ and $P_i \in \reals^5$.

We begin this case, as before, by examining the Taylor expansions of $\pi$
and $\pi \circ f$ about a point $P=\pi(p)$ in $\reals^5$ for which (B3)
and (B4) hold.  With $h=(h_1,h_2) \in \reals^2$, any point $P+\del P$ in
$\pi(u)$ near $P$ can be written in the form
\begin{eqnarray*}
P + \del P &=& \pi\left(p+h\right) \\
&=& P + h_1 \pi_x(p) + h_2 \pi_y(p) + \half h_1^2 \pi_{xx}(p) \\
& & {} + h_1 h_2 \pi_{xy}(p) + \half h_2^2 \pi_{yy}(p) + Rem(h, p) 
\end{eqnarray*}
where $Rem(h, p)$ is the Taylor remainder term, now a vector in
$\reals^5$.  Again, there is a constant $\Ctay>0$ such that $\|Rem(h, p)\|
\leq \Ctay \|h\|^3$ for sufficiently small $h$.  The {\bf canonical
embedding basis} (for $\reals^5$) at $P$ will consist of the five
derivative vectors from (B4): $\pi_x(p)$, $\pi_y(p)$, $\pi_{xx}(p)$,
$\pi_{xy}(p)$, and $\pi_{yy}(p)$.  With respect to this basis, we can
write
\begin{equation}\label{EQN2.5}
\del P = \left( h_1, h_2, \half h_1^2, h_1 h_2, \half h_2^2 \right)_P +
Rem(h, p).
\end{equation}
Next, look at the image $F(P)=\pi(f(p))$.  The Taylor expansion of
$F(P+~\del P) = F(\pi(p+h)) = \pi(f(p+h))$ is given by
\begin{eqnarray*}
F(P+\del P) 
&=& F(P) + h_1 (\pi \circ f)_x(p) + h_2 (\pi \circ f)_y(p) + \half h_1^2
(\pi \circ f)_{xx}(p) \\
& &\quad {}+ h_1h_2 (\pi \circ f)_{xy}(p) + \half h_2^2 (\pi \circ
f)_{yy}(p) + Rem_f(h,p) 
\end{eqnarray*}
Here, $Rem_f$ is the Taylor remainder vector associated with $\pi \circ
f$, and without loss of generality, it too satisfies $\|Rem_f(h, p)\| \leq
\Ctay \|h\|^3$ for small $h$.

\begin{defn}
Assuming (B1) -- (B4), we define the {\bf Eckmann-Ruelle linearization}
$M=M(P)$ to be the unique linear map on $\reals^5$ satisfying
\begin{equation}\label{EQN2.6}
\begin{matrix}
M \pi_x(p) &=& (\pi \circ f)_x(p) \\
M \pi_y(p) &=& (\pi \circ f)_y(p) 
\end{matrix}
 \qquad \qquad 
\begin{matrix}
M \pi_{xx}(p) &=& (\pi \circ f)_{xx}(p) \\
M \pi_{xy}(p) &=& (\pi \circ f)_{xy}(p) \\
M \pi_{yy}(p) &=& (\pi \circ f)_{yy}(p)
\end{matrix} 
\end{equation}
\end{defn}

We will prove in Theorem \ref{JJ3} that this linear map $M(P)$ provides
the best local linearization of $F$ in $\reals^5$ about the base point
$P$.  The proof of Theorem \ref{JJ3} will be virtually identical to that
of Theorem \ref{JJ1} whose main idea was the systematic cancellation of
low order terms between the Taylor expansions about a point and its image.  
Up to second order, there are five distinct terms to be eliminated, each
requiring a basis element.  Thus, we embed into $\reals^5$.  Likewise, to
kill off all third-order terms in a longer expansion, we would need a
total of nine terms, and so we would embed in $\reals^9$.  This is more
restrictive than the one-dimensional case above since we cannot embed
into an arbitrary $\reals^m$.

Theorem \ref{JJ3} requires a slightly stronger notion of limit point.  
Since the underlying phase space has more than one dimension, we should be
able to approach any underlying base point from more than two directions.  
This leads to the definition of an ``approach direction.''  The idea is
that there is a line extending out from the base point and some
subsequence of points from the trajectory approach the base point along
this line.

\begin{defn}
Let $l$ be a unit vector in $\reals^m$.  A subset $S$ of $\reals^m$ has
the {\bf approach direction} $l$ at the base point $P \in \reals^m$
provided there is a sequence $\left\{Q_{k}\right\}_{k=1}^\infty$ from $S$
such that:
\begin{enumerate}
\item $Q_{k} \to P$ as $k \to \infty$, and
\item ${{\del Q_{k}} \over {\left\|\del Q_{k}\right\|}} \to l$ as $k \to
\infty$, where $\del Q_k := Q_k - P$.
\end{enumerate}
We call a collection of approach directions at $P$ {\bf distinct} provided
that no two are the same and no two are reflections through the origin.  
Multiple approach directions are not required to be linearly independent.
\end{defn}

Figure 2.2 shows three distinct approach directions at a point $P$ on a
fractal attractor in $\reals^2$.  One can find trajectory points
converging to $P$ which are on or near the intersections of the lines
$l_1$, $l_2$, and $l_3$ with the attractor.  This behavior was automatic
in the one-dimensional case, and points in hyperbolic systems will have
multiple approach directions.  It may even be the case that typical points
in arbitrary (multi-dimensional) chaotic systems have infinitely-many
distinct approach directions.

%figure2_2.ps

\begin{figure} 
\begin{center}
\scalebox{.5}{\begin{turn}{-90}\includegraphics{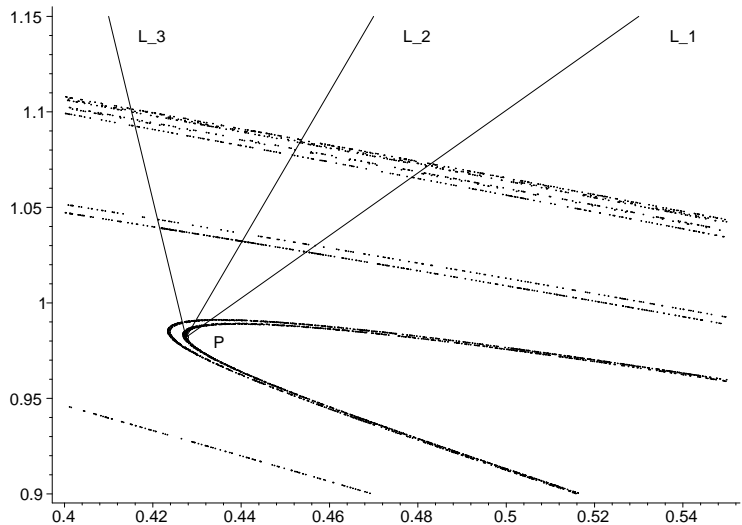}\end{turn}}
\end{center} 
\caption[Three approach directions for a fractal attractor.]{The
trajectory has three approach directions $L_1$, $L_2$, and $L_3$ at the
base point $p$.  The trajectory is dense in the fractal attractor shown,
so data points are available arbitrarily close to the intersections of
the three lines with the attractor.}
\end{figure}

It is important to be able to relate the (often fractal) geometry of the
reconstructed attractor back to the geometry of the underlying attractor.  
One reason for this, of course, is that we can only observe the geometry
of the reconstruction.  Another reason is that we need access to facts
about the underlying system in order to extract information from the
observations.  These reasons provide motivation for the next lemma whose
proof is straightforward when $\pi$ is a diffeomorphism of a neighborhood
of $P$.

\begin{lemma}\label{JJ2}
Assume (B1) and (B2).  Let $P=\pi(p)$ be a point of
$\pi([0,1]\times~[0,1])$ for which (B3) and (B4) hold.  If $P$ is in the
closure of a trajectory of $F$, $P_0, P_1, \dots \in~\reals^5$ that has
$d$ distinct approach directions at $P$, then the underlying trajectory
$p_0, p_1, \dots \in \reals^2$, where $P_i = \pi(p_i)$, has $d$ distinct
approach directions at the point $p \in [0,1]\times[0,1]$.
\end{lemma}

We prove the Local Linearization Theorem for the case where we reconstruct
a two-dimensional underlying system in $\reals^5$.  In the statement
below, we require the reconstructed trajectory in $\reals^5$ to have three
distinct approach directions at the point $P$.  We need three distinct
approach directions in order to determine the Eckmann-Ruelle
linearization.  Suppose, for instance, that the available data points
approach the base point $P \in \reals^5$ only along two directions.  By
Lemma \ref{JJ2}, the underlying attractor in $\reals^2$ will also have two
distinct approach directions.  Suppose that the data lies exactly on the
$x$- and $y$-axes (in local coordinates in $\reals^2$).  We can compute
$\pi_x(p)$, $\pi_{xx}(p)$, \dots from points of the form $(h_1,0)$.  
Likewise, we can compute $\pi_y(p)$, $\pi_{yy}(p)$, \dots from points of
the form $(0,h_2)$.  However, at each of these points, all terms involving
mixed partial derivatives vanish from the Taylor expansions, making it
impossible to compute $\pi_{xy}(p)$.  Of course, without knowledge of
$\pi_{xy}(p)$ (and where it maps), we cannot uniquely determine a best
local linearization at $P$.  This is why we require three distinct
approach directions.

\begin{thm}[Local Linearization, $\reals^2 \to \reals^5$]\label{JJ3}
Assume (B1) and (B2).  Let $P=\pi(p)$ be a point of
$\pi([0,1]\times[0,1])$ for which (B3) and (B4) hold.  If $P$ is in the
closure of a trajectory of $F$, $P_0, P_1, \dots \in \reals^5$ that has
three distinct approach directions at $P$, then the Eckmann-Ruelle
linearization $M=M(P)$ defined by (\ref{EQN2.6}) is the unique linear map
such that
\begin{eqnarray}\label{EQN2.7}
F(P + \del P) - F(P) - M \del P &=& O\!\left(\left\|\del P\right\|^3 
\right) \\
\hbox{as $\left\|\del P\right\| \to 0$}, & & \hbox{where $P+\del P \in
\pi([0,1]\times[0,1])$} \nonumber
\end{eqnarray}
\end{thm}

\PROOF
We use the same notation as in the proof of Theorem \ref{JJ1}, except that
we write $h=(h_1,h_2) \in \reals^2$.  The first thing to do is show that
the Eckmann-Ruelle linearization $M=M(P)$ satisfies (\ref{EQN2.7}).  As in
the proof of Theorem \ref{JJ1}, we use the definition of $M$ in equation
(\ref{EQN2.6}) to cancel terms from the Taylor expansions.  Thus,
\begin{eqnarray*}
\left\|F(P+\del P) - F(P) - M \del P\right\|
&=& \left\|Rem_f(h,p) - M Rem(h,p)\right\| \\
&\leq& \Ctay (1 + \|M\|) \left\|h\right\|^3
\end{eqnarray*}
for small $h$.  Therefore, by (\ref{EQN2.1}), for small $\del P$ with
$P+\del P \in \pi( [0,1] \times [0,1])$,
$$
\left\| F(P+\del P) - F(P) - M \del P \right\| \leq \Ctay (1 + \|M\|)
C_\pi^3 \left\|\del P\right\|^3,
$$
establishing (\ref{EQN2.7}).

It remains to show that $M$ is the only matrix that satisfies
(\ref{EQN2.7}).  Suppose to the contrary there is another matrix $\Mt$
such that
$$
\left\| F(P+\del P) - F(P) - \Mt \del P \right\| \leq C_1 \|\del P\|^3 
\qquad \hbox{as $\del P \to 0$, $P+\del P \in \pi(u)$}
$$
for some constant $C_1$.  Set $A:=\Mt - M$.  For small $\del P$ with
$P+\del P \in \pi(u)$,
\begin{eqnarray*}
\left\|A \del P \right\|
&\leq& \left\| F(P+\del P) - F(P) - M\del P \right\| + \left\| F(P+\del P)
- F(P) - \Mt\del P \right\| \\
&\leq& C_2 \|\del P\|^3.
\end{eqnarray*}
With respect to the canonical embedding basis at $P$, the small vectors
$\del P$ have the form shown in (\ref{EQN2.5}), and so
\begin{eqnarray*}
\left\| A \left(h_1, h_2, \half h_1^2, h_1 h_2, \half h_2^2\right)_P \right\|
&\leq& \left\| A \del P \right\| + \left\|A Rem(h, p) \right\| \\
&\leq& C_2 \|\del P\|^3 + \|A\| \left\| Rem(h, p) \right\| \\
&\leq& C_2 C_\pi^3 \left\|h\right\|^3 + \|A\| \Ctay \left\|h\right\|^3 \\
& :=& C_3 \left\|h\right\|^3 
\end{eqnarray*}
Thus, for all tangent vectors $h \in u$ sufficiently small
$$
\left\| A \left(h_1, h_2, \half h_1^2, h_1 h_2, \half h_2^2\right)_P
\right\|^2 \leq C_3^2 \left\|h\right\|^6.
$$
If we represent the matrix $A$ with respect to the canonical embedding
basis at $P$, then the left-hand side is a polynomial $p(h_1,h_2)$ with
degree at most $4$.  At last, we use the assumption that we have three
distinct approach directions.  It follows from Lemma \ref{JJ2}, that the
underlying trajectory $p_0, p_1, \dots \in \reals^2$ also has three
distinct approach directions.  This fact, together with the inequality
above, gives us precisely the hypotheses needed to apply Proposition
\ref{PP2} of Appendix A.  Thus, $p(h_1,h_2)$ must be identically zero, and
it follows that all of the coefficients of the matrix $A$ must also be
zero.  Therefore, $\Mt = M(P)$ proving that $M(P)$ is indeed the unique
matrix satisfying (\ref{EQN2.7}).
\QED

This proof extends further to the case when we embed a two-dimensional
system into higher dimensions.  To do this, we first need to embed into an
appropriate dimension.  In the proof, we cancelled all of the first-order
and second-order terms in a Taylor expansion.  In the general setting, we
want to cancel all terms of order up to and including order $D$.  There
are $\sum_{d=1}^D {{d+1} \choose 1} = \half D(D+3)$ such terms.  For each
term to be cancelled, we need a basis vector in the canonical embedding
basis.  Therefore, we can embed our two-dimensional system into any
$m$-dimensional Euclidean space where $m=\half D(D+3)$ for some $D>0$.  
For $D=2$ we have $m=5$, and for $D=3$ we have $m=9$.

We now assume that we are given some $D>1$ with $m = \half D(D+3)$ the
corresponding dimension.  We make the following specific assumptions about
$f$, $\pi$, and a point $P=\pi(p) \in \reals^m$.

\begin{description}
\item[(B5)] {\it $f$ maps the unit square $[0,1] \times [0,1]$ into
itself, and $f$ is $C^{D+1}$.}
\item[(B6)] {\it $\pi$ maps $[0,1] \times [0,1]$ into $\reals^m$, and
$\pi$ is $C^{D+1}$.}
\item[(B7)] {\it For this $p \in [0,1]\times[0,1]$, $P=\pi(p) \in
\reals^m$, and the set $\pi^{-1}(P)$ contains only one point, namely $p$.}
\item[(B8)] {\it The various partial derivative vectors for $\pi$ at $p$
of order at most $D$, namely 
\begin{equation*}
\begin{matrix}
\pi_x(p) &:=& \dpidx \\
\pi_y(p) &:=& \dpidy 
\end{matrix}
\quad
\begin{matrix}
\pi_{xx}(p) &:=& \ddpidxx \\
\pi_{xy}(p) &:=& \ddpidxy \\
\pi_{yy}(p) &:=& \ddpidyy \\
\end{matrix}
\quad \dots \quad
\begin{matrix}
\pi_{xx \ldots xx}(p) &:=& {{\partial^D \pi} \over {\partial x^D}}(p) \\
\pi_{xx \ldots xy}(p) &:=& {{\partial^D \pi} \over {\partial x^{D-1}
\partial y}}(p) \\
& \vdots & \\
\pi_{yy \ldots yy}(p) &:=& {{\partial^D \pi} \over {\partial y^D}}(p) 
\end{matrix}
\end{equation*}
are linearly independent in $\reals^m$.}
\end{description}

\noindent Properties (B7) and (B8) are generic properties in the space of
$C^{D+1}$ functions from $[0,1] \times [0,1]$ into $\reals^m$.  The $m$
vectors in (B8) form the canonical embedding basis at $P$.  As before, we
define the {\bf Eckmann-Ruelle linearization} at $P$ to be the unique
linear map $M=M(P)$ defined by the relations :
\begin{equation*}
\begin{matrix}
M \pi_x(p) &=& (\pi \circ f)_x(p) \\
M \pi_y(p) &=& (\pi \circ f)_y(p)
\end{matrix}
\qquad
\begin{matrix}
M \pi_{xx}(p) &=& (\pi \circ f)_{xx}(p) \\
M \pi_{xy}(p) &=& (\pi \circ f)_{xy}(p) \\
M \pi_{yy}(p) &=& (\pi \circ f)_{yy}(p) 
\end{matrix}
\end{equation*}
\begin{equation}\label{EQN2.8}
\cdots \qquad
\begin{matrix}
M \pi_{xx \ldots xx}(p) &=& (\pi \circ f)_{xx \ldots xx}(p) \\
M \pi_{xx \ldots xy}(p) &=& (\pi \circ f)_{xx \ldots xy}(p) \\
& \vdots & \\
M \pi_{yy \ldots yy}(p) &=& (\pi \circ f)_{yy \ldots yy}(p)
\end{matrix}
\end{equation}
Here is the general theorem for two-dimensional underlying dynamics.

\begin{thm}[Local Linearization, $\reals^2 \to \reals^m$]\label{JJ4}
Let $m=\half D(D+3)$ where $D>1$.  Assume (B5) and (B6).  Let $P=\pi(p)$
be a point of $\pi([0,1]\times[0,1])$ for which (B7) and (B8) hold.  If
$P$ is in the closure of a trajectory of $F$, $P_0, P_1, \dots
\in~\reals^m$, and if the trajectory has $D+1$ distinct approach
directions at $P$, then the Eckmann-Ruelle linearization $M=M(P)$ defined
by (\ref{EQN2.8}) is the unique linear map such that
\begin{eqnarray}\label{EQN2.9}
F(P + \del P) - F(P) - M \del P & = & O\!\left(\left\|\del P\right\|^{D+1} 
\right) \\
\hbox{as $\left\|\del P\right\| \to 0$}, & &
\hbox{where $P+\del P \in \pi([0,1]\times[0,1])$} \nonumber
\end{eqnarray}
\end{thm}

The proof of this is virtually identical to the proof of Theorem \ref{JJ3}
and is omitted.  Note that the polynomial proposition \ref{PP2} is already
stated generally enough to be used for this proof.

The Eckmann-Ruelle linearization will also exist in certain other cases,
such as when reconstructing a three-dimensional underlying system into
$\reals^9$.  The arguments above extend in a natural way to certain cases
where we have a dynamical system $f$ on $\reals^n$ being reconstructed in
a larger dimensional space $\reals^m$.  Assume that $D>1$ and $f$ and
$\pi$ each have at least $D+1$ continuous derivatives in each coordinate.  
Note that there are $\sum_{d=1}^D {{d+n-1} \choose {n-1}} = {{D+n} \choose
{n}} -1$ terms of order at most $D$ in the Taylor series of $f$ and $\pi$.  
Then, when $m = {{D+n} \choose {n}} -1$, we can construct the canonical
embedding basis as in assumption (B8) and define the Eckmann-Ruelle
linearization $M(P)$ as in (\ref{EQN2.8}).  Because (\ref{EQN2.8})
guarantees that the Taylor series will collapse nicely, it is easy to see
that $M(P)$ satisfies (\ref{EQN2.9}).  We will also need some mild
geometric condition similar to those given in the previous theorems to
guarantee that $M(P)$ is the only matrix satisfying (\ref{EQN2.9}).  We
will not pursue these ideas further at this time.

% CHAPTER 3  --  chap3.tex
%\documentclass{report}
%\begin{document}

%\input{erdmacro.tex}
\chapter{Lyapunov Exponent Formulas}

\newcommand{\lyapprod}[1]{\displaystyle\prod_{j=0}^{n-1}{#1}(p_j)}
\newcommand{\Vand}{V\!and_{\fiverm 2,5}}

Once the local linearization matrices have been computed at each point of
the reconstructed trajectory, we must extract Lyapunov exponents from
them.  Mimicking the standard definition of Lyapunov exponents (see for
example, \cite[pp.\ 31]{OSY}), we define the {\bf Eckmann-Ruelle-Lyapunov
(ERL) exponents} of the reconstructed trajectory $P_0, P_1, \dots$ in
$\reals^m$ to be the values obtained by the limit
\begin{equation}\label{EQN3.1}
h_{ER}(P_0,\nu) := \lim_{n \to \infty} {1 \over n} \ln \left\|
M_{n-1} M_{n-2} \cdots M_1 M_0\nu \right\|
\end{equation}
for unit vectors $\nu \in \reals^m$, where $M_i=M(P_i)$ is the best local
linearization (i.e., Eckmann-Ruelle linearization) at the point
$P_i=\pi(p_i)$ in the trajectory.  We need a separate definition here
because the standard definition of Lyapunov exponent uses the Jacobian
derivative, which need not exist along the trajectory.  In this chapter,
we show that the matrix product in (\ref{EQN3.1}) can be written as an
upper triangular matrix.  A straightforward calculation will then produce
a formula for the limiting values of (\ref{EQN3.1}).

In practice, the limit in (\ref{EQN3.1}) can be computed using the
treppen-iteration algorithm described in \REF{ER, EKRC} and elsewhere.  
This method uses QR matrix decomposition to convert the matrix product
$M_{n-1} \cdots M_0$ into a product of upper triangular matrices $R_{n-1}
\cdots R_0$.  The diagonal elements of the latter upper triangular matrix
are the products of the corresponding diagonal elements of the $R_i$.  
Then, we can read the Lyapunov exponents right from the diagonals of the
intermediate matrices $R_i$:
$$
\lambda_j = \lim_{n \to \infty} {1 \over n} \sum_{i=0}^{n-1} \ln \left|
\left(R_i\right)_{jj} \right|.
$$
Theorem \ref{TS2} justifies the ability to read the exponents directly
from the diagonal in this way.  A proof of the theorem will be given in
Appendix B.

\begin{defn}
A sequence of real numbers $\left\{r_n\right\}$ has {\bf (geometric)
growth rate $\gamma$} provided
$$
\lim_{n \to \infty} {{\ln |r_n|} \over n} = \gamma.
$$
\end{defn}

\setcounter{timchap}{\value{chapter}}
\setcounter{timthm}{\value{thm}}

\begin{thm}\label{TS2}
For $k=1,2 \dots$, let $A_k$ be $m \times m$ upper triangular matrices,
and define $S_n = A_n \cdots A_1$.  Assume the magnitudes of the entries
of $A_k$ are bounded independent of $k$, and that the diagonal entries of
$S_n$ have growth rates $\gamma_1, \dots, \gamma_m$ as $n \to \infty$.  
Then there exist vectors $v_1, \dots, v_m \in \reals^m$ such that for each
$i=1, \dots, m$, $\left\| S_n v_i \right\|$ has growth rate $\gamma_i$.
\end{thm}

The next theorem and its two-dimensional analog, Theorem \ref{JJ9}, are
the main results of this paper.  In both theorems, most of the hypotheses
are used to guarantee that the Eckmann-Ruelle linearization exists at each
point of the trajectory.

\begin{thm}[Lyapunov Exponent Formula, $\reals^1 \to \reals^m$]
\label{JJ6} \ 
Assume (A1), (A2) and that the trajectory of $f$ in $[0,1]$,
$p_0,
p_1,
\dots$, has Lyapunov exponent $\lambda$.  If each point of the
reconstructed trajectory, $P_i=\pi(p_i) \in \reals^m$, satisfies (A3) and
(A4) and is a limit point of the trajectory, then the reconstructed
trajectory has Eckmann-Ruelle-Lyapunov exponents $\lambda, 2\lambda,
3\lambda, \dots, m\lambda$.
\end{thm}

In order to prove this, a few lemmas will be useful.  Lemma \ref{JJ7} is a
technical lemma leading to the matrix representation for $M(P)$ given in
Lemma \ref{JJ8}.  With this matrix representation, we will prove Theorem
\ref{JJ6}.

\begin{lemma}\label{JJ7}
Let $J$ be an interval, and let $k,d$ be positive integers.  If $f:J \to
J$ is a $C^{k+1}$ function, then for $1 \leq j \leq i \leq k$ there are
differentiable scalar functions $a_{ij}(t)$, such that for any $C^{k+1}$
function $\phi:J \to \reals^d$, $\phi(t)=\left( \phi_1(t), \dots,
\phi_d(t)\right)$, we have
\begin{equation}\label{EQN3.2}
\left(\phi \circ f\right)^{(i)}(t) = \sum_{j=1}^{i} a_{ij}(t)
\phi^{(j)}(f(t)), \qquad \hbox{ for $i=1,\dots,k$}
\end{equation}
where $\phi^{(j)}(t) = \left({{d^j \phi_1} \over {dt^j}}(t),\dots, {{d^j
\phi_d} \over {dt^j}}(t)\right)$ represents the $j$-th derivative vector
of $\phi(t)$.
\end{lemma}

Note that the scalar functions $a_{ij}(t)$ depend only on $f$ and not on
the other function $\phi$.  As will be seen in the proof, these functions
$a_{ij}(t)$ are the results of collecting terms involving the derivatives
of $\phi$.

\THM{Proof of Lemma \ref{JJ7}}
Since vector-valued functions of a real-variable are naturally
differentiated componentwise, it is enough to prove the lemma for any one
component.  Thus, we may assume $d=1$ and $\phi:J \to \reals$.

The proof is by induction.  For $i=1$, we apply the Chain Rule:
$$
\left(\phi \circ f\right)^{(1)}(t) = f'(t) \phi^{(1)}(f(t))
$$
and note that $a_{11}(t):=f'(t)$ is $C^k$.  For $i=2$, we apply the chain
rule again:
$$
\left(\phi \circ f\right)^{(2)}(t) = f'(t)^2 \phi^{(2)}(f(t)) +
f''(t)\phi^{(1)}(f(t)).
$$
Note that $a_{21}(t) := f''(t)$ and $a_{22}(t) := \left(f'(t)\right)^2$
are both $C^{k-1}$.

For induction, suppose that (\ref{EQN3.2}) holds for $i$ where the
$a_{ij}(t)$ are $C^{k+1-i}$, depending only on $f$, not on $\phi$.  We
differentiate (\ref{EQN3.2}) and collect together terms involving the
derivatives of $\phi$.  Specifically, if we define the functions
$a_{(i+1)j}(t)$ using
\begin{eqnarray}\label{EQN3.3}
a_{(i+1)1}(t) &:=& a'_{i1}(t) \nonumber \\
a_{(i+1)j}(t) &:=& a'_{ij}(t) + a_{i(j-1)}(t)f'(t), \qquad j=2,\dots,i{-}1
\nonumber \\
a_{(i+1)i}(t) &:=& i f'(t)^{i-1}f''(t) + a_{i(i{-}1)}(t)f'(t) \\
a_{(i+1)(i+1)}(t) &:=& f'(t)^{i+1} \nonumber
\end{eqnarray}
then we obtain from differentiating (\ref{EQN3.2}):
\begin{equation*}
\begin{split}
\left(\phi \circ f\right)^{(i+1)}(t)
&= {d \over {dt}} \left(f'(t)^i \phi^{(i)} (f(t)) + \sum_{j=1}^{i-1}
a_{ij}(t) \phi^{(j)}(f(t))\right) \\
&= i f'(t)^{i-1} f''(t) \phi^{(i)}(f(t)) + f'(t)^{i+1} \phi^{(i+1)}(f(t))\\
& \qquad + \sum_{j=1}^{i-1} \left(a_{ij}'(t) \phi^{(j)}(f(t)) +a_{ij}(t)
f'(t) \phi^{(j+1)}(f(t)) \right) \\
&= \sum_{j=1}^{i+1} a_{(i+1)j}(t) \phi^{(j)}(f(t)).
\end{split}
\end{equation*}
Note that the functions $a_{(i+1)j}(t)$ are $C^{k+1-(i+1)}$ and that they
depend only on the function $f$, not on the other function $\phi$.  This
completes the proof.
\QED

\begin{lemma}\label{JJ8}
Assume (A1) and (A2).  Let $P=\pi(p) \in \reals^m$ and let $Q = F(P)$ be
points satisfying (A3) and (A4) in the closure of a trajectory of $F$.
Let $\alpha$ and $\beta$ represent the canonical embedding bases at $P$
and $Q$, respectively.  With respect to these bases, the Eckmann-Ruelle
linearization $[M(P)]_{\alpha}^{\beta}$ is upper triangular with diagonal
elements $M(P)_{jj}=f'(p)^j$.
\end{lemma}

\PROOF
Recall that the canonical embedding basis at a point $P=\pi(p)$ in
$\reals^m$ consists of the vectors $\pi^{(1)}(p), \dots, \pi^{(m)}(p)$ in
$\reals^m$.  Set $M=M(P)$ and $q=f(p)$.  By the definition of the
Eckmann-Ruelle linearization (\ref{EQN2.3}), we have $M \pi^{(i)}(p) =
(\pi \circ
f)^{(i)} (p)$.  We must write the vectors $(\pi \circ f)^{(i)} (p)$ in
terms of the canonical embedding basis at $Q = \pi(q)$.  Applying Lemma
\ref{JJ7} with $k=d=m$ and $\phi=\pi$, we have
$$
M \pi^{(i)}(p) = \left(\pi \circ f\right)^{(i)}(p)
= f'(p)^i \pi^{(i)} (q) + \sum_{j=1}^{i-1} a_{ij}(p) \pi^{(j)} (q)
\qquad \hbox{for $i=1, \dots, m$}
$$
because $f(p)=q$.  This provides a representation for $M$ with respect to
the canonical embedding bases $\alpha$ and $\beta$:
\begin{equation*}
[M]_{\alpha}^{\beta} = 
\begin{pmatrix}
 f'(p) & a_{21}(p) & a_{31}(p) & \cdots & a_{k1}(p) \\
  0    &  f'(p)^2  & a_{32}(p) & \cdots & a_{k2}(p) \\
  0    &     0     &  f'(p)^3  & \cdots & a_{k3}(p) \\
\vdots &  \vdots   &  \vdots   & \ddots &  \vdots   \\
  0    &     0     &     0     & \cdots &  f'(p)^m  
\end{pmatrix}.
\end{equation*}
\QED

It is important to note that this matrix form for the Eckmann-Ruelle
linearization is completely independent of the embedding $\pi$.  
Information about $\pi$ is used to form the coordinate system with respect
to which we view the dynamics; however, once inside that coordinate
system, we only see the action of $f$.

\THM{Proof of Theorem \ref{JJ6}}
For each point $P_i \in \reals^m$, the Eckmann-Ruelle linearization $M_i =
M(P_i)$ exists.  To compute the ERL exponents, we must evaluate the limit
(\ref{EQN3.1}).  Let $\beta_i$ denote the canonical embedding basis at
$P_i$.  Notice that the canonical embedding bases for the upper triangular
matrix representation fit together perfectly:
$$
\left[M_{i+1}\right]_{\beta_{i+1}}^{\beta_{i+2}}
\left[M_i\right]_{\beta_i}^{\beta_{i+1}} =
\left[M_{i+1}M_i\right]_{\beta_i}^{\beta_{i+2}}.
$$
It follows that the matrix product $M_{n-1} M_{n-2} \cdots M_1 M_0$ is
upper triangular when expressed with respect to the canonical embedding
bases $\beta_0$ and $\beta_n$.  That is, the matrix product $\left[M_{n-1}
M_{n-2} \cdots M_1 M_0\right]_{\beta_0}^{\beta_n}$ can be written
$$
\begin{pmatrix}
\lyapprod{f'} &  b_{12}  &  b_{13}  & \cdots &  b_{1k} \\
0  & \left(\lyapprod{f'}\right)^2 &  b_{23}  & \cdots &  b_{2k} \\
0  &  0  & \left(\lyapprod{f'}\right)^3 & \cdots &  b_{3k} \\
\vdots & \vdots  &  \vdots  & \ddots & \vdots  \\
0  &  0  &   0   &  \cdots  & \left(\lyapprod{f'}\right)^m 
\end{pmatrix}
$$
where the $b_{ij}$ are numbers which depend solely on the underlying
dynamical system $f$ and the first $n$ points of the trajectory of $p_0$
in $[0,1]$.

We apply Theorem \ref{TS2} to complete the proof.  Note that the elements
of each $M_i$ (with respect to the appropriate coordinate systems) are
combinations of the first $m$ derivatives of $f$, each of which is
continuous.  These elements will be bounded independent of $i$ on any
compact set containing the entire trajectory.  By hypothesis, the Lyapunov
exponent $\lambda$ of $f$ is given by
$$
\lambda = \lim_{n \to \infty} {1 \over n} \ln \left|\lyapprod{f^\prime}\right|,
$$
and it is clear that the diagonal entries of the matrix product $M_{n-1}
\cdots M_0$ have growth rates $\lambda, 2\lambda, \dots, m\lambda$.  
Theorem \ref{TS2} now provides the vectors which grow at these
characteristic rates, and we conclude that the ERL exponents are $\lambda,
2\lambda, \dots, m\lambda$.
\QED

With this theorem, we expect that whenever the underlying system is
one-dimensional with Lyapunov exponent $\lambda$, we will compute
exponents $\lambda, 2\lambda, \dots, m\lambda$ in the absence of noise.  
Note that if the system has a positive Lyapunov exponent (and hence is
chaotic), then the true exponent will be the {\it smallest} of the
computed numbers!

Next, we consider the case where the underlying system is two-dimensional
and reconstructed in $\reals^5$.  The basic arguments are similar though
the situation is more involved, as we shall see.  The formula for the
computed Lyapunov exponents in this case is given in the next theorem.

\begin{thm}[Lyapunov Exponent Formula, $\reals^2 \to \reals^5$]
\label{JJ9} \ 
Assume (B1), \ \linebreak[3] (B2), and that the trajectory of $f$ in
$[0,1]\times[0,1]$,
$p_0, p_1, \dots$, has Lyapunov exponents $\lambda$ and $\mu$.  Assume
each point of the reconstructed trajectory, $P_i=\pi(p_i)$ in $\reals^5$,
satisfies (B3) and (B4), and the set $\left\{P_0, P_1, P_2,
\dots
\right\}$ in $\reals^5$ has at least three distinct approach directions at
each $P_i$.  Then, the reconstructed trajectory has
Eckmann-Ruelle-Lyapunov exponents $\lambda$, $\mu$, $2\lambda$, $\lambda +
\mu$, and $2\mu$.
\end{thm}

\PROOF
Recall that the canonical embedding basis given in assumption (B4) from
Chapter 2 consists of the first and second order partial derivative
vectors of $\pi$, namely $\pi_x(p)$, $\pi_y(p)$, $\pi_{xx}(p)$,
$\pi_{xy}(p)$, and $\pi_{yy}(p)$.  In fact, the construction of the
Eckmann-Ruelle linearization from the Taylor series of $f$ and $\pi$ can
be carried out with respect to any orthonormal set of coordinates.  The
uniqueness part of the Local Linearization Theorem \ref{JJ3} ensures that
the resulting linear map will be the same.  Thus, we may introduce
convenient local coordinate systems at each point $p_i \in \reals^2$ of
the trajectory of $p_0$.

Without loss of generality, we assume $\lambda \geq \mu$.  We begin by
choosing a unit Lyapunov vector $v_0 \in \reals^2$ corresponding to the
exponent $\mu$ at the initial point $p_0 \in \reals^2$ of the underlying
trajectory.  That is, we choose a unit vector $v_0$ such that
$$
\lim_{n \to \infty} {1 \over n} \ln \left\| Df_{p_{n-1}} Df_{p_{n-2}}
\cdots Df_{p_1} Df_{p_0} v_0 \right\| = \mu.
$$

By Oseledec's Multiplicative Ergodic Theorem \REF{ER}, almost every other
vector in $\reals^2$ has growth rate $\lambda$.  Choose a unit vector
$w_0$ perpendicular to $v_0$.  This gives us an orthonormal basis
$\left\{v_0, w_0\right\}$ for $\reals^2$ based at $p_0$.  In general,
given the basis $\left\{v_n, w_n\right\}$ at $p_n \in \reals^2$, we
construct the basis at $p_{n+1}$ by setting $v_{n+1} := Df_{p_n} v_n /
\|Df_{p_n} v_n\|$ and choosing $w_{n+1}$ to be the unit vector
perpendicular to $v_{n+1}$ that satisfies $\langle w_{n+1}, Df_{p_n} w_n
\rangle > 0$.  Thus, at each point $p_n$, we have an orthonormal basis for
$\reals^2$.  For each $n$, we write points $p' \in \reals^2$ near $p_n$ as
$p'=(x,y)$ provided $p' = p_n + x v_n + y w_n$.  Now, we can describe the
underlying dynamics $f$ near $p_n$ in terms of these bases at $p_n$ and
$p_{n+1}$ by $f(x,y) = (g(x,y), h(x,y))$.  Though we will not explicitly
show it, this representation for $f$ depends on the base point $p_n$, and
the reader should keep in mind that the component functions $g$ and $h$
may look very different as we vary $p_n$.  Note that $h_x(p_n) = 0$
because $Df_{p_n} v_n = \|Df_{p_n} v_n\| v_{n+1}$ has no $y$-component at
$p_{n+1}$.  Thus, we can write the Jacobian derivative of $f$ at $p_n$ as:
$$
Df_{p_n} = 
\begin{pmatrix} g_x(p_n) & g_y(p_n) \\ 0 & h_y(p_n) \end{pmatrix}.
$$

Following the ideas in the previous case, a calculation using the chain
rule produces the next set of equations as the analog of (\ref{EQN3.2}),
where we write \mbox{$q_n=f(p_n)$} and the partial derivatives of $g$ and
$h$ are evaluated at $p_n$:
\begin{eqnarray*}
(\pi \circ f)_x(p_n)    &=& g_x \pi_x(q_n) \\
(\pi \circ f)_y(p_n)    &=& g_y \pi_x(q_n) + h_y \pi_y(q_n) \\
(\pi \circ f)_{xx}(p_n) &=& g_{xx} \pi_x(q_n) + h_{xx} \pi_y(q_n) + g_x^2
\pi_{xx}(q_n) \\
(\pi \circ f)_{xy}(p_n) &=& g_{xy} \pi_x(q_n) + h_{xy} \pi_y(q_n) + g_x
g_y \pi_{xx}(q_n) + g_x h_y \pi_{xy}(q_n) \\
(\pi \circ f)_{yy}(p_n) &=& g_{yy} \pi_x(q_n) + h_{yy} \pi_y(q_n) + g_y^2
\pi_{xx}(q_n) + 2g_y h_y \pi_{xy}(q_n) + h_y^2 \pi_{yy}(q_n)
\end{eqnarray*}
For each $n$, let $\beta_n$ denote the canonical embedding basis for
$\reals^5$ at $P_n=\pi(p_n)$.  Representing the Eckmann-Ruelle
linearization with respect to $\beta_n$ and $\beta_{n+1}$ as we did in
Lemma \ref{JJ8}, we obtain this matrix representation for $M_n=M(P_n)$:
$$
[M_n]_{\beta_n}^{\beta_{n+1}} = 
\begin{pmatrix}
g_x & g_y &  g_{xx} &  g_{xy} &  g_{yy} \\
 0  & h_y &  h_{xx} &  h_{xy} &  h_{yy} \\
 0  &  0  &  g_x^2  & g_x g_y &  g_y^2  \\
 0  &  0  &    0    & g_x h_y & 2g_yh_y \\
 0  &  0  &    0    &    0    &  h_y^2  
\end{pmatrix}
$$
All of the terms above the diagonal come from combinations of first and
second derivatives of $f$, and, because $f$ is $C^3$ by hypothesis, they
will be bounded independent of $p_n$.  Note that the upper left $2 \times
2$ block is the Jacobian $Df_{p_n}$ of the underlying dynamics and that
the lower $3 \times 3$ block contains only combinations of terms from
$Df_{p_n}$.

As in the proof of Theorem \ref{JJ6}, the product $\left[M_{n-1} \cdots
M_0\right]_{\beta_0}^{\beta_n}$ is upper triangular when written with
respect to the canonical embedding bases:
$$
\begin{pmatrix}
\lyapprod{g_x} & b_{12} & b_{13} & b_{14} & b_{15} \\
0 & \lyapprod{h_y} & b_{23} & b_{24} & b_{25} \\
0 & 0 & \left(\lyapprod{g_x}\right)^2 & b_{34} & b_{35} \\
0 & 0 & 0 & \lyapprod{g_x(p_j)h_y} & b_{45} \\
0 & 0 & 0 & 0 & \left(\lyapprod{h_y}\right)^2 
\end{pmatrix}
$$
where the $b_{ij}$ are numbers which depend solely on the underlying
dynamical system $f$ and the first $n$ points of the trajectory of $p_0$
in $[0,1]\times[0,1]$.  Now, we need to compute the growth rates of the
diagonal terms.  To do this, we compute the Lyapunov exponents $\lambda$
and $\mu$ of the underlying dynamical system in terms of the components of
the Jacobian matrices.  Since $Df_{p_n} v_n = \|Df_{p_n} v_n\| v_{n+1}$,
we have $g_x(p_n) = \|Df_{p_n} v_n\|$.  It follows that
\begin{equation}\label{EQN3.4}
\begin{split}
\lim_{n \to \infty} {1 \over n} \ln\left(\prod_{i=0}^{n-1} g_x(p_n)\right)
&= \lim_{n \to \infty} {1 \over n} \ln\left(\prod_{i=0}^{n-1} \|Df_{p_i}
v_i\|\right) \\
&= \lim_{n \to \infty} {1 \over n} \ln\left\|Df_{p_{n1}} \cdots Df_{p_0}
v_0 \right\| \\
&= \mu 
\end{split}
\end{equation}
because $v_0$ was chosen to be a Lyapunov vector for $\mu$.  Recall from
\cite[pp.\ 632]{ER} that the growth rate of areas is given by the
sum of the
Lyapunov exponents.  Thus,
\begin{equation*}
\begin{split}
\mu + \lambda
&= \lim_{n \to \infty} {1 \over n} \ln
\left|\det\left(Df^n{}_{p_0}\right)\right| \\
&= \lim_{n \to \infty} {1 \over n} \ln \left(\prod_{i=0}^{n-1}
\left|\det\left(Df_{p_i}\right)\right|\right) \\
&= \lim_{n \to \infty} {1 \over n} \ln \left(\prod_{i=0}^{n-1}
\left|g_x(p_n) h_y(p_n)\right|\right) \\
&= \mu + \lim_{n \to \infty} {1 \over n}
\ln\left(\prod_{i=0}^{n-1}\left|h_y(p_n)\right|\right)
\end{split}
\end{equation*}
and it follows that
\begin{equation}\label{EQN3.5}
\lambda=\lim_{n \to \infty} {1 \over
n}\ln\left(\prod_{i=0}^{n-1}\left|h_y(p_n)\right|\right).
\end{equation}

Finally, we see that the diagonal terms have growth rates of $\lambda$,
$\mu$, $2\lambda$, $\lambda + \mu$, and $2\mu$.  It follows from Theorem
\ref{TS2} that the reconstructed trajectory, $P_0, P_1, \dots$ in
$\reals^5$ has these ERL exponents, completing the proof of Theorem
\ref{JJ9}.
\QED

%\end{document}

% CHAPTER 4  --  chap4.tex
%\documentclass{report}
%\begin{document}

%\input{erdmacro.tex}
\chapter{Convergence Theory}

\newcommand{\Vandm}{V\!and_m}
\newcommand{\ETE}{E^T\!E}
\newcommand{\Cpert}{C_{Pert}}

In Chapter 2, we showed that, at least in theory, the local linearization
matrices are not derivatives.  The formulas derived in Chapter 3 for the
Lyapunov exponents are meaningful only if the numerically determined
linearization matrices are close to the Eckmann-Ruelle linearizations
$M(P)$.  Because $M(P)$ is guaranteed to be the best linearization {\it in
the limit} as the neighborhood radius shrinks to zero (see Theorems
\ref{JJ1} and \ref{JJ3}), it is possible that for a particular radius, a
different matrix will do ``better'' than the Eckmann-Ruelle linearization,
though we hope the ``better'' matrix will still be close to $M(P)$.  In
this chapter, we prove that, over small neighborhoods around the base
point $P$, the best linearization matrix will in fact be close to $M(P)$.

To proceed, we must specify what we mean by the ``best'' linearization
over a small neighborhood.  That is, we need a means of measuring the
error involved in the linearization process.  The following definition
serves this purpose by finding the worst-case error committed by a matrix
$L$ used as the local linearization.

\begin{defn}
Given $F:\reals^m \to \reals^m$, an trajectory of $F$, $P_0, P_1, \dots
\in \reals^m$, and $P \in \reals^m$ in the closure of the trajectory.  
For an $m \times m$ matrix $L$ and $\epsilon > 0$, define
$$
W(L,P,\epsilon) := \sup_{\|P_i-P\| \leq \epsilon} \left\| P_{i+1} - F(P)
- L(P_i - P) \right\|
$$
where the supremum is taken over those values of $i$ for which $\| P_i - P
\| \leq \epsilon$.  For matrices $L_1$ and $L_2$, we say that $L_1$ is a
{\bf better linearization} than $L_2$ over the $\epsilon$-ball about $P$
provided $W(L_1,P,\epsilon) < W(L_2,P,\epsilon)$.
\end{defn}

In practice, one needs to avoid false-nearest-neighbors by taking the
supremum over $i$ where both $\| P_i - P \| \leq \epsilon$ and $\|
P_{i+1}-F(P) \| \leq \epsilon$.  However, this is not necessary for the
theory because $\pi$ is one-to-one in a neighborhood of $p$ (by either
(A3') or (B3')) and we will let $\epsilon \to 0$.  Hence, we can assume
$\epsilon$ is small enough that the false-nearest-neighbor problem never
arises.

We point out that the matrix minimizing $W(\cdot, P, \epsilon)$ is not
necessarily the best least squares fit but rather the best ``minimax''
fit.  The least squares problem seems harder to formulate theoretically.  
At the end of this chapter, we present preliminary results for a
least squares theory that incorporates infinite data.

In the statements of Theorems \ref{JJ10} and \ref{JJ15} below, we assume
that we have a sequence of ``best'' linearizations $M_k$ over a shrinking
set of $\epsilon_k$-balls.  This is the interpretation we give to the
condition $W(M_k,P,\epsilon_k) \leq W(M(P),P,\epsilon_k)$. Thus, each
matrix $M_k$ does better than $M(P)$ over the $\epsilon_k$-ball around $P$
though not necessarily over any other ball around $P$.

As before, we begin with the one-dimensional case where $f:[0,1] \to
[0,1]$ and $\pi:[0,1] \to \reals^m$.  We say the {\bf reconstructed
trajectory is dense in the curve} at the point $P \in \pi([0,1])$ provided
there is a neighborhood $U$ of $P$ such that the trajectory $\{P_i: i \geq
0\}$ is dense in the curve $\pi([0,1]) \cap U$.

\begin{thm}[Convergence Theorem, $\reals^1 \to \reals^m$]\label{JJ10}
Assume (A\!1) and (A2).  Let $P \in \pi([0,1])$ be a point satisfying (A3)
and (A4) where the reconstructed trajectory is dense in the curve.  Let
$\{\epsilon_k\}_{k=1}^\infty$ be a decreasing sequence of positive
numbers, $\epsilon_k \to 0$.  Let $\{M_k\}_{k=1}^\infty$ be a sequence of
matrices for which we have $W(M_k,P,\epsilon_k) \leq W(M(P),P,\epsilon_k)$
for all $k$.  Then $M_k \to M(P)$ as $k \to \infty$. 
\end{thm}

The key point for the proof is to control the size of $\left\| M_k - M(P)
\right\|$ as $\epsilon_k \to 0$.  This control will be achieved via the
next proposition.  Recall that a convex set contains every line segment
connecting any two of its points.  That is, $S$ is {\bf convex} provided
that whenever $u,v \in S$ and $t \in [0,1]$ we have $tu + (1-t)v \in S$.  
The {\bf convex hull} of a set $S$ is the smallest convex set containing
$S$.

\begin{prop}\label{JJ11}
Let $S$ be a subset of $\reals^m$, and let $Hull(S)$ denote its convex
hull.  Assume that $Hull(S)$ contains a closed $m$-dimensional ball of
radius $r$.  If $A$ is an $m \times m$ matrix such that $\left\|Av\right\|
\leq B$ for all $v \in S$, then $\left\|A\right\| \leq {{2B} \over r}$.
\end{prop}

\PROOF
Note that $\left\|Av\right\| \leq B$ actually holds for all $v \in
Hull(S)$.  Let $\eta$ be a unit vector such that $\left\|A\eta\right\| =
\left\|A\right\|$.  Since $Hull(S)$ contains the closed ball
$\overline{B(c,r)}$ for some $c \in Hull(S)$, there exists $z \in Hull(S)$
on the surface of this ball for which the vector $z-c$ points in the
direction of $\eta$.  Therefore, $z-c = r\eta$, and we have
$$
r\left\|A\right\| = r\left\|A\eta\right\| = \left\|A(z-c)\right\| \leq
\left\|Az\right\| + \left\|Ac\right\| \leq 2B.
$$
\QED

Here is how we will use the proposition.  We let $A_k$ be the matrix
$M_k - M(P)$ and $S_k$ the set of small tangent vectors from the
$\epsilon_k$-ball about $P$.  The bound on $\left\|Av\right\|$ will come
from the condition $W(M_k,P,\epsilon_k) \leq W(M(P),P,\epsilon_k)$.  We
will then be able to conclude a bound on $\left\|M_k - M(P)\right\|$ in
terms of $\epsilon_k$.  In order to proceed, then, we must examine the
convex hull of the tangent vectors.

Recall that when written in the canonical embedding basis at the base
point $P$, little tangent vectors $\del P_i:=P_i - P$ have the form shown
in (\ref{EQN2.2}).  This suggests that we consider curves in $\reals^m$ of
the form
\begin{equation}\label{EQN:v1m}
V(h)=\left(h, {1 \over 2}h^2, \dots, {1 \over {m!}}h^m\right) + Pert(h)
\end{equation}
where $Pert(h)$ is a continuous vector-valued function with $Pert(0)=0$.  
Let $Hull_\epsilon(V)$ denote the convex hull of the set $\{V(h): h \in
[-\epsilon,\epsilon]\}$.

\begin{figure}   
\begin{center}
\scalebox{.7}{\begin{turn}{-90}\includegraphics{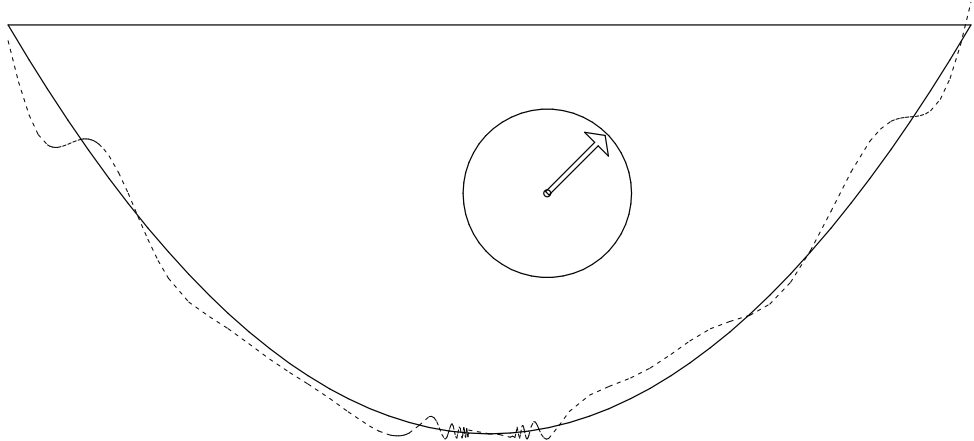}\end{turn}}  
\end{center}
\caption[The convex hull of the curve contains a small ball.]{The solid
curve outlines the convex hull of the curve $(h,\half h^2)$ for $h \in
[-1,1]$.  No small perturbation $V(h)$ (dashed curve) can significantly
change the interior of this convex hull.  Thus, the convex hull of $V(h)$
will always contain a ball of radius $C_{rad}$ .}
\end{figure}

We need some additional notation.  For $i=0,1,\dots,m$, define $x_i = {i
\over m}$.  Let $\Vandm$ denote the $m \times m$ Vandermonde matrix:
$$
\Vandm = \begin{pmatrix}
1      & x_1    & x_1^2  & \dots  & x_1^{m-1} \\
1      & x_2    & x_2^2  & \dots  & x_2^{m-1} \\
\vdots & \vdots & \vdots & \vdots & \vdots    \\
1      & x_m    & x_m^2  & \dots  & x_m^{m-1} 
\end{pmatrix}.
$$
Since determinants vary continuously with the elements of the matrix,
there is $\epsilon_0>0$ such that if we perturb the elements of $\Vandm$
by no more than $\epsilon_0$ then the determinant of the perturbed matrix,
$\Vandm + E$, is still close to the determinant of the unperturbed matrix
$\Vandm$:
$$
\hbox{if $|E_{ij}| \leq \epsilon_0$ for all $i$ and $j$, then }
\det\left(\Vandm + E\right) \geq \half \det\left(\Vandm \right) > 0.
$$

\begin{lemma}\label{JJ12}
There exists a constant $C_{rad}>0$ such that for any curve $V(h)$ of the
form (\ref{EQN:v1m}) satisfying $\max_{-1 \leq h \leq 1}
\|Pert(h)\| \leq {{\epsilon_0} \over {m \ m!}}$, the convex hull
$Hull_1(V)$ contains a ball of radius $C_{rad}$.
\end{lemma}

\PROOF
The $m+1$ points $V(x_i)$ in $\reals^m$, $i=0,1,\dots,m$, form an
$m$-simplex entirely contained (along with its interior) inside
$Hull_1(V)$.  Note that $V(x_0)$ is the zero vector.  Treating the
$V(x_i)$ as row-vectors and writing $\xi_{ij}$ for the $j$th component of 
$Pert(x_i)$, the volume of this simplex is given by
\cite[pp.\ 44]{Aitken}:
\begin{multline*}
{1 \over {m!}} \det 
\begin{vmatrix}V(x_1) \\ V(x_2) \\ \vdots \\ V(x_m) \end{vmatrix}
= {1 \over {m!}} \det
\begin{vmatrix}
x_1 + \xi_{11} & {1 \over 2}x_1^2 + \xi_{12} & \dots & {1 \over
{m!}}x_1^m + \xi_{1m} \\
x_2 + \xi_{21} & {1 \over 2}x_2^2 + \xi_{22} & \dots & {1 \over
{m!}}x_2^m + \xi_{2m} \\
\vdots & \vdots & \vdots & \vdots \\
x_m + \xi_{m1} & {1 \over 2}x_m^2 + \xi_{m2} & \dots & {1 \over
{m!}}x_m^m + \xi_{mm}
\end{vmatrix}
\\
\begin{split}
&= {1 \over {m!}} {{m!} \over {m^m}} \left( \prod_{k=1}^m {1 \over {k!}}
\right) \det
\begin{vmatrix}
1 + {1 \over {x_1}}\xi_{11} & x_1 + {2 \over {x_1}}\xi_{12} & \dots &
x_1^{m-1} + {{m!} \over {x_1}}\xi_{1m} \\
1 + {1 \over {x_2}}\xi_{21} & x_2 + {2 \over {x_2}}\xi_{22} & \dots &
x_2^{m-1} + {{m!} \over {x_2}}\xi_{2m} \\
\vdots & \vdots & \vdots & \vdots  \\
1 + {1 \over {x_m}}\xi_{m1} & x_m + {2 \over {x_m}}\xi_{m2} & \dots &
x_m^{m-1} + {{m!} \over {x_m}}\xi_{mm}
\end{vmatrix} \\
&\geq {1 \over {m^m}} \left( \prod_{k=1}^m {1 \over {k!}} \right) \half
\det(\Vandm) > 0
\end{split}
\end{multline*}
because the perturbations of the elements of $\Vandm$ satisfy:
$$
\left| {{j!} \over {x_i}}\xi_{ij} \right| \leq {{m!} \over {x_1}} \left|
\xi_{ij} \right| \leq m \ m!  \|Pert(x_i)\| \leq \epsilon_0.
$$
Since the vectors $V(x_i)$ cannot vary too far from their ``unperturbed
positions'' at the points $\left(x_i, {1 \over 2}x_i^2, \dots, {1 \over
{m!}}x_i^m\right)$, and since the $m$-simplex they form has an absolute
lower bound on its volume, there must be some radius $C_{rad}>0$ such that
this $m$-simplex always contains a ball of that radius, no matter what the
perturbation function $Pert(h)$.
\QED

Note that the vectors $V(x_i)$ and the Vandermonde matrix used in this
proof produce a simplex with very small volume.  Indeed, other choices for
the $x_i$ could produce larger volumes and hence larger ``guaranteed''
radii $C_{rad}$.  Fortunately, we do not need the radius to be large, only
that such a guaranteed radius does in fact exist.

\begin{prop}\label{JJ13}
Let $\Cpert > 0$ and $0 < \epsilon \leq \min\left(1, {{\epsilon_0} \over
{m \ m! \Cpert}}\right)$.  For any curve $V(h)$ of the form
(\ref{EQN:v1m}) with
$\left\|Pert(h)\right\| \leq \Cpert |h|^{m+1}$ defined on $[-\epsilon,
\epsilon]$, $Hull_\epsilon(V)$ contains a ball of radius $C_{rad}
\epsilon^m$, where $C_{rad}$ is independent of $\epsilon$ and $V(h)$.
\end{prop}

\THM{Sketch of Proof}
This is a consequence of changing coordinate systems between the standard
basis $\left\{e_1,e_2,\dots,e_m\right\}$ and the basis $\left\{\epsilon
e_1,\epsilon^2 e_2,\dots,\epsilon^m e_m\right\}$.  Then, as $h$ ranges
over $[-\epsilon,\epsilon]$, ${h \over \epsilon}$ ranges over $[-1,1]$.  
Together, the conditions on $\epsilon$ and $Pert(h)$ imply that the
perturbation term is still small enough to use Lemma \ref{JJ12}.  This
guarantees the existence of a ball of radius $C_{rad}$ contained within
the convex hull of $V\left({h \over \epsilon}\right)$ over $[-1,1]$.  
Naturally, this ball has axes of length $C_{rad}$ in each of the principal
directions of the second basis.  When we convert coordinates back to the
standard basis, the ball becomes an ellipsoid with axes of length $C_{rad}
\epsilon^i$ for $i=1,2,\dots,m$.  In particular, the ellipsoid contains a
ball of radius $C_{rad} \epsilon^m$.
\QED

\THM{Proof of Theorem \ref{JJ10}}
The hypotheses guarantee that the Eckmann-Ruelle linearization $M(P)$
exists and is the unique best linearization as $\epsilon_k \to 0$.  We
will show that there is a constant $C$ (independent of $k$) such that
$\|M_k - M(P)\| \leq C \epsilon_k$ for large $k$.

By Theorem \ref{JJ1}, there are $\epsilon^\prime > 0$ and $C_{er}
>0$ such that for every $0 < \epsilon < \epsilon^\prime$ we have
$W(M(P),P,\epsilon) \leq C_{er} \epsilon^{m+1}$.  Without loss of
generality, $\epsilon_1 < \min\{\epsilon^\prime,1\}$.  Then,
\begin{equation}\label{EQN4.1}
W(M_k,P,\epsilon_k) \leq W(M(P),P,\epsilon_k) \leq C_{er}
\epsilon_k^{m+1} \quad \hbox{ for $k=1,2,\dots$}
\end{equation}

First, we construct a suitable set $S_k$.  With respect to the canonical
embedding basis at $P$, $\del P_i := P_i - P = \left(h_i, \half h_i^2,
\dots, {1 \over {m!} }h_i^m\right)_P + Rem(h_i).$ Without loss of
generality we may assume that $Rem(h)$ is defined on $[-\epsilon_1,
\epsilon_1]$.  Define the function $V(h)$ on $[-\epsilon_1, \epsilon_1]$
by equation (\ref{EQN:v1m}) with $Pert(h_i) = Rem(h_i)$, and note that
$\del P_i = V(h_i)$.  Since the reconstructed trajectory is dense in the
curve at $P$, we may assume that the points $P_i$ in $B(P,\epsilon_1)$
trace out the curve \mbox{$\pi([0,1]) \cap B(P,\epsilon_1)$}.  Because
$\pi$ is a diffeomorphism, these points $P_i \in \reals^m$ pull back to
scalars $p_i \in \reals$ that are dense in some subinterval of $[0,1]$.  
By (\ref{EQN2.1}), this interval contains the subinterval
$\left[p-{{\epsilon_1} \over {C_\pi}},p+{{\epsilon_1} \over
{C_\pi}}\right]$.  It follows that the $h_i = p_i-p$ are dense in the
interval $I_1 = \left[-{{\epsilon_1} \over {C_\pi}},{{\epsilon_1} \over
{C_\pi}}\right]$.  By similar reasoning, for each $k$, the preimages of
$P_i \in B(P,\epsilon_k) \cap \pi([0,1])$ are dense in the interval
$\left[p-{{\epsilon_k} \over {C_\pi}},p+{{\epsilon_k} \over
{C_\pi}}\right]$, and the corresponding $h_i$ are dense in $I_k =
\left[-{{\epsilon_k} \over {C_\pi}},{{\epsilon_k} \over {C_\pi}}\right]$.  
Define the set \mbox{$S_k := \left\{V(h): h \in I_k\right\}$} in
$\reals^m$.  The convex hull of $S_k$ is $Hull_{{\epsilon_k} \over
{C_\pi}}(V)$, and by Proposition \ref{JJ13}, it contains a ball of radius
${{C_{rad}} \over {C_\pi^m}} \epsilon_k^m$ whenever $\epsilon_k \leq
\min\left(1, {{\epsilon_0} \over {m \ m! \Ctay}}\right)$.

For each $k$, let $A_k:=M_k - M(P)$ be the $m \times m$ error matrix.  
Then, for each $i$ with $\left\|\del P_i\right\| \leq \epsilon_k$, we have
by (\ref{EQN4.1})
\begin{equation*}
\begin{split}
\left\|A_k \del P_i\right\| 
& \leq \left\|P_{i+1} - F(P) - M(P) \del P_i\right\| + \left\|P_{i+1} -
F(P) - M_k \del P_i\right\| \\
& \leq W(M(P),P,\epsilon_k) + W(M_k,P,\epsilon_k) \\
& \leq 2C_{er} \epsilon_k^{m+1}
\end{split}
\end{equation*}
Because $\del P_i = V(h_i)$, the $h_i$ are dense in $I_k$, and $V(h)$ is
continuous on this interval, this inequality extends to all $h \in I_k$.  
Therefore,
$$
\left\|A_k v\right\| \leq 2C_{er}\epsilon_k^{m+1} 
\qquad
\hbox{ for all $v \in S_k$,}
$$
and, by Proposition \ref{JJ11},
$$
\left\|A_k\right\| \leq {{4 C_{er} C_\pi^m} \over {C_{rad}}}
\epsilon_k.
$$
This completes the proof.
\QED

Before we proved Theorem \ref{JJ10} in the one-dimensional case, we added
an assumption about the distribution of the orbit within the reconstructed
attractor that was not needed in previous chapters.  In particular, we
assumed that, in a neighborhood of the base point, the orbit was dense
along the curved segment of the reconstructed attractor.  Unfortunately, a
density assumption is much too strong for the case of two-dimensional
underlying dynamics because our reconstructed attractor could be fractal
in nature.

It turns out that we need only add a third condition to the definition of
``approach direction.'' This condition ensures that the distances between
the base point and points approaching along the direction vector shrink
exponentially fast.

\begin{defn}
Let $l$ be a unit vector in $\reals^m$.  A subset $S$ of $\reals^m$ has
the {\bf fractal approach direction} $l$ at the base point $P \in
\reals^m$ if there is a sequence $\left\{Q_{k}\right\}_{k=1}^\infty$ from
$S$ such that:
\begin{enumerate}
\item $Q_{k} \to P$ as $k \to \infty$,
\item ${{\del Q_{k}} \over {\left\|\del Q_{k}\right\|}} \to l$ as $k \to
\infty$, where $\del Q_k := Q_k - P$, and 
\item there are constants $0 < C^- \leq C^+ < 1$ such that for all $k$
\begin{equation}\label{EQN4.2}
C^- \leq {{\left\|\del Q_{k+1}\right\|} \over {\left\|\del Q_k\right\|}}
\leq C^+.
\end{equation}
\end{enumerate}
\noindent A collection of fractal approach directions at $P$ is {\bf
distinct} provided that no two are the same and no two are reflections
through the origin.  Multiple fractal approach directions are not
required to be linearly independent.
\end{defn}

As noted in Chapter 2 (see Figure 2.2), this kind of behavior occurs for
points in hyperbolic systems and may occur in all chaotic systems with two
or more dimensions.

We need a lemma like \ref{JJ2} that translates fractal approach directions
in the reconstruction space down to fractal approach directions in the
underlying phase space.

\begin{lemma}\label{JJ14}
Assume (B1) and (B2).  Let $P=\pi(p)$ be a point of
\mbox{$\pi([0,1]\times[0,1])$} for which (B3) and (B4) hold.  If $P$ is in
the
closure of a trajectory of $F$, $P_0, P_1, \dots \in \reals^5$ that has
$d$ distinct fractal approach directions at $P$, then the underlying
trajectory $p_0, p_1, \dots \in \reals^2$, where $P_i = \pi(p_i)$, has $d$
distinct fractal approach directions at the point $p \in
[0,1]\times[0,1]$.
\end{lemma}

\PROOF
By Lemma \ref{JJ2}, we get $d$ distinct approach directions in $\reals^2$.  
Thus, we need only verify the third condition for each one.  Let $l$ be an
approach direction in $\reals^2$ coming from a fractal approach direction
in $\reals^5$.  Let $Q_k=\pi(q_k)$, $k=1,2,\dots$, be the points in
$\reals^5$ forming the subsequence that converges to this fractal approach
direction.  By the previous comment, $q_k \to p$ and ${{q_k-p} \over
{\left\|q_k-p\right\|}} \to l$.  We can use the constants $C^-, C^+$ from
condition (3) together with (\ref{EQN2.1}) to translate (3) to:
$$
{{C^-} \over {C_\pi^2}} \leq {{\left\|q_{k+1} - p\right\|} \over
{\left\|q_k-p\right\|}} \leq C^+ C_\pi^2.
$$
If $C^+ C_\pi^2 <1$, we are done.  Otherwise, we have to come up with new
bounds.  We form a new subsequence from $\{q_k\}_{k=1}^\infty$ as follows.  
Start with $\hat q_1 = q_1$.  Then, having picked $\hat q_k = q_i$ for
some $i$, we must pick $\hat q_{k+1}$.  Look back to the original sequence
$\{q_k\}$ and set $\hat q_{k+1}$ to be the first term after $q_i$, say
$q_{i+j}$ with $j>0$, for which $\left\|q_{i+j} - p\right\| < C^+
\left\|q_i - p\right\|$.  Note that $\left\|q_{i+j-1} - p\right\| \geq C^+
\left\|q_i - p\right\|$.  Then,
$$
C^+ > {{\left\|\hat q_{k+1} - p\right\|} \over {\left\|\hat q_k -
p\right\|}} = {{\left\| q_{i+j} - p\right\|} \over {\left\| q_{i+j-1} -
p\right\|}} \ {{\left\| q_{i+j-1} - p\right\|} \over {\left\| q_i -
p\right\|}} \geq {{C^-} \over {C_\pi^2}} \ C^+.
$$
These bounds now satisfy (3) of the definition.  Of course, since $\{\hat
q_k\}$ is a subsequence of $\{q_k\}$, we have $\hat q_k \to p$ and ${{\hat
q_k - p} \over {\left\|\hat q_k -p \right\|}} \to l$ as well.
\QED

The Convergence Theorem below applies only for the case of a
reconstruction in $\reals^5$, but the extensions to appropriate
higher-dimensional cases should be clear.

\begin{thm}[Convergence Theorem, $\reals^2 \to \reals^5$]\label{JJ15}
Assume (B1) and (B2).  Let $P=\pi(p)$ be a point of
$\pi([0,1]\times[0,1])$ for which (B3) and (B4) hold.  Assume that $P$ is
in the closure of a trajectory of $F$, $P_0, P_1, \dots \in \reals^5$ that
has three distinct fractal approach directions at $P$.  Let
$\{\epsilon_k\}_{k=1}^\infty$ be a decreasing sequence of positive
numbers, $\epsilon_k \to 0$.  Let $\{M_k\}_{k=1}^\infty$ be a sequence of
matrices such that $W(M_k,P,\epsilon_k) \leq W(M(P),P,\epsilon_k)$ for all
$k$.  Then $M_k \to M(P)$ as $k \to \infty$.
\end{thm}

This proof will be similar to that given in the one-dimensional case over
the course of Lemmas \ref{JJ11}, \ref{JJ12}, and \ref{JJ13}.  The present
situation will be far more technically involved, however, because the
geometry is no longer simple.  In the previous case, we had only a single
approach direction and it was reasonable to assume the trajectory was
dense along the line.  These properties are not generic when the
underlying dynamics have two or more dimensions.  (See Figure 2.2.)  In
the present case, the local distribution of the trajectory can be sparse,
approaching only along three fractal approach directions.  Recall from the
discussion in Chapter 2 that three approach directions are necessary to
ensure that we can uniquely determine the Eckmann-Ruelle linearization.

In Lemmas \ref{JJ12} and \ref{JJ13}, we singled out a configuration of
points in the underlying space (easily chosen thanks to the density
assumption) whose reconstructed images had a convex hull that could be
guaranteed to contain a small ball.  We will do the same thing here,
though we have to be more careful.  Our points are no longer nicely
constrained to a line (they converge along three approach directions in
the plane), and we cannot choose any points we want (there is no density
assumption).  We will choose five points $z_i$ situated near our three
fractal approach directions $l_i$ at $P$ as in Figure 4.2.  We will use
the spacing between points $z_i$ that (\ref{EQN4.2}) guarantees in the
same way that we used the guaranteed spacing between $x_i = {i \over m}$
and $x_{i-1} = {{i-1} \over m}$ in the one-dimensional case.

\begin{figure}   
\begin{center}
\scalebox{.6}{\begin{turn}{-90}\includegraphics{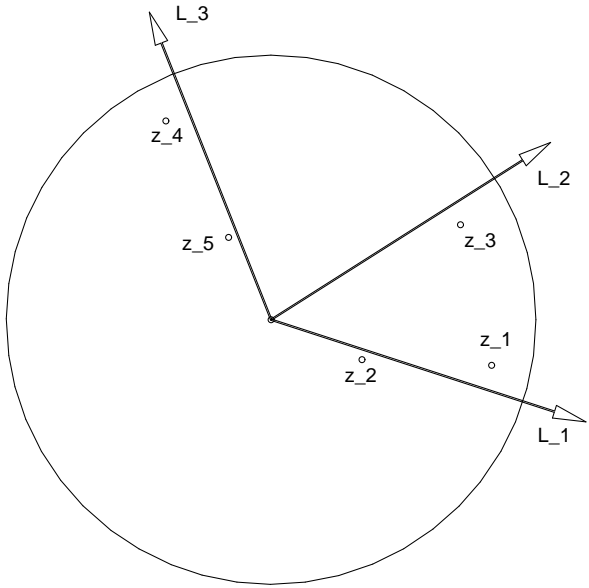}\end{turn}}  
\end{center}
\caption[A useful configuration of points along fractal approach
directions.]{For this configuration of points $z_1, \dots, z_5$ in
$\reals^2$, the convex hull in $\reals^5$ of the set $\left\{\pi(z_1),
\dots, \pi(z_5)\right\}$ is guaranteed to contain a ball of radius
$C_{rad}$.}
\end{figure}

To be more specific, for $i=1,2,3$, let $l_i =
\left(\cos(\alpha_i),\sin(\alpha_i)\right)$ be the fractal approach
directions.  Without loss of generality, we can label the $l_i$ to satisfy
\begin{gather}\label{EQN:alpha}
0 \leq \alpha_i < 2\pi \qquad \hbox{for $i=1,2,3$} \notag \\
0 < \alpha_2 - \alpha_1 < \pi \\
0 < \alpha_3 - \alpha_2 < \pi \notag
\end{gather}
Each $l_i$ will have associated constants $C_i^-$ and $C_i^+$ from
(\ref{EQN4.2}), but we can easily find constants $0 < C^- \leq C^+ < 1$
that work for all three $l_i$ simultaneously.  We want to consider
quintuples of points $z_i = \left(r_i \cos(\theta_i), r_i
\sin(\theta_i)\right) \in \reals^2$, $1 \leq i \leq 5$, that satisfy these
constraints for some $\eta>0$:
\begin{equation}\label{EQN4.3}
\begin{matrix}
C^- \leq r_i \leq 1 & i=1,3,4 \\
C^- \leq {{r_{i+1}} \over {r_i}} \leq C^+ & i=1,4
\end{matrix}
\qquad\qquad\qquad
\begin{matrix}
\left| \theta_i - \alpha_1 \right| \leq \eta & i=1,2 \\
\left| \theta_3 - \alpha_2 \right| \leq \eta & \\
\left| \theta_i - \alpha_3 \right| \leq \eta & i=4,5
\end{matrix}
\end{equation}
We will also consider functions $V:\reals^2 \to \reals^5$ of the form 
\begin{equation}\label{EQN:v25}
V(x,y) = \left(x,y,\half x^2, xy, \half y^2\right) + Pert(x,y)
\end{equation}
where $Pert(x,y)$ is a continuous vector-valued function.  In the
following lemma, the function $V$ plays the role of our measurement
function $\pi$, written in canonical embedding coordinates.

\begin{lemma}\label{JJ16}
Given constants $0< C^- \leq C^+ <1$ and $\alpha_1, \alpha_2, \alpha_3$
satisfying (\ref{EQN:alpha}), there exist positive constants $\eta_0$,
$\rho$, and $C_{rad}$ such that for \par
\begin{enumerate}
\item  any five points $z_i \in \reals^2$, $1 \leq i \leq 5$,
satisfying
(\ref{EQN4.3}) with $0 < \eta \leq \eta_0$, and 
\item  any function $V:\reals^2 \to \reals^5$ of the form (\ref{EQN:v25}) 
with $Pert(0,0)=0$ and $\left\|Pert(x,y)\right\| \leq \rho$ for all
$\left\|(x,y)\right\| \leq 1$,
\end{enumerate}
the convex hull in $\reals^5$ of $\left\{V(0), V(z_1),
\dots, V(z_5)\right\}$ contains a ball of radius $C_{rad}$.
\end{lemma}

\PROOF
Given the five points $z_i \in \reals^2$ satisfying (\ref{EQN4.3}), we
define
the $5 \times 5$ matrix $N = N(r_1, \dots, r_5, \theta_1, \dots,
\theta_5)$:
$$
N = \begin{pmatrix}
r_1\cos(\theta_1) & r_1\sin(\theta_1) & \half r_1^2\cos^2(\theta_1) &
r_1^2\cos(\theta_1)\sin(\theta_1) & \half r_1^2\sin^2(\theta_1) \\
r_2\cos(\theta_2) & r_2\sin(\theta_2) & \half r_2^2\cos^2(\theta_2) &
r_2^2\cos(\theta_2)\sin(\theta_2) & \half r_2^2\sin^2(\theta_2) \\
r_3\cos(\theta_3) & r_3\sin(\theta_3) & \half r_3^2\cos^2(\theta_3) &
r_3^2\cos(\theta_3)\sin(\theta_3) & \half r_3^2\sin^2(\theta_3) \\
r_4\cos(\theta_4) & r_4\sin(\theta_4) & \half r_4^2\cos^2(\theta_4) &
r_4^2\cos(\theta_4)\sin(\theta_4) & \half r_4^2\sin^2(\theta_4) \\
r_5\cos(\theta_5) & r_5\sin(\theta_5) & \half r_5^2\cos^2(\theta_5) &
r_5^2\cos(\theta_5)\sin(\theta_5) & \half r_5^2\sin^2(\theta_5) 
\end{pmatrix}$$
and a function $\mu$:
$$
\mu(r_1, \dots, r_5, \theta_1, \dots, \theta_5) = {1 \over {5!}} \det
N(r_1, \dots, r_5, \theta_1, \dots, \theta_5).
$$
Note that the constraints in (\ref{EQN4.3}) imply
\begin{alignat*}{2}
r_2 &\geq C^- r_1   &   r_1 - r_2 &\geq (1 - C^+)r_1 \\
r_5 &\geq C^- r_4   &   r_4 - r_5 &\geq (1 - C^+)r_4
\end{alignat*}
so that 
\begin{equation*}
\begin{split}
\lefteqn{\mu(r_1,r_2,r_3,r_4,r_5,\alpha_1,\alpha_1,\alpha_2, 
\alpha_3,\alpha_3)} \\
& = {1 \over {4 \cdot (5!)}} r_1 r_2 r_3^2 r_4 r_5 (r_1-r_2)
(r_4-r_5) \sin(\alpha_2-\alpha_1) \sin^2(\alpha_3-\alpha_1)
\sin(\alpha_3-\alpha_2) \\
& \geq {1 \over {480}} \left(C^-\right)^{10} (1-C^+)^2
\sin(\alpha_2-\alpha_1) \sin^2(\alpha_3-\alpha_1)
\sin(\alpha_3-\alpha_2)\\
& := B(C^-,C^+,\alpha_1,\alpha_2,\alpha_3)
\end{split}
\end{equation*}
Since determinants vary continuously with the matrix elements, there
exists $\eta_0$ such that if the $z_i$ satisfy (\ref{EQN4.3}) with
$0 < \eta \leq \eta_0$, we have:
\begin{equation*}
\begin{split}
\mu(r_1, r_2, r_3, r_4, r_5, \theta_1, \theta_2, \theta_3,
\theta_4, \theta_5)
& \geq \half \mu(r_1,r_2,r_3,r_4,r_5,
\alpha_1,\alpha_1,\alpha_2,\alpha_3,\alpha_3) \\
& \geq \half B(C^-,C^+,\alpha_1,\alpha_2,\alpha_3)
\end{split}
\end{equation*}
In particular, when $0 < \eta \leq \eta_0$, we can bound $\mu$ from below
independent of $r_i$, $\theta_i$, and $\eta$.  It follows that there
exists $\rho>0$ such that if we perturb any matrix
$N=N(r_1,\dots,r_5,\theta_1,\dots,\theta_5)$ by an error matrix
$E=(E_{ij})$ with $|E_{ij}| \leq \rho$, then $\det (N+E) \geq \half
\det(N)$.

Since $V(0,0)=0$, the simplex formed in $\reals^5$ from the six points
$V(0)$, $V(z_1)$, \dots, $V(z_5)$ has five-dimensional volume given by the
determinant \cite[pp.\ 44]{Aitken}:
\begin{equation*}
{1 \over {5!}} \det \begin{pmatrix}V(z_1) \\ \vdots \\ V(z_5) \end{pmatrix}
\geq \half {1 \over {5!}} \det \left(N(r_1,\dots,r_5, \theta_1, \dots,
\theta_5)\right)
\geq {1 \over 4} B(C^-,C^+,\alpha_1,\alpha_2,\alpha_3)
\end{equation*}
whenever the $z_i$ satisfy (\ref{EQN4.3}) with $0<\eta \leq \eta_0$ and
$\|Pert(x,y)\| \leq \rho$.  Thus, the convex hull of these points has a
guaranteed minimum volume.  Since the positions of the points $V(z_i)$ are
constrained in space, there is a constant $C_{rad}>0$ such that there will
always be a small ball of radius $C_{rad}$ contained somewhere within this
convex hull.
\QED

\begin{lemma}\label{JJ17}
Let $C^-$, $C^+$, $\alpha_1$, $\alpha_2$, $\alpha_3$, $C_{rad}$, $\eta_0$,
and $\rho$ be as in Lemma \ref{JJ16}.  Let $\Cpert >0$ and $0 < \epsilon
\leq \min\left(1, {\rho \over {\Cpert}}\right)$.  Then, for 
\begin{enumerate}
\item any five points $z_i = \left(\epsilon r_i \cos(\theta_i), \epsilon
r_i \sin(\theta_i)\right)$ in $\reals^2$, $1 \leq i \leq 5$, with $r_i$
and $\theta_i$ satisfying (\ref{EQN4.3}) for some $0 < \eta \leq \eta_0$,
and
\item any function $V:\reals^2 \to \reals^5$ of the form (\ref{EQN:v25}) 
with 
$$
\left\|Pert(x,y)\right\| \leq \Cpert \left\|(x,y)\right\|^3 
\qquad \hbox{ for all $\left\|(x,y)\right\| \leq \epsilon$},
$$
\end{enumerate}
the convex hull in $\reals^5$ of $\left\{V(0), V(z_1),
\dots, V(z_5)\right\}$ contains a ball of radius $C_{rad}\epsilon^2$.
\end{lemma}

\SKETCH
This proof is similar to that of Lemma \ref{JJ13} and follows from
changing coordinate systems between the standard basis $\{e_1, e_2, e_3,
e_4, e_5\}$ and the basis $\{\epsilon e_1, \epsilon e_2, \epsilon^2 e_3,
\epsilon^2 e_4, \epsilon^2 e_5\}$.
\QED

\THM{Proof of Theorem \ref{JJ15}}
The hypotheses guarantee that the Eckmann-Ruelle linearization $M(P)$
exists and is the unique best linearization as $\epsilon_k \to 0$.  We
will show that there is a constant $C$ (independent of $k$) such that
$\left\|M_k - M(P)\right\| \leq C \epsilon_k$ for large $k$.

By Theorem \ref{JJ3}, there are $\epsilon^\prime > 0$ and $C_{er}
>0$ such that for every $0 < \epsilon < \epsilon^\prime$ we have
$W(M(P),P,\epsilon) \leq C_{er} \epsilon^3$.  Without loss of
generality, $\epsilon_1 < \min\{\epsilon^\prime,1\}$.  Then,
$$
W(M_k,P,\epsilon_k) \leq W(M(P),P,\epsilon_k) \leq C_{er} \epsilon_k^3 
\quad \hbox{ for $k=1,2,\dots$}
$$
For each $k$, let $A_k := M_k - M(P)$ and let $S_k$ be the set of small
tangent vectors at $P$:
$$
S_k := \left\{P_j - P \in \reals^5 : P_j=\pi(p_j) \hbox{ and }
\left\|p_j-p\right\| \leq {{\epsilon_k} \over {C_\pi}}\right\}.
$$
For each $P_j \in S_k$, $\left\|P_j-P\right\| \leq \epsilon_k$ by
(\ref{EQN2.1}).  As in the proof of Theorem \ref{JJ10}, we see that
$$
\left\|A_k v\right\| \leq 2 C_{er} \epsilon_k^3 \qquad \hbox{for
each $v \in S_k$.}
$$
Thus, we need only show that the convex hull $Hull(S_k)$ contains a
ball in $\reals^5$.  We will use Lemma \ref{JJ17} to accomplish this.

With respect to the canonical embedding basis at $P$, we have for $P_i$
near $P$:
$$
P_i - P = \left( h_{i1}, h_{i2}, \half h_{i1}^2, h_{i1} h_{i2}, \half
h_{i2}^2 \right)_P + Rem(h_{i1}, h_{i2}, p)
$$
where $p_i-p = (h_{i1},h_{i2}) \in \reals^2$ and $Rem(h_{i1}, h_{i2}, p)$
is a vector-valued function continuous in a neighborhood containing the
closed ball of radius ${{\epsilon_1} \over {C_\pi}}$ about (0,0).  Define
the function $V(h_1,h_2)$ as in Lemma \ref{JJ17}, with $Pert(h_1,h_2) =
Rem(h_1,h_2,p)$.  Note that $P_i-P=V(h_{i1},h_{i2})$ and we can take
$\Cpert = \Ctay$ since $\left\|Rem(h_{i1},h_{i2},p)\right\| \leq \Ctay
\left\|(h_{i1},h_{i2})\right\|^3$ (see Chapter 2).  It remains to
construct the five points in $\reals^2$ that Lemma \ref{JJ17} requires.

By hypothesis, the attractor has three distinct fractal approach
directions, and by Lemma \ref{JJ14}, these translate down to distinct
fractal approach directions $l_1, l_2, l_3$ at $p \in \reals^2$ in the
two-dimensional underlying attractor.  These approach directions in
$\reals^2$ are unit vectors, so we write $l_i = \left(\cos(\alpha_i),
\sin(\alpha_i)\right)$ for $i=1,2,3$.  Since they are distinct approach
directions, the angles $\alpha_i$ are distinct and no two differ by a
multiple of $\pi$.  Moreover, we can arrange the $l_i$ so that they
satisfy (\ref{EQN:alpha}).  For each $l_i$, there is a sequence of trajectory
points in $\reals^2$, $q_{ij} = p_{n(i,j)}, j=1,2,\dots$, satisfying the
definition of fractal approach direction for that $l_i$.  Thus, there are
constants $0 < C^- \leq C^+ < 1$ such that
\begin{equation}\label{EQN4.4}
C^- \leq {{\left\| q_{i(j+1)}-p \right\|} \over {\left\| q_{ij}-p
\right\|}} \leq C^+
\qquad \hbox{for each $i=1,2,3$ and $j=1,2,\dots$} 
\end{equation}
where the constants $C^-$ and $C^+$ work for all three fractal approach
directions.  At this point, we have constants $C^-$, $C^+$, $\alpha_1$,
$\alpha_2$, and $\alpha_3$, and we obtain the constants $C_{rad}$,
$\eta_0$ and $\rho$ from Lemma \ref{JJ16}.

It will be convenient to write the points $q_{ij}$ in coordinates with
origin at $p$: $q_{ij}-p = \left(r_{ij} \cos(\theta_{ij}),r_{ij}
\sin(\theta_{ij})\right)$.  From condition (2) of the definition of
fractal approach direction, we have ${{q_{ij}-p} \over
{\left\|q_{ij}-p\right\|}} \to l_i$ as $j \to \infty$, which translates
naturally to $\theta_{ij} \to \alpha_i$.  We may now choose a number $J$
large enough that for any $i=1,2,3$ and $j \geq J$, $\left|\theta_{ij} -
\alpha_i\right| \leq \eta_0$.  Let $K$ denote the least index $k$ for
which ${{\epsilon_k} \over {C_\pi}} \leq \min\left(1, {\rho \over
{\Cpert}}\right)$ and such that for each $i=1,2,3$, we have
$\max\left\{\left\|q_{ij}-p\right\| : j \geq J\right\} > {{\epsilon_k}
\over {C_\pi}}$.

Now, let $k \geq K$ and consider the ${{\epsilon_k} \over {C_\pi}}$-ball
about $p$.  Let $q_{1j}$, $j \geq J$, be the first point in the sequence
converging to $l_1$ for which $\|q_{1j} - p\| \leq {{\epsilon_k} \over
{C_\pi}}$.  Set \mbox{$z_1 = q_{1j}-p = \left(r_1 \cos(\theta_1), r_1
\sin(\theta_1)\right)$}.  Since $\|q_{1(j-1)} -p\| > {{\epsilon_k} \over
{C_\pi}}$, it follows from (\ref{EQN4.4}) that $r_1 = \|q_{1j}-p\| \geq
{{\epsilon_k} \over {C_\pi}} C^-$.  Let $z_2 = q_{1(j+1)}-p = \left(r_2
\cos(\theta_2), r_2 \sin(\theta_2)\right)$ correspond to the next point in
the sequence for $l_1$.  Thus, $C^- \leq {{r_2} \over {r_1}} \leq C^+$.  
Similarly, we choose points $z_4$ and $z_5$ for $l_3$.  We choose $z_3$
from $l_2$ in the same way we chose $z_1$ and $z_4$, i.e., to be the first
point along the subsequence converging to $l_2$ within the radius of
${{\epsilon_k} \over {C_\pi}}$.  We do not need a closer point along
$l_2$.  Figure 4.2 shows the configuration we have constructed.

Let $\hat r_i = {{C_\pi} \over {\epsilon_k}} r_i$ and write
$z_i = \left( {{\epsilon_k} \over {C_\pi}} \hat r_i \cos(\theta_i),
{{\epsilon_k} \over {C_\pi}} \hat r_i \sin(\theta_i) \right)$ for
$i=1,\dots,5$. By construction, the $\hat r_i$ and $\theta_i$ satisfy
(\ref{EQN4.3}) with $\eta=\eta_0$.  Thus, we can apply Lemma \ref{JJ17} to
conclude that the convex hull of the points $V(0) = P$, $V(z_1)$, \dots,
$V(z_5)$ contains a five-dimensional ball of radius ${{C_{rad}} \over
{C_\pi^2}} \epsilon_k^2$.  Of course, each point $V(z_i)$ corresponds to
some vector $P_j-P$ in $S_k$, and it follows that $Hull(S_k)$ contains
this small ball.  Finally, by Proposition \ref{JJ11}, we conclude that $$
\left\|A_k\right\| \leq {{4 C_{er} C_\pi^2} \over {C_{rad}}} \epsilon_k
\qquad \hbox{ for $k \geq K$.} $$ \QED

We turn our attention now to formulating our convergence theorems in terms
of least squares estimates rather than minimax estimates.  As before, we
need a way of comparing matrices, i.e., a way of measuring the error
involved in the linearization process over a particular $\epsilon$-ball.  
In practice, this is done by summing the squares of the error terms.  The
techniques of least squares can then determine the matrix which minimizes
this sum of squared errors.  Usually the data set is finite.  However, we
are interested in a convergence question which, by its very nature,
requires an infinite data set.  Thus, we must find a reasonable notion for
least squares that allows us to consider infinite data.  The arguments we
give now are intended to motivate the definition that will follow.

Imagine that we have exactly $N$ trajectory points $P_0, \dots, P_{N-1}$
in $\reals^m$ where $F(P_i)=P_{i+1}$, and we want to determine the best
local linearization in the least squares sense over an
$\epsilon$-neighborhood about $P$.  We would then be looking for a matrix
$M$ which minimizes
$$
\sum_{\left\|P_i-P\right\| \leq \epsilon} \left\| P_{i+1}-F(P) - M (P_i-P)
\right\|^2
$$
where the sum is taken over only those trajectory points with
$\left\|P_i-P\right\| \leq \epsilon$.  Any matrix which minimizes the
quantity above will also minimize
$$
{1 \over {N^\prime}} \sum_{i=0}^{N-1} \left\| P_{i+1}-F(P) - M (P_i-P)
\right\|^2 \chi_{B(P,\epsilon)}(P_i)
$$
where $N^\prime$ is the number of data points $\epsilon$-close to $P$.  
This quantity represents the average of the squared errors.  The
characteristic function $\chi_{B(P,\epsilon)}$ is used to selectively pick
off just those values which are close to the base point $P$.  For the
infinite data case, we take the limit as $N \to \infty$, being careful to
adjust the value of $N^\prime$ for each value of $N$:
\begin{equation}\label{EQN4.5}
\lim_{N \to \infty} {1 \over {N^\prime}} \sum_{i=0}^{N-1} \left\|
P_{i+1}-F(P) - M (P_i-P)\right\|^2 \chi_{B(P,\epsilon)}(P_i)
\end{equation}
where
$$
N^\prime = \sum_{i=0}^{N-1} \chi_{B(P,\epsilon)}(P_i).
$$
The number $N^\prime$ counts those trajectory points $P_i$ (among the
first $N$ iterates) which are $\epsilon$-close to $P$.  We will rewrite
(\ref{EQN4.5}) using the Birkhoff Ergodic Theorem \REF{KH}.  
In what follows, we let $A$ be the attractor of the underlying system
and $\mu$ 
its natural measure (an invariant, ergodic probability measure).  Set
\mbox{$I(P,\epsilon) = A \cap~\pi^{-1}(B(P,\epsilon))$}.  For a
matrix $L$, we define the differentiable function
$$
\Phi_L(x,p):= \left\|\pi(f(x)) - \pi(f(p)) - L\left(\pi(x) - \pi(p)\right)
\right\|^2.
$$
For almost every $p_0$ (with respect to $\mu$), the limit (\ref{EQN4.5})
can be written:
\begin{equation*}
\begin{split}
\lim_{N \to \infty} {{{1 \over N}\sum_{i=0}^{N-1} \Phi_L(f^i(p_0),p)
\chi_{B(P,\epsilon)}(\pi(f^i(p_0)))} \over {{1 \over N}\sum_{i=0}^{N-1}
\chi_{B(P,\epsilon)}(\pi(f^i(p_0)))}}
&= {{\int_A \Phi_L(x,p) \chi_{B(P,\epsilon)}(\pi(x)) d\mu(x)}
\over {\int_A \chi_{B(P,\epsilon)}(\pi(x)) d\mu(x)}} \\
&= {1 \over {\mu(I(P,\epsilon))}} \int_{I(P,\epsilon)}
\Phi_L(x,p) d\mu(x)
\end{split}
\end{equation*}
Since the denominator $\mu(I(P,\epsilon))$ is independent of the matrix
$L$, any matrix which minimizes the limit in (\ref{EQN4.5}) also minimizes
the quantity
$$
LS(L,P,\epsilon) := \left(\int_{I(P,\epsilon)} \Phi_L(x,\pi^{-1}(P))
d\mu(x)\right)^\half.
$$
We say the matrix $L_1$ is a {\bf better linearization} of $F$ over the
$\epsilon$-ball about $P$ than the matrix $L_2$ provided
$LS(L_1,P,\epsilon) < LS(L_2,P,\epsilon)$.

For the Eckmann-Ruelle matrix $M(P)$ of a reconstruction from
$\reals^1$ into $\reals^m$, it follows from the proof of the Local
Linearization Theorem \ref{JJ1} that
\begin{multline*}
\int_{I(P,\epsilon)} \Phi_{M(P)}(x,\pi^{-1}(P)) d\mu(x) \\
\begin{split}
& = \int_{I(P,\epsilon)} \left\|Rem_f(x-p) - M(P) Rem(x-p)\right\|^2
d\mu(x) \\
& \leq \Ctay^2 \left(1+\|M(P)\|\right)^2 \int_{I(P,\epsilon)} |x-p|^{2m+2}
d\mu(x) \\
& \leq \Ctay^2 \left(1+\|M(P)\|\right)^2 C_\pi^{2m+2} \epsilon^{2m+2}
\mu\left(I(P,\epsilon)\right)
\end{split}
\end{multline*}
since $|x-p|\leq C_\pi\left\|\pi(x)-P\right\| \leq C_\pi \epsilon$ by
(\ref{EQN2.1}).  Because the natural measure $\mu$ is a probability measure,
\begin{equation}\label{EQN4.6}
LS(M(P),P,\epsilon) \leq C \epsilon^{m+1}.
\end{equation}
We can do better still if $\mu$ is a uniform measure on the unit interval
$[0,1]$, that is $\mu$ is absolutely continuous with respect to Lebesgue
measure and has a positive continuous density function ${\cal F}(x)$.  In
this case, since $I(P,\epsilon)$ is contained in the subinterval
$\left[p-C_\pi \epsilon, p+C_\pi \epsilon\right]$ by (\ref{EQN2.1}), 
we have
$$
\mu(I(P,\epsilon)) \leq \int_{p-C_\pi \epsilon}^{p+C_\pi \epsilon} d\mu(x) \leq
\max_{0 \leq x \leq 1} \left|{\cal F}(x)\right| \cdot 2C_\pi \epsilon.
$$
Thus, we can improve (\ref{EQN4.6}) in this case to 
\begin{equation}\label{EQN4.7}
LS(M(P),P,\epsilon) \leq C \epsilon^{m+{3 \over 2}}.
\end{equation}

The goal for the rest of this chapter is to prove that, in the case where
$\mu$ is a uniform measure on [0,1], any sequence of matrices will
converge to $M(P)$ if each matrix in the sequence is a better
linearization than $M(P)$ over some small ball.  Specifically, we prove:

\begin{thm}[Least Squares Convergence, $\reals^1\!\to\!
\reals^m$]\label{JJ18}
Assume (A\!1) and (A2).  Let $A$ be a compact attractor for $f$ with
natural
measure $\mu$.  Assume that $\mu$ is invariant, ergodic, and uniform, with
positive density function ${\cal F}$.  Let $P \in \reals^m$ be a limit
point of the reconstructed attractor $\pi(A)$ for which (A3) and (A4)
hold.  Let $\{\epsilon_k\}_{k=1}^\infty$ be a decreasing sequence of
positive numbers, $\epsilon_k \to 0$.  Let $\{M_k\}_{k=1}^\infty$ be a
sequence of matrices such that $LS(M_k,P,\epsilon_k) \leq
LS(M(P),P,\epsilon_k)$ for all $k$.  Then $M_k \to M(P)$ as $k \to
\infty$.
\end{thm}

Before giving the proof of \ref{JJ18}, we prove a few lemmas.  In what
follows, $\left< \cdot, \cdot \right>$ will denote the usual inner product
of vectors in $\reals^m$.  We start with a definition.

\begin{defn}
For a function $f$ mapping the interval [-1,1] into $\reals^m$, and for $0
\leq \epsilon \leq 1$, we define
$$
\delta(f,\epsilon) := \min_{\left\|\eta\right\|=1}
\int_{-\epsilon}^\epsilon \left< \eta, f(h) \right>^2 dh.
$$
\end{defn}

\begin{lemma}\label{JJ19}
If $C(h) = \left(h, \half h^2, \dots, {1 \over {m!}} h^m\right)$ for
$h \in [-1,1]$, then $\delta(C,1) > 0$.
\end{lemma}

\PROOF
Suppose $\delta(C,1) = 0$.  Since the map $\eta \mapsto \displaystyle
\int_{-1}^1 \left<\eta, C(h)\right>^2 dh$ is continuous on the unit sphere
in $\reals^m$, there must be some unit vector $\nu$ for which
$\displaystyle\int_{-1}^1 \left<\nu, C(h)\right>^2 dh = 0$.  Hence,
$\left<\nu, C(h)\right>=0$ for all $h \in [-1,1]$.  However, $\nu \not =
0$, and so $\left<\nu, C(h)\right>=\displaystyle\sum_{n=1}^m {{\nu_n}
\over {n!}}h^n$ is a nontrivial polynomial, a contradiction.
\QED

\begin{lemma}\label{JJ20}
Let $C(h)$ be as above, and let $Pert(h)$ be a continuous vector-valued
function into $\reals^m$ such that $\left\|Pert(h)\right\| \leq \Cpert
|h|^{m+1}$ for $h \in [-1,1]$.  If $V(h):=C(h)+Pert(h)$, then
$\delta(V,\epsilon) \geq \half \delta(C,1) \epsilon^{2m+1}$ as $\epsilon
\to 0$.
\end{lemma}

\PROOF
Note that $\left<\eta, V(h)\right>^2 \geq \left<\eta, C(h)\right>^2 + 2
\left<\eta, C(h)\right> \left<\eta, Pert(h)\right>$.  Thus,
\begin{equation*}
\begin{split}
\delta(V,\epsilon)
&\geq \min_{\left\|\eta\right\|=1} \int_{-\epsilon}^\epsilon \left<\eta,
C(h)\right>^2 dh + 2 \min_{\left\|\eta\right\|=1}
\int_{-\epsilon}^\epsilon \left<\eta, C(h)\right> \left<\eta,
Pert(h)\right> dh \\
&\geq \delta(C, \epsilon) - 2 \min_{\left\|\eta\right\|=1}
\int_{-\epsilon}^\epsilon \big| \left<\eta, C(h)\right> \left<\eta,
Pert(h)\right> \big| dh 
\end{split}
\end{equation*}
We bound this minimum as follows, where $e_m$ is the unit vector $(0,
\dots, 0, 1)$:
\begin{equation*}
\begin{split}
\min_{\left\|\eta\right\| = 1} \int_{-\epsilon}^\epsilon \big| \left<\eta,
C(h)\right> \left<\eta, Pert(h)\right> \big| dh 
&\leq \int_{-\epsilon}^\epsilon \big| \left<e_m, C(h)\right> \left<e_m,
Pert(h)\right> \big| dh \\
&\leq {{\Cpert} \over {m!}} \int_{-\epsilon}^\epsilon |h|^{2m+1} dh \\
&= {{\Cpert} \over {(m+1)!}} \epsilon^{2m+2}
\end{split}
\end{equation*}
Next, we determine a lower bound for $\delta(C,\epsilon)$.  Let
$E_\epsilon$ denote the diagonal matrix $E_\epsilon = diag(\epsilon,
\epsilon^2, \dots, \epsilon^m)$, and note that
$$
\left\|E_\epsilon \eta\right\| \geq \epsilon^m \left\|\eta\right\| =
\epsilon^m \qquad \hbox{ for every unit vector $\eta$.}
$$  
Setting $h=\epsilon t$ and $\gamma = E_\epsilon \eta /
\left\|E_\epsilon \eta\right\|$, we change variables in the integral
defining $\delta(C,\epsilon)$:
\begin{equation*}
\begin{split}
\int_{-\epsilon}^\epsilon \left<\eta, C(h)\right>^2 dh
& = \epsilon \int_{-1}^1 \left<E_\epsilon\eta, C(t)\right>^2 dt \\
& = \epsilon \left\|E_\epsilon \eta\right\|^2 \int_{-1}^1 \left<\gamma,
C(t)\right>^2 dt \\
& \geq \epsilon^{2m+1} \delta(C,1)
\end{split}
\end{equation*}
and taking the minimum over all unit vectors $\eta$, we obtain
$$
\delta(C, \epsilon) \geq \delta(C,1) \epsilon^{2m+1}.
$$
Since $\delta(C,1)>0$, we have
$$
\delta(V,\epsilon) \geq \delta(C,1) \epsilon^{2m+1} - {{2\Cpert} \over
{(m+1)!}} \epsilon^{2m+2} \geq \half \delta(C,1) \epsilon^{2m+1}
$$
for sufficiently small $\epsilon$.
\QED

\begin{lemma}\label{JJ21}
There is a constant $C_1>0$, independent of $\epsilon$, such that for all
small $\epsilon>0$
$$
\min_{\left\|\eta\right\| = 1} \int_{p-\epsilon}^{p+\epsilon} \left< \eta,
\pi(x)-P \right>^2 dx \geq C_1 \epsilon^{2m+1}.
$$
\end{lemma}

\PROOF
Recall that $\pi(x)-P = T(x-p) + Rem(x-p)$, where $T(x-p)$ is the
order-$m$ Taylor polynomial:
$$
T(x-p) = \sum_{n=1}^m {1\over {n!}}\pi^{(n)}(p) (x-p)^n
$$
and $Rem(x-p)$ is the Taylor remainder term satisfying
$$
\left\|Rem(x-p)\right\| \leq \Ctay |x-p|^{m+1}.
$$
Let $Q$ be the change-of-basis matrix such that with respect to the
canonical embedding basis, $Q^{-1}T(h) = \left( h, \half h^2, \dots, {1
\over {m!}}h^m \right)_P = C(h).$ Then, we can set $\gamma:= Q^T\eta /
\left\|Q^T\eta\right\|$ and $V(x-p) := C(x-p) + Q^{-1}Rem(x-p)$, and we
have
$$
\left< \eta, \pi(x)-P \right>
= \left< Q^T\eta, C(x-p)+Q^{-1}Rem(x-p) \right>
= \left\|Q^T\eta\right\| \left< \gamma, V(x-p) \right>.
$$
In order to apply Lemma \ref{JJ20} to $V$, we note that because $Q$ is a
fixed nonsingular matrix, $\left\|Q^{-1} Rem(x-p)\right\| \leq
\left\|Q^{-1}\right\| \Ctay |x-p|^{m+1}$.  We will also need to examine
$\left\|Q^T\eta\right\|$.  Observe that $Q^T$ acts on
the unit circle to produce an ellipse with axes of length
$\sqrt{\sigma_i}$, where $\sigma_i>0$ are the singular values of $Q$.  
Thus, for each unit vector $\eta$, we have $\left\|Q^T\eta\right\| \geq  
\sqrt{\sigma_{min}}>0$ where $\sigma_{min} = \min
\sigma_i$.  Thus, setting $h=x-p$, we have for each unit vector $\eta$
$$
\int_{p-\epsilon}^{p+\epsilon} \left< \eta, \pi(x)-P \right>^2 dx
= \left\|Q^T\eta\right\|^2 \int_{-\epsilon}^\epsilon \left<
\gamma, V(h) \right>^2 dh
\geq \half \sigma_{min} \delta(C,1) \epsilon^{2m+1}
$$
by Lemma \ref{JJ20}.  The conclusion follows by taking the minimum over
all unit vectors $\eta$.
\QED

\begin{lemma}\label{JJ22}
For each $m \times m$ matrix $E$, there is a unit vector $\eta_0 \in
\reals^m$ such that
$$
\left\|Ev\right\| \geq \left\|E\right\| \left|\left<\eta_0, v\right>\right|
$$
for all vectors $v \in \reals^m$.
\end{lemma}

\PROOF
Let $\eta_0$ be a unit eigenvector for $\ETE$ corresponding to the largest
eigenvalue $\alpha$ of $\ETE$.  It follows from the theory of the singular
value decomposition that $\alpha = \left\|\ETE\right\| =
\left\|E\right\|^2$ \cite[pp.\ 266]{ND}.  Moreover, $\ETE$ is
self-adjoint,
hence diagonalizable with an orthonormal basis of eigenvectors.  For any
vector $v$ in $\reals^m$, we can write $v = \left<\eta_0, v\right>\eta_0 +
w$ for some vector $w$ perpendicular to $\eta_0$.  Since the various
eigenspaces are orthogonal and invariant under $\ETE$, we have
$\left<w, \eta_0\right> = \left<\ETE w, \eta_0\right> = \left<w,
\ETE\eta_0\right> = 0.$ Therefore,
\begin{equation*}
\begin{split}
\left\|Ev\right\|^2 
&= \left<\ETE v, v\right> \\
&= \left<\eta_0, v\right>^2 \left<\ETE \eta_0, \eta_0\right> + \left<\ETE
w, w\right> \\
&= \left<\eta_0, v\right>^2 \left<\alpha \eta_0, \eta_0\right> +
\left\|Ew\right\|^2 \\
&\geq \left<\eta_0, v\right>^2 \alpha
\end{split}
\end{equation*}
The conclusion follows by taking square roots.
\QED

\THM{Proof of Theorem \ref{JJ18}}
The hypotheses guarantee that the Eckmann-Ruelle linearization $M=M(P)$
exists and is the unique best linearization as $\epsilon_k \to 0$.  We
prove that $\left\|M_k - M\right\| \to 0$ as $k \to \infty$.  For any $k$,
we have
\begin{multline*}
\int_{I(P,\epsilon_k)}
\left\|\left(M_k-M\right)\left(\pi(x)-P\right)\right\|^2 d\mu(x) \\
\begin{split}
&\leq \int_{I(P,\epsilon_k)} \left(\Phi_{M_k}(x,\pi^{-1}(P))^\half +
\Phi_M(x,\pi^{-1}(P))^\half\right)^2 d\mu(x) \\
&\leq \left(LS(M_k,P,\epsilon_k) + LS(M,P,\epsilon_k)\right)^2 
\end{split}
\end{multline*}
where we have used the H\"older Inequality for the last step.  By
hypothesis and (\ref{EQN4.7}), it follows that
$$
\int_{I(P,\epsilon_k)}
\left\|\left(M_k-M\right)\left(\pi(x)-P\right)\right\|^2 d\mu(x) \leq 4C^2
\epsilon_k^{2m+3}.
$$
On the other hand, by Lemma \ref{JJ22}, there is a unit vector $\eta_0$
such that
\begin{multline*}
\int_{I(P,\epsilon_k)}
\left\|\left(M_k-M\right)\left(\pi(x)-P\right)\right\|^2 d\mu(x) \\
\geq \left\|M_k - M\right\|^2 \int_{I(P,\epsilon_k)} \left\langle
\eta_0,
\pi(x)-P \right\rangle^2 d\mu(x).
\end{multline*}
We need a lower bound in terms of $\epsilon$ for the integral on the
right.  Note that the interval $I(P,\epsilon_k)$ contains the interval
$\left[p-{{\epsilon_k} \over {C_\pi}}, p+{{\epsilon_k} \over
{C_\pi}}\right]$.  Using the continuity and positivity of the density
function ${\cal F}$ and Lemma \ref{JJ21}, we have for sufficiently small
$\epsilon_k$:
\begin{equation*}
\begin{split}
\int_{I(P,\epsilon_k)} \left< \eta_0, \pi(x)-P \right>^2 d\mu(x) 
& \geq \half {\cal F}(p) \int_{p-{{\epsilon_k} \over
{C_\pi}}}^{p+{{\epsilon_k} \over {C_\pi}}} \left< \eta_0, \pi(x)-P
\right>^2 dx \\
& \geq \half {\cal F}(p) C_1 \left({{\epsilon_k} \over
{C_\pi}}\right)^{2m+1}
\end{split}
\end{equation*}
for some constant $C_1>0$, independent of $\epsilon_k$.  It follows that 
$$
\left\|M_k - M\right\|^2 {{{\cal F}(p) C_1} \over {2 C_\pi^{2m+1}}} \
\epsilon_k^{2m+1} \leq 4C^2 \epsilon_k^{2m+3}
$$
and therefore $\left\|M_k - M\right\| = O\!\left(\epsilon_k\right)$ as
$\epsilon_k \to 0$, completing the proof.
\QED

%\end{document}

% CHAPTER 5  --  chap5.tex
%\documentclass{report}
%\usepackage{amstex}
%\begin{document}

%\input{erdmacro.tex}
\chapter{Numerical Computations}

\newcommand{\GRAPH}[1]{\medskip\centerline{\bf INSERT FIGURE {#1}
HERE!!}\medskip\noindent}

In this chapter, we describe numerical experiments illustrating the theory
developed in the previous chapters.  As we shall see, we can compute
numerically the Eckmann-Ruelle linearization in (noise-free) examples.  
The algorithm we use is similar to that outlined in \REF{EKRC}, though the
specific program was written to handle any function as a measurement
function not just time-delay measurement functions.  We briefly outline
the algorithm here.

Recall that the Eckmann-Ruelle procedure has three basic steps.  In the
first step, one reconstructs the attractor with a measurement function.  
In the second step, one determines a local linearization matrix at each
point of the reconstructed trajectory.  In the final step, one extracts
the Lyapunov exponents from the linearization matrices obtained in the
previous step.

For our numerical experiments, we assume that we are given the underlying
dynamical system $f:\reals^n \to \reals^n$ and a measurement function
$\pi:\reals^n \to \reals^m$ for reconstructing the attractor.  We iterate
the function $f$ a number of times to remove transients.  Then, continuing
to iterate $f$, we produce a trajectory $\{p_i\} \in \reals^n$.  Instead of
storing the points $p_i$, we apply the measurement function to each $p_i$
and store the vectors $P_i = \pi(p_i) \in \reals^m$.  We then sort our list
of $P_i$ by their first coordinates, keeping track of where each $P_i$ ends
up in the sorted list.

The procedure for computing the local linearization matrix at a base point
$P$ is straightforward and based on least-squares methods.  The goal is to
determine the $m \times m$ matrix $M$ which best satisfies $M(P_i-P)
\approx P_{i+1}-F(P)$ for all $P_i$ close to $P$.  Given a neighborhood
radius $\epsilon$, we quickly search our sorted list for those $P_i$ whose
first coordinate lies within a distance $\epsilon$ of the first coordinate
of $P$.  Any other point $P_i$ certainly lies outside the ball in
$\reals^m$ of radius $\epsilon$ about $P$.  This first search often
reduces the number of points we need to check carefully to less than 10\%
of the trajectory.  For each $P_i$ in our shortened list, we construct the
vectors $P_i - P$ and $P_{i+1} - F(P)$ and check their lengths.  If both
of these vectors have length less than or equal to $\epsilon$, then we
keep them.  Since the linearization we seek satisfies $M
\left(P_i-P\right) \approx P_{i+1} - F(P)$, we build matrices $A$ and $B$
for which the $k$-th column of $A$ is a vector $P_i-P$ and the $k$-th
column of $B$ is the corresponding vector $P_{i+1}-F(P)$.  Then, our local
linearization matrix $M$ will be the $m \times m$ solution of the matrix
equation $MA=B$.  We solve this equation using a singular value
decomposition routine taken from \REF{PTVF}.

Once the local linearization matrices have been computed, we must compute
the exponents.  Recall from Chapter 3 that the Eckmann-Ruelle-Lyapunov
(ERL) exponents of the reconstructed trajectory $P_0, P_1, \dots$ in
$\reals^m$ are the values obtained by the limit
$$
h_{ER}(P_0,\nu) := \lim_{n \to \infty} {1 \over n} \ln \left\|
M_{n-1} M_{n-2} \cdots M_1 M_0\nu \right\|
$$
for unit vectors $\nu \in \reals^m$, where $M_i=M(P_i)$ is the $m
\times m$ local linearization matrix (i.e., the Eckmann-Ruelle
linearization) at the point $P_i$ in the trajectory.  To evaluate this
limit, we employ the treppen-iteration algorithm suggested in \REF{ER,
EKRC}.  Given our sequence $M_i$ of linearization matrices, we use the QR
matrix decomposition to find orthogonal matrices $Q_i$ and
upper-triangular matrices $R_i$ (with non-negative diagonal elements) such
that
$$
M_i Q_{i-1} = Q_i R_i \qquad \hbox{ for $i=0,1,2,\dots$}
$$
where we take $Q_{-1}$ to be the $m \times m$ identity matrix.  Then, we
can write
$$
M_{n-1} M_{n-2} \cdots M_1 M_0 = Q_{n-1} R_{n-1} R_{n-2} \cdots R_0.
$$
The orthogonal matrix $Q_{n-1}$ won't affect the matrix norm.  Thus,
$$
h_{ER}(P_0,\nu) = \lim_{n \to \infty} {1 \over n} \ln \left\|
R_{n-1} R_{n-2} \cdots R_1 R_0\nu \right\|.
$$
The product $R_{n-1} \cdots R_0$ is upper-triangular, and its diagonal
elements are the eigenvalues of the matrix, which we expect will grow like
the Lyapunov exponents.  As mentioned in Chapter 3, Theorem \ref{TS2}
justifies reading the exponents directly from the diagonal:
$$
\lambda_k = \lim_{n \to \infty}{1 \over n} \sum_{i=0}^{n-1} \ln
\left((R_i)_{kk}\right)
\qquad \hbox{for $k=1,\dots,m$}.
$$
Thus, we are able to compute the ERL exponents from our linearizations by
using the diagonal elements from the QR decomposition.

We shall now discuss several examples.

\begin{figure}
\begin{center}
\begin{turn}{90}
\scalebox{.6}{\includegraphics{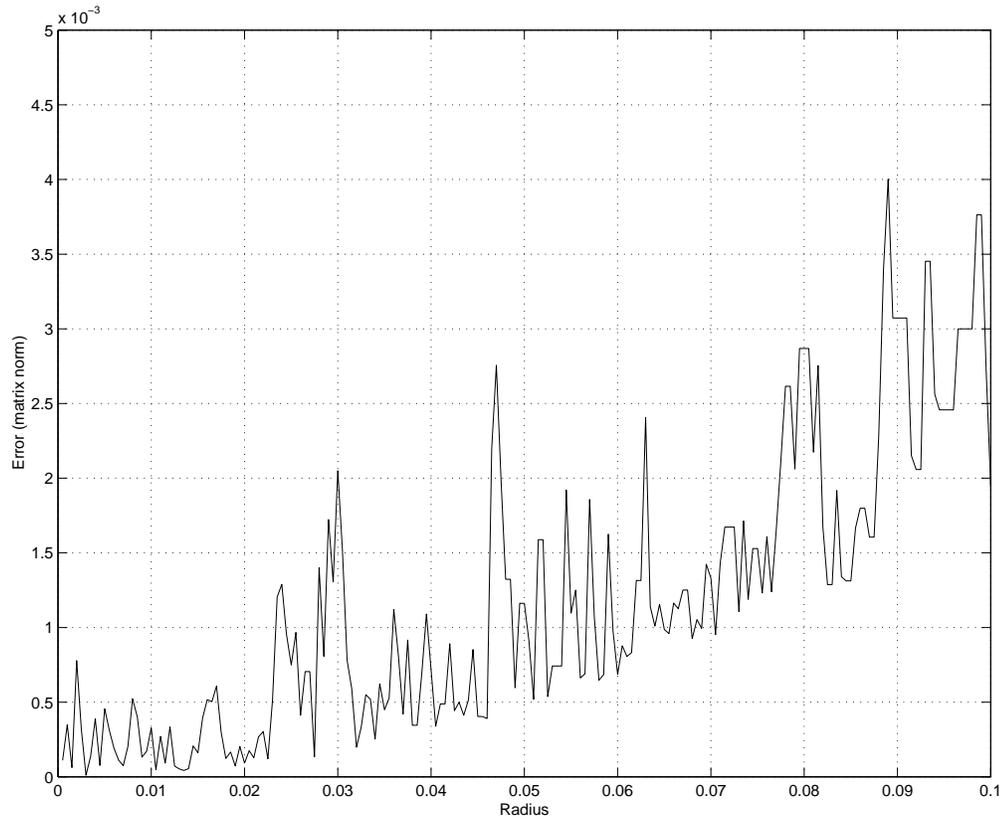}}
\end{turn}
\end{center}
\caption[Convergence with radius of local linearizations for the doubling
map at a point.]{Graph of the difference (in matrix norm) between the
computed linearization and the Eckmann-Ruelle linearization at the fixed
point (1,0) for the doubling map $f(p) = 2p (mod \ 2\pi)$ reconstructed on
the unit circle.  The calculation used 100,000 data points.  Note the
downward trend of the graph, indicating the convergence of the
linearizations to the Eckmann-Ruelle matrix.  The spikes in the graph
appear to be numerical artifacts.  For radii below about 0.01, very few
data points lie sufficiently close to the base point to use in the
calculation.}
\label{FIG5.1}
\end{figure}

In Chapter 2, we considered the doubling map on an interval reconstructed
on the unit circle in $\reals^2$ where the underlying dynamics
was \mbox{$f(p) = 2p \ (mod \ 2\pi)$} on $[0,2\pi]$, and the
measurement function was $\pi(p) = (\cos(p),\sin(p))$.  We computed the
Eckmann-Ruelle linearization in this case to be
\begin{equation}\label{EQN5.1}
M(x,y) = 
\begin{pmatrix} 
4x^3                  & -4y^3   \\ 
2y\left(2x^2+1\right) & 2x\left(2y^2+1\right) 
\end{pmatrix}
\end{equation}
as a function of the points $(x,y)$ on the unit circle.  In numerical
experiments, our routines to compute the best local linearization find
good approximations to this matrix.  For example, we can compute the
linearization matrix at the fixed point (1,0) of the reconstructed
dynamics $F$.  Using a trajectory of 100,000 data points, the
linearization matrix for a neighborhood of radius $\epsilon = 0.05$ is
computed to be
$$
\begin{pmatrix}3.999601 & 0.000000 \\ 0.001048 & 1.999675 \end{pmatrix} 
\approx
\begin{pmatrix}4 & 0 \\ 0 & 2 \end{pmatrix} = M(1,0).
$$
Theorems \ref{JJ10} and \ref{JJ18} suggest that as the neighborhood
radius decreases, the computed linearizations should converge in
matrix norm to $M(1,0)$.  Figure \ref{FIG5.1} shows a graph of this
convergence as the radius shrinks to zero.  To generate this graph,
we first compute and store a reconstructed trajectory of 100,000 data
points on the unit circle. Then, for each radius $\epsilon$, we find
those points in the trajectory within the specified radius of our
base point and use them to compute the linearization matrix.

\begin{figure}
\begin{center}
\begin{turn}{90}
\scalebox{.6}{\includegraphics{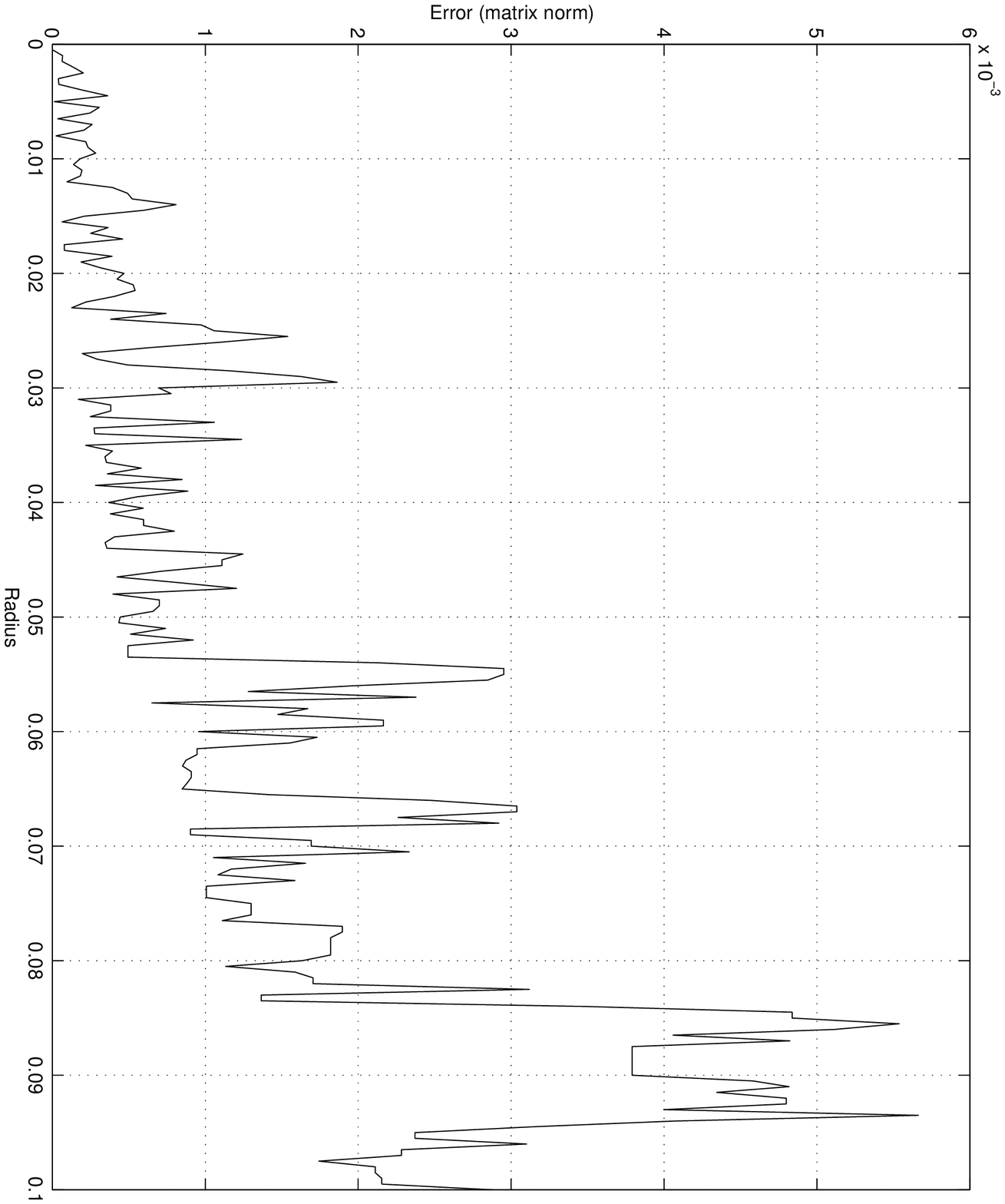}}
\end{turn}
\end{center}
\caption[Convergence with radius of local linearizations for the doubling
map at a point.]{Graph of the difference (in matrix norm) between the
computed linearization and the Eckmann-Ruelle linearization at the point
$\left({\sqrt{2} \over 2},{\sqrt{2} \over 2}\right)$ for the doubling map
reconstructed on the unit circle.  The
calculation used 100,000 data points.  Note the downward trend of the
graph, indicating the convergence of the linearizations to the
Eckmann-Ruelle matrix.  The spikes in the graph appear to be numerical
artifacts.  For radii below about 0.01, very few data points lie
sufficiently close to the base point to use in the calculation.}
\label{FIG5.2}
\end{figure}

Likewise, the local linearizations converge at other points on the unit
circle.  For a neighborhood radius of $\epsilon=0.05$, the linearization
matrix computed at $\left({\sqrt{2} \over 2},{\sqrt{2} \over 2}\right)
\approx (.707,.707)$ is
$$
\begin{pmatrix} 1.414398 & -1.413499 \\ 2.828112 & 2.828112 \end{pmatrix} 
\approx
\begin{pmatrix} \sqrt{2} & -\sqrt{2} \\ 2\sqrt{2} & 2\sqrt{2}\end{pmatrix} 
= M\left({\sqrt{2} \over 2},{\sqrt{2} \over 2}\right).
$$
Figure \ref{FIG5.2} shows the convergence with radius of the linearization
matrices at this point.

\begin{figure}
\begin{center}
\begin{turn}{90}
\scalebox{.6}{\includegraphics{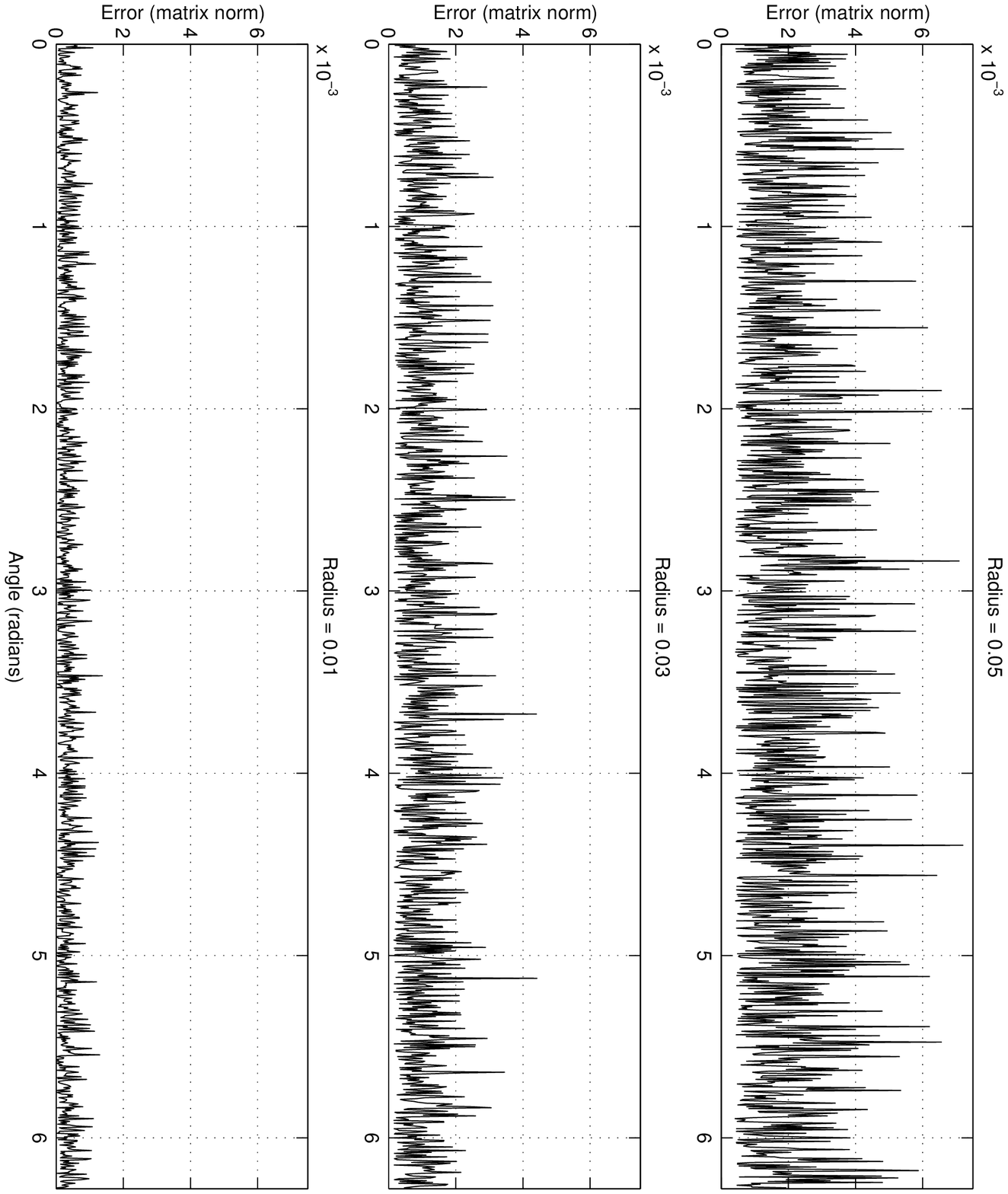}}
\end{turn}
\end{center}
\caption[Uniform convergence of local linearizations for the doubling map
on the unit circle.]{These graphs show, for different radii, the error in
matrix norm between the computed linearization and the Eckmann-Ruelle
linearization for the doubling map at each point on the unit circle
(expressed as an angle). As
the radius decreases, the error decreases uniformly.  The graphs were
created using a single trajectory of 200,000 data points on the unit
circle.}
\label{FIG5.3}
\end{figure}

For a given neighborhood radius, we can determine at each point on the
circle the error between the computed linearization and the Eckmann-Ruelle
linearization.  Figure \ref{FIG5.3} shows this graph for several different
radii.

With close agreement at each point in the interval between the
numerically-determined linearization and the Eckmann-Ruelle linearization,
we expect that the computed Lyapunov exponents will match those given by
the Lyapunov exponent formula of Theorem \ref{JJ6}.  To test this, we
computed the Lyapunov exponents for this example using 100,000 data points
on the circle.  We found values of $0.693173 \approx \ln(2)$ and $1.386244
\approx 2\ln(2)$, as predicted.  The first value is the true Lyapunov
exponent of the doubling map, while the second value is a spurious
exponent.  Notice that the largest computed value is not a Lyapunov
exponent of the underlying dynamical system.

The Lyapunov exponent formula of Theorem \ref{JJ6} holds for any generic
map and specifically for the one-to-one maps with linearly independent
derivative vectors.  Since the formula itself is independent of the
measurement function $\pi$, we expect to compute the same exponents when
different measurement functions are used.  Changing the measurement
function will likely change the speed of convergence somewhat, but we
still expect the results to be close.  With this in mind, we performed a
Lyapunov exponent calculation for the doubling map reconstructed on an
ellipse using the measurement function $\pi(p) = (2\cos(p),
\sin(p)-\half\cos(p))$ and found exponents of $0.693151 \approx \ln(2)$
and $1.386267 \approx 2\ln(2)$.

\begin{figure}
\begin{center}
\scalebox{.4}{\begin{turn}{-90}\includegraphics{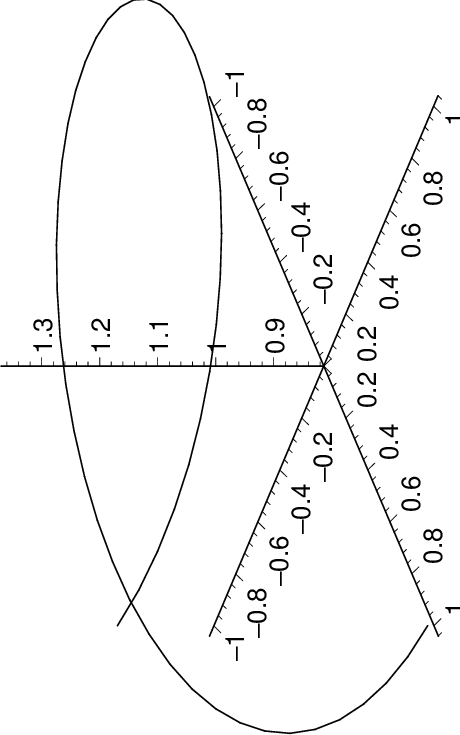}\end{turn}}
\end{center}
\caption[A curve in $\reals^3$.]{The curve $\pi(p) =
(\cos(2\pi p), \sin(2\pi p),
\half e^{1-\half p})$ for $p \in [0,1]$.}
\label{FIG5.4} 
\end{figure}

Next, we examine the case of a reconstruction from $\reals^1$ into
$\reals^3$.  We use the logistic map $f(p) = 4p(1-p)$ on $[0,1]$
reconstructed onto a curve by the measurement function $\pi(p) =
(\cos(2\pi p), \sin(2\pi p), \half e^{1-\half p})$.  Figure \ref{FIG5.4}
shows this curve in $\reals^3$.

\begin{figure}
\begin{center}
\begin{turn}{90}
\scalebox{.6}{\includegraphics{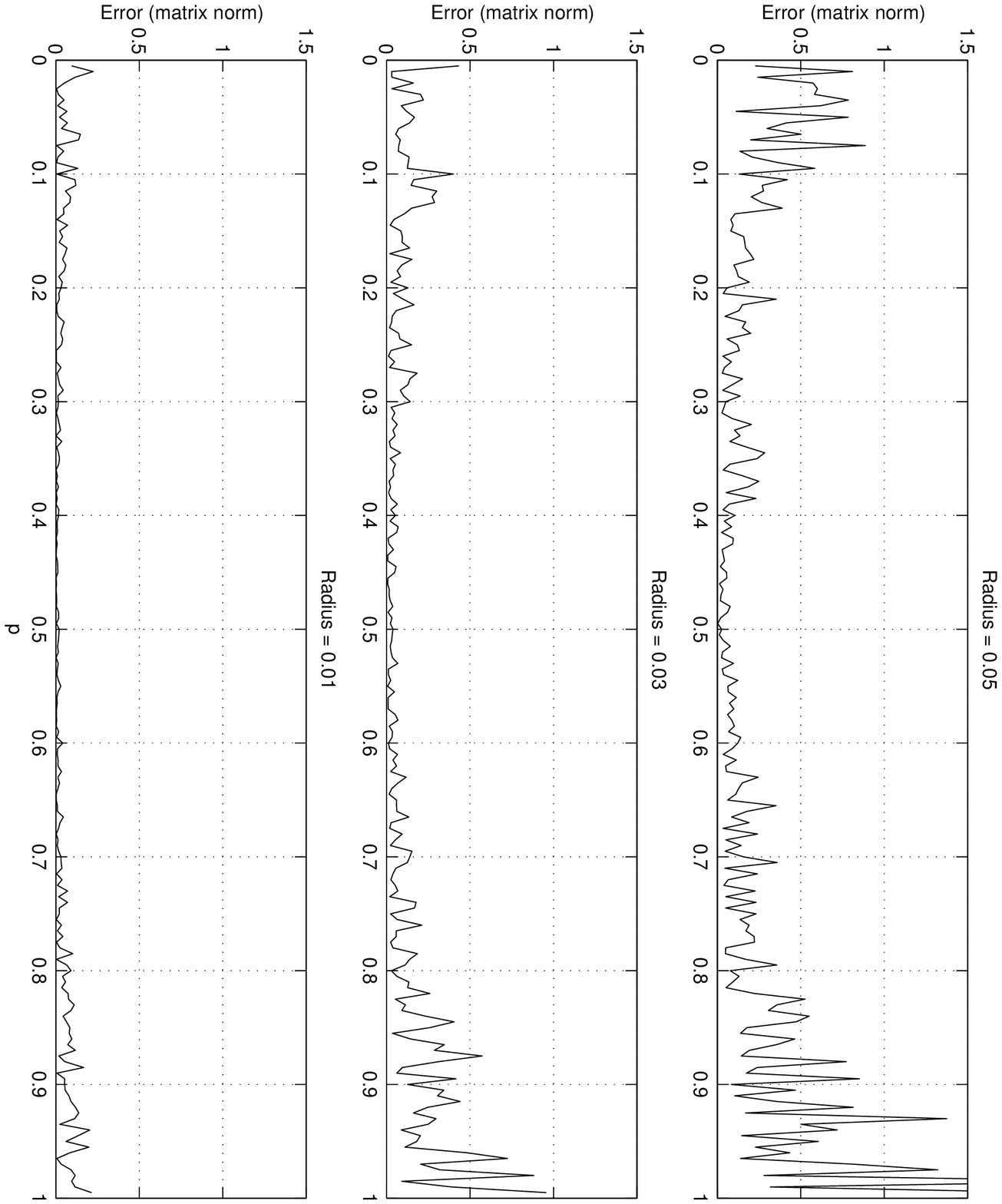}}
\end{turn}
\end{center}
\caption[Uniform convergence of local linearizations for the logistic map
on a curve in $\reals^3$.]{These graphs show, for different radii, the
error in matrix norm between the computed linearization and the
Eckmann-Ruelle linearization of the logistic map at each point on the
curve in $\reals^3$.  As the radius decreases, the error decreases
uniformly.  The graphs were created using a single trajectory of 200,000
data points on the curve in $\reals^3$. }
\label{FIG5.5}
\end{figure}

In this case, the formula for the Eckmann-Ruelle linearization is more
complicated than (\ref{EQN5.1}), but for any individual base point, it is
easily computed from the definition given in Chapter 2.  For example, at
the point $\pi(0.125) \approx (0.707,0.707,1.277)$ on the curve in
$\reals^3$, the Eckmann-Ruelle linearization is
$$
M(\pi(0.125)) =
\begin{pmatrix}
-0.395832 & 2.313014 & -18.147168 \\ 
-6.279704 & -3.080283 & 65.170183 \\
-0.000276 & -0.098277 & 1.550445 
\end{pmatrix}
$$
and, computing with neighborhood radius $\epsilon = 0.05$, the local
linearization matrix is
$$
\begin{pmatrix} 
-0.391945 & 2.313262 & -18.187396 \\
-6.273483 & -3.079227 & 65.103345 \\
-0.000214 & -0.098268 & 1.549787
\end{pmatrix}.
$$
Figure \ref{FIG5.5} shows the error in the linearizations over the whole
curve. To compute the Lyapunov exponents for this example, we used 100,000
data points on the curve.  We found values $0.6884401 \approx \ln(2)$,
$1.392024 \approx 2\ln(2)$, and $2.056767 \approx 3\ln(2)$.  This is
consistent with Theorem \ref{JJ6} since the Lyapunov exponent for the
logistic map is $\ln(2)$.  Note that the largest computed exponent is
spurious.

We also computed the Lyapunov exponents of several other reconstructions of
the logistic map in $\reals^3$.  The results are listed in Table
\ref{TA5.6}.
\begin{table}
\caption[The computed exponents of several reconstructions of the
logistic map into $\reals^3$.]{The Eckmann-Ruelle-Lyapunov exponents of
several reconstructions of the logistic map $f(p)=4p(1-p)$ into
$\reals^3$.  Each computation involved 100,000 data points.  The computed
exponents are roughly $\ln(2)$, $2\ln(2)$, and $3\ln(2)$.
}\label{TA5.6}
\begin{center}
\begin{tabular}{|c|ccc|} 
\hline
Measurement Function $\pi(p)$ & \multicolumn{3}{c|}{Computed ERL
Exponents} \\
\hline
$(p, p^2, p^3)$ & 0.664767 & 1.411537 & 2.082149 \\
$(p, \sqrt{p}-2p^3, \half p\cos(2.5\pi p))$ & 0.678497 &
1.391379 & 2.075309 \\
$(\cos(2\pi p), p+\sin(2\pi p), p+p\sin(5\pi p^2))$ &
0.687375 & 1.385631 & 2.066531 \\
\hline
\end{tabular}
\end{center}
\end{table}

In addition to the somewhat arbitrary measurement functions we have used
thus far, we can also construct time-delay embeddings in $\reals^3$ of the
logistic map.  At each iteration, we record some quantity based on the
current state of the system.  In an actual experiment, this choice would be
realized by the apparatus used to measure and record the data.  The
results of computing Lyapunov exponents for several time-delay examples
are given in Table \ref{TA5.7}.

\begin{table}
\caption[The computed exponents of several time-delay reconstructions of
the logistic map into $\reals^3$.]{The Eckmann-Ruelle-Lyapunov exponents
of several time-delay reconstructions of the logistic map $f(p)=4p(1-p)$
into $\reals^3$.  For each iteration, we recorded a scalar determined by
the current state $p$ of the system.  Each computation used 100,000 data
points.  The computed exponents are roughly $\ln(2)$,
$2\ln(2)$, and $3\ln(2)$.
}\label{TA5.7}
\begin{center}
\begin{tabular}{|c|ccc|} 
\hline
Recorded Quantity & \multicolumn{3}{c|}{Computed ERL Exponents} \\
\hline
$p$          & 0.691530 & 1.386031 & 2.083390 \\
$\cos(2p)-p$ & 0.692866 & 1.386281 & 2.080733 \\
$\ln(1+p)$   & 0.691398 & 1.386725 & 2.079474 \\
\hline
\end{tabular}
\end{center}
\end{table}

\begin{figure}
\begin{center}
\begin{turn}{-90}
\scalebox{.4}{\includegraphics{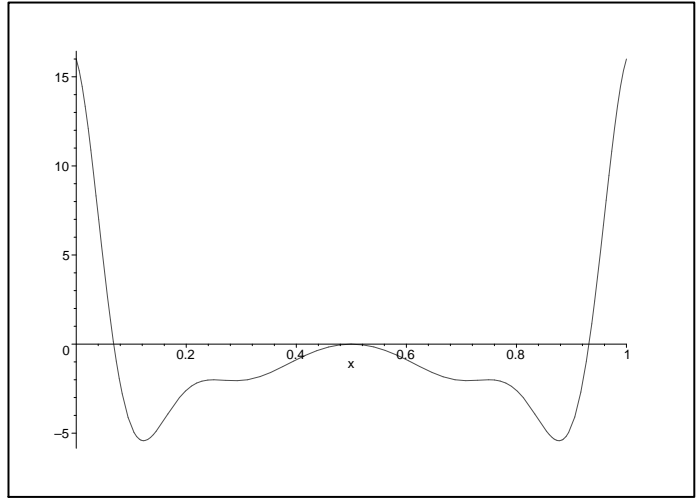}}
\end{turn}
\end{center}  
\caption[Graph of the first coordinate of an Eckmann-Ruelle matrix.]{A
graph of the first coordinate of the Eckmann-Ruelle matrix for the
logistic map $f(p)=4p(1-p)$ on [0,1] reconstructed on the unit circle in
$\reals^2$ by the measurement function $\pi(p) = (\cos(2\pi p), \sin(2\pi
p) )$.  Note the large derivatives near the ends of the interval.}
\label{FIG5.8} 
\end{figure}

The Eckmann-Ruelle linearization described in (\ref{EQN5.1}) may be
misleadingly simple.  It is possible to compute an explicit formula for
the Eckmann-Ruelle linearization for the case of the logistic map $f(p) =
4p(1-p)$ on $[0,1]$ reconstructed onto the unit circle by the measurement
function $\pi(p) = (\cos(2\pi p), \sin(2\pi p))$.  One then finds that
some of the component functions of that matrix can have derivatives as
large as 200 in absolute value.  See Figure \ref{FIG5.8}.

We move on to discuss the case of two-dimensional underlying dynamics.  
Here, we base our experiments on the \henon map $f(x,y) = (1.4 -x^2 +
0.3y, x)$.  We begin by considering the measurement function $\pi(x,y) =
(x,y,x^2,y^2,xy)$.  For this measurement function, we can compute the
Eckmann-Ruelle linearization explicitly:
$$
M(\pi(x,y)) = 
\begin{pmatrix}
0 & 0.3 & -1 & 0 & 0 \\
1 &  0  &  0 & 0 & 0 \\
1.2xy-8x^3 & 0.84 + 0.6x^2 & -2.8+6x^2-0.6y & .09 & -1.2x \\
0 &  0  &  1 & 0 & 0 \\
1.4+3x^2 &  0  & -3x & 0 & 0.3 
\end{pmatrix}
$$
where for convenience, we refer to $M$ using coordinates in the underlying
space $\reals^2$ instead of coordinates in the reconstruction space
$\reals^5$.  We observe the same phenomena when the underlying dynamics
are two-dimensional as we did in the previous cases.  For example, at the
point $\pi(1.555478,0.398567) \in \reals^5$ on the reconstructed \henon
attractor, the linearization matrix for a neighborhood of radius $\epsilon
= 0.01$ is computed to be
$$
\begin{pmatrix}
  0.000000 &  0.300000 & -1.000000 & 0.000000 &  0.000000 \\
  1.000000 &  0.000000 &  0.000000 & 0.000000 &  0.000000 \\
-29.365725 &  2.290703 & 11.478416 & 0.090305 & -1.866085 \\
  0.000000 &  0.000000 &  1.000000 & 0.000000 &  0.000000 \\
  8.657187 & -0.001385 & -4.666094 & 0.000326 &  0.300723 
\end{pmatrix}
$$
At another point, $\pi(-1.741541,1.753985)$, on the reconstructed \henon
attractor in $\reals^5$, the local linearization for $\epsilon = 0.01$ is
computed to be
$$
\begin{pmatrix}
 0.000000 &  0.300000 & -1.000000 & 0.000000 & 0.000000 \\
 1.000000 &  0.000000 &  0.000000 & 0.000000 & 0.000000 \\
38.590278 &  2.658478 & 14.345224 & 0.090317 & 2.089740 \\
 0.000000 &  0.000000 &  1.000000 & 0.000000 & 0.000000 \\
10.498785 & -0.000234 &  5.224592 & 0.000067 & 0.300000 
\end{pmatrix}.
$$

\begin{figure}
\begin{center}
\begin{turn}{90}
\scalebox{.6}{\includegraphics{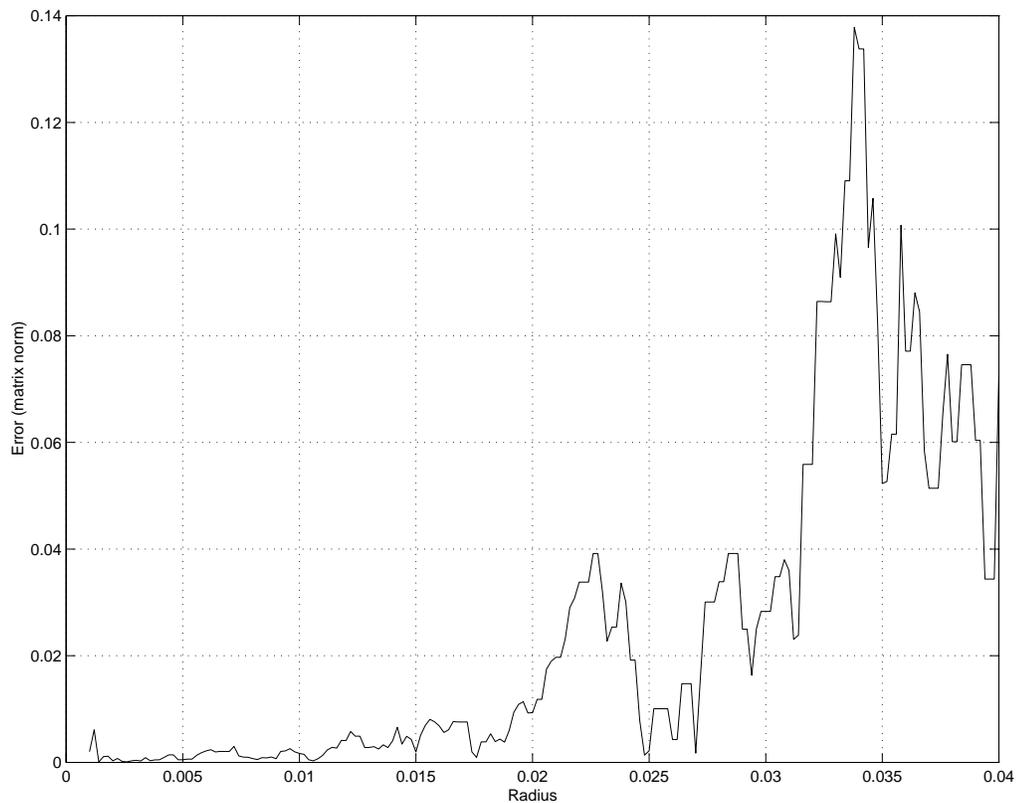}}
\end{turn}
\end{center}
\caption[Convergence with radius of local linearizations for the \henon
map at a point.]{Graph of the difference (in matrix norm) between the   
computed linearization and the Eckmann-Ruelle linearization at the point
(-1.741541, 1.753985) for the \henon map reconstructed in $\reals^5$.  The
calculation used 1,000,000 data points.  The graph does not show radii
larger than about 0.04 because this base point lies about 0.045 units
away from a significant bend in the \henon attractor.}
\label{FIG5.9}
\end{figure}

Both of these examples are in good agreement with the general formula
above.  As before, we can graph the convergence of the local
linearizations to the Eckmann-Ruelle linearization as the neighborhood
radius shrinks to zero.  See Figure \ref{FIG5.9}.  Using $300,000$ data
points, we computed the Lyapunov exponents for this example.  Recall that
the true Lyapunov exponents of the \henon map are approximately $\lambda =
0.42$ and $\mu = -1.62$.  We found values of $0.886200 \approx 2\lambda$,
$0.422459 \approx \lambda$, $-1.195450 \approx \lambda + \mu$, $-1.636780
\approx \mu$, and $-3.183227 \approx 2\mu$.  These values are consistent
with the Lyapunov exponent formula given in Theorem \ref{JJ9}.

We also computed the Lyapunov exponents from time-delay reconstructions of
the \henon system.  Table \ref{TA5.10} shows the results.  As one can see
in Table \ref{TA5.10}, we do not find all of the exponents predicted by
the Lyapunov exponent formula in Theorem \ref{JJ9}.  Three of the
exponents match nicely with the predicted formula, namely $\lambda \approx
0.42$, $2\lambda \approx 0.84$, and $\mu \approx -1.62$, but the other two
vary somewhat from their expected values.  With more data and smaller
neighborhood radii in the computations, those last two exponents would
likely converge to their proper values.  Note that the true Lyapunov
exponents appear in each of these examples, suggesting that they converge
to their correct values rather quickly.  This is not surprising if one
looks back at the proofs of Theorem \ref{JJ3} in Chapter 2 and Theorem
\ref{JJ9} in Chapter 3.  Any matrix that does not map the first-order
terms of the Taylor series correctly will have $O(\|\del P\|)$ error,
instead of the $O(\|\del P\|^3)$ error of the Eckmann-Ruelle
linearization.  Fortunately, algorithms to compute local linearizations
will instead find matrices with error $O(\|\del P\|^2)$ or better, and
these matrices will map the first-order Taylor terms correctly.  This
ensures that the true Lyapunov exponents will be among the first of the
computed exponents to converge.

\begin{table}
\caption[The computed exponents of several time-delay reconstructions of
the \henon map into $\reals^5$.]{The Eckmann-Ruelle-Lyapunov exponents of
several time-delay reconstructions of the \henon map into $\reals^5$.  
For each iteration, we recorded a scalar determined by the current state
$(x,y)$ of the system.  Each computation involved 300,000 data points.  
Three of the computed exponents match the Lyapunov exponent formula:
$0.84 \approx 2\lambda$, $0.42 \approx \lambda$, and $-1.62 \approx \mu$.
The other exponents will probably converge for longer time series.
}\label{TA5.10}
\begin{center}
\begin{tabular}{|c|ccccc|} 
\hline
Recorded Quantity & \multicolumn{5}{c|}{Computed ERL Exponents} \\
\hline
$x$    & 0.838426 & 0.418513 & -1.070009& -1.619577 & -2.288856 \\
$y-xy$ & 0.839063 & 0.418518 & -1.064124 & -1.619594 & -2.426907 \\
$\half\left(x^2+y^2\right)$ & 0.837956 & 0.418331 & -0.846145 & -1.621547
& -2.404775 \\
$\arctan\left({y \over x}\right)$ & 0.840121 & 0.418186 &
-1.062045 & -1.618537 & -2.226276  \\
$\cos(x)+\sin(y)$ & 0.839557 & 0.418119 & -1.019653 & -1.621305 &
-2.378717 \\
$\cos(y-1)-{3 \over 4}x^2$ & 0.839295 & 0.418780 & -1.095620 & -1.621523 &
-2.624901 \\
\hline
\end{tabular}
\end{center}
\end{table}

%\end{document}

% Appendices
\appendix
% APPENDIX A  --  polyapp.tex
%\documentclass{report}
%\usepackage{amstex}
%\begin{document}

%\input{erdmacro.tex}
\chapter{Polynomial Results}

In this appendix, we derive the facts about polynomials that we need in
the text.

\begin{prop}\label{PP1}
Let $p(x)$ be a polynomial of degree $d \geq 0$, and suppose there are
infinitely many values $x_i \to 0$ for which $|p(x_i)| \leq C |x_i|^q$ for
some $q>d$.  Then $p(x)=0$ for all $x$.
\end{prop}

\PROOF
Write $p(x)=a_dx^d+ \dots + a_1x + a_0$.  By the continuity of $p(x)$,
$$
0 \leq |a_0| = |p(0)| = \lim_{x_i \to 0}|p(x_i)| \leq \lim_{x_i \to 0}
C|x_i|^q = 0
$$
since $q > d \geq 0$.  Thus, $a_0=0$.

For induction, suppose that $a_0 = a_1 = \dots = a_{k-1} = 0$ where $k
\leq d < q$.  Write $p(x)=x^k(a_dx^{d-k} + \dots + a_{k+1}x + a_k) := x^k
q_k(x)$.  Then, by the continuity of $q_k(x)$,
$$
0 \leq |a_k| = |q_k(0)| = \lim_{x_i \to 0}|q_k(x_i)| = \lim_{x_i \to 0}
{{|p(x_i)|} \over {|x_i|^k}} \leq \lim_{x_i \to 0} {{C|x_i|^q} \over
{|x_i|^k}} = \lim_{x_i \to 0} C|x_i|^{q-k} = 0
$$
since $q-k \geq q-d > 0$.  Thus, $a_k=0$, and the lemma follows from
induction on $k$.
\QED

We would like to prove a similar result for polynomials with more
variables.  Unfortunately, the result is not true, even for two variables,
without extra hypotheses.  For example, if $p(x,y) = x-y$, then $p(x_i,
x_i) = 0$ for any sequence $x_i \to 0$.  Thus, $p(x,y)$ need not be zero,
even though there is a sequence going to $(0,0)$ satisfying $|p(x_i,y_i)|
\leq C \|(x_i,y_i)\|^q$ for every $q>0$.  The added hypotheses in the next
lemma guarantee that the polynomial must be small in many directions, not
just one unlucky choice.

Recall the definition of an approach direction from our discussion in
Chapter 2 of the case where the underlying dynamics is two-dimensional.

\begin{defn}
Let $l$ be a unit vector in $\reals^m$.  A subset $S$ of $\reals^m$ has
the {\bf approach direction} $l$ at the base point $P \in \reals^m$
provided there is a sequence $\left\{Q_{k}\right\}_{k=1}^\infty$ from $S$
such that:
\begin{enumerate}
\item $Q_{k} \to P$ as $k \to \infty$, and 
\item ${{\del Q_{k}} \over {\left\|\del Q_{k}\right\|}} \to l$ as $k \to
\infty$, where $\del Q_k := Q_k - P$.
\end{enumerate}
We call a collection of approach directions at $P$ {\bf distinct} provided
that no two are the same and no two are reflections through the origin.  
Multiple approach directions are not required to be linearly independent.
\end{defn}

\begin{prop}\label{PP2}
Let $p(x,y)$ be a polynomial of (total) degree $d$ (that is, each monomial
in $p$ has total degree at most $d$).  Assume that there is a sequence of
points $(x_k, y_k) \to (0,0)$ in $\reals^2$ such that $|p(x_k,y_k)| \leq C
\|(x_k,y_k)\|^q$ for some $q>d$.  If the set $\left\{ (x_k,y_k) : k \geq 0
\right\}$ has $d+1$ distinct approach directions
at (0,0), then $p(x,y)=0$ for all $(x,y) \in \reals^2$.
\end{prop}

\PROOF
Write $p(x,y) = \sum_{j=0}^d p_j(x,y)$, where $p_j(x,y) = \sum_{e=0}^j
a_{je} x^e y^{j-e}$ is a polynomial of degree $j$ with only monomials of
total degree $j$.  We will prove inductively that each polynomial
$p_l(x,y)$ is identically zero for $l=0,1, \dots, d$.

We begin with the $l=0$ case.  Note that
$$
0 \leq |a_{00}| = |p(0,0)| = \lim_{k \to \infty} |p(x_k,y_k)| \leq \lim_{k
\to \infty} C\|(x_k,y_k)\|^q = 0
$$
since $q > d \geq 0$.  Thus, $p_0(x,y) = a_{00} = 0$, and so $p$ has no
constant term.

Without loss of generality, we may now assume that $d \geq 1$.

Next, we show that $p_1(x,y) = a_{11}x + a_{10}y = 0$ by proving that
$a_{11} = a_{10} = 0$.  This is the $l=1$ case and it will demonstrate the
basic idea for the induction that follows.  For an approach direction
$(\alpha, \beta)$, let $\{(\hat x_k, \hat y_k)\}$ be a subsequence such
that $(\hat x_k, \hat y_k) / \|(\hat x_k, \hat y_k)\| \to (\alpha,
\beta)$.  For convenience, set $n_k := \|(\hat x_k, \hat y_k)\|$.  Then,
\begin{equation*}
\begin{split}
0 
&\leq |a_{11}\alpha + a_{10}\beta | 
= \lim_{k \to \infty} \left|a_{11} {{\hat x_k} \over {n_k}} + a_{10}
{{\hat y_k} \over {n_k}}\right|
= \lim_{k \to \infty} {{|p_1(\hat x_k, \hat y_k)|} \over {n_k}} \cr
&\leq \limsup_{k \to \infty} \left( {{|p(\hat x_k, \hat y_k)|} \over
{n_k}} + \sum_{j=2}^d \sum_{e=0}^j |a_{je}| {{|\hat x_k|^e |\hat
y_k|^{j-e}} \over {n_k}} \right) \\
&\leq \limsup_{k \to \infty} \left( {{C n_k^q} \over {n_k}} + \sum_{j=2}^d
|a_{jj}| |\hat x_k|^{j-1} {{|\hat x_k|} \over {n_k}} + \sum_{j=2}^d
\sum_{e=0}^{j-1} |a_{je}| |\hat x_k|^e |\hat y_k|^{j-e-1} {{|\hat y_k|}
\over {n_k}} \right) \\
&= 0
\end{split}
\end{equation*}
because $q > d \geq 1$, $|\hat x_k| / n_k \to \alpha$, $|\hat y_k| / n_k \to \beta$, and $|\hat x_k|, |\hat y_k| \to 0$.  Therefore, $a_{11}\alpha + a_{10}\beta = 0$ for each approach vector $(\alpha, \beta)$.  When $\alpha \not= 0$, we can divide by $\alpha$ to get:
$$
a_{11} + a_{10}\left({{\beta} \over {\alpha}}\right) = 0
$$
We must now consider several cases.

%\smallskip\noindent\underbar{CASE 1:}  
\CASE {1}
{\it Assume that either $d \geq 2$, or $d=1$ but there are 2 distinct
approach vectors with $\alpha \not= 0$.}
In this case, there are at least two approach vectors $(\alpha,\beta)$
with $\alpha \not= 0$.  If we write $f(t)=a_{10}t + a_{11}$, then for each
such vector, we have $f\left({\beta \over \alpha}\right)=0$.  As there are
at least two such values and $f(t)$ is linear, we must have $f(t)$
identically zero.  Hence, $a_{11} = a_{10} = 0$ and therefore $p_1(x,y) =
0$ for all $(x,y)$.

%\smallskip\noindent\underbar{CASE 2:}  
\CASE {2}
{\it Assume that $d = 1$ and that one of the 2 approach directions is $(0,
\pm 1)$.}
In this case, we see directly that $|a_{10}| = |a_{11}0 + a_{10}(\pm 1)| =
0$.  Then, for the approach vector $(\alpha, \beta)$ with $\alpha \not=
0$, we have $0 = |a_{11}\alpha + 0\cdot \beta | = |a_{11} \alpha|$.  We
conclude that $a_{11} = 0$ as well.  Therefore $p_1(x,y) = 0$ for all
$(x,y)$.

In all cases, $p_1(x,y) = 0$ as desired.  The main induction step that
follows is similar.

Suppose we have shown that $p_0(x,y) = p_1(x,y) = \dots = p_{l-1}(x,y) =
0$ are all identically zero for some $l \leq d$.  We want to show that
$p_l(x,y) = 0$ is also identically zero.  For any approach direction
$(\alpha, \beta)$ (using the same notation as above):
\begin{equation*}
\begin{split}
0 
&\leq \left|\sum_{e=0}^l a_{le}\alpha^e\beta^{l-e} \right| 
= \lim_{k \to \infty} \left|\sum_{e=0}^l a_{le} \left({{\hat x_k} \over
{n_k}}\right)^e \left({{\hat y_k} \over {n_k}}\right)^{l-e}\right|
= \lim_{k \to \infty} {{|p_l(\hat x_k, \hat y_k)|} \over {n_k^l}} \cr
&\leq \limsup_{k \to \infty} \left( {{|p(\hat x_k, \hat y_k)|} \over
{n_k^l}} + \sum_{j=l+1}^d {{|p_j(\hat x_k, \hat y_k)|} \over {n_k^l}}
\right) \\
&\leq \limsup_{k \to \infty} \left( C n_k^{q-l} + \sum_{j=l+1}^d
\sum_{e=0}^{j} T_{je}^k \right) \\
&= 0
\end{split}
\end{equation*}
where the terms $T_{je}^k$ look like
$$
T_{je}^k = |a_{je}| |\hat x_k|^{e-p_1} |\hat y_k|^{j-e-p_2} \left({{|\hat
x_k|} \over {n_k}}\right)^{p_1} \left({{|\hat y_k|} \over
{n_k}}\right)^{p_2}
$$
for some appropriate choice of $p_1 + p_2 = l < j$.  The lim sup above is
zero because
$$
q-l \geq q-d > 0, \quad
{1 \over {n_k}}(\hat x_k, \hat y_k) \to (\alpha,\beta), \quad
\hbox{ and }
|\hat x_k|, |\hat y_k| \to 0
$$
together imply that for each $j$ and $e$, the terms $T_{je}^k \to 0$ as $k
\to \infty$.  Thus,
$$
\sum_{e=0}^l a_{le} \alpha^e \beta^{l-e} = 0
\qquad \hbox{ for each $(\alpha, \beta)$.}
$$
For those approach vectors with $\alpha \not= 0$, we can divide through by
$\alpha^l$:
\begin{equation}\label{EQNA.1}
\sum_{e=0}^l a_{le} \left({\beta \over \alpha}\right)^{l-e} = 0.
\end{equation}
We now must consider a few cases.

%\smallskip\noindent\underbar{CASE 1:}
\CASE {1}
{\it Assume that either $d > l$, or $d=l$ but there are $d+1$ distinct
approach vectors with $\alpha \not= 0$.}
In this case, there are at least $l+1$ approach vectors $(\alpha,\beta)$
with $\alpha \not= 0$.  If we write $f(t)=\sum_{e=0}^l a_{le} t^{l-e}$,
then for each such approach vector, we have $f\left({\beta \over
\alpha}\right)=0$ by (\ref{EQNA.1}).  As there are at least $l+1$ such
values and $\deg\left(f(t)\right) \leq l$, we must have $f(t)$ identically
zero.  Hence, $a_{l0} = \dots = a_{ll} = 0$ and therefore $p_l(x,y) = 0$
for all $(x,y)$.

%\smallskip\noindent\underbar{CASE 2:}  
\CASE {2}
{\it Assume that $d = l$, and that one of the $l+1$ approach directions is
$(0, \pm 1)$.}
In this case, we see directly that 
$$
|a_{l0}| = \left|\sum_{e=0}^l a_{le} 0^e (-1)^{l-e}\right| = 0
$$
Thus, $a_{l0} = 0$ and the polynomial $f(t)$ defined as in the previous
case is actually of degree at most $l-1$.  Then, for the $d=l$ approach
vectors $(\alpha, \beta)$ with $\alpha \not= 0$, we have $f\left({\beta
\over \alpha}\right)=0$.  We conclude, as before, that $f(t)$ is
identically zero.  Hence, $a_{l0} = \dots = a_{ll} = 0$ and $p_l(x,y) = 0$
for all $(x,y)$.

In all cases, $p_l(x,y) = 0$ as desired.  This completes the induction
step and the proof.
\QED

%\end{document}

% APPENDIX B  --  timsapp.tex

%\input{erdmacro.tex}
\chapter{Lyapunov Exponents of Upper Triangular Matrix Products}

In this appendix, we prove Theorem \ref{TS2} that when a sequence of
upper-triangular $m \times m$ matrices are multiplied together, we can
find vectors whose lengths grow at the same rate as the diagonal elements
of the matrix product.  This result can be found elsewhere in the
literature, for example in \REF{JPS}, though it usually is presented in a
more general context.  This presentation of the theorem roughly follows
\REF{JPS} but can also be considered a distillation of a piece of
Oseledec's original proof.

\begin{lemma}\label{TS1}
For $k=1,2, \dots$, let $A_k$ be an $m \times m$ matrix written in block
structure $A_k = 
\left(\begin{smallmatrix} a_k & b_k \\ 0 & B_k \end{smallmatrix}\right)$, 
where $a_k \in \reals$, $b_k \in \reals^{1 \times (m-1)}$, and $B_k \in
\reals^{(m-1) \times (m-1)}$.  Set $S_0=I_m$, the $m \times m$ identity
matrix, and for $n=1,2,\dots$ define $S_n = A_n \cdots A_1 = 
\left(\begin{smallmatrix}s_n & t_n \\ 0 & T_n \end{smallmatrix}\right)$, 
using the same block structure. Then
$$
t_n = a_n \cdots a_1 \sum_{k=1}^n {{b_kT_{k-1}} \over {a_k \cdots a_1}}.
$$
\end{lemma}

\PROOF
The proof is by induction on $n$.  For $n=1$, $t_1 = a_1
\displaystyle \sum_{k=1}^1 {{b_1} \over {a_1}} = b_1$.  In general,
\begin{equation*}
\begin{split}
t_{n+1}
&= a_{n+1}t_n + b_{n+1}T_n \\
&= a_{n+1} a_n \cdots a_1 \sum_{k=1}^n {{b_kT_{k-1}} \over {a_k \cdots
a_1}} + a_{n+1} \cdots a_1 {{b_{n+1} T_n} \over {a_{n+1} \cdots a_1}} \\
&= a_{n+1} \cdots a_1 \sum_{k=1}^{n+1} {{b_kT_{k-1}} \over {a_k \cdots a_1}}. 
\end{split}
\end{equation*}
\QED

%\begin{defn}
\noindent {\bf Definition 3.1.} {\it
A sequence of real numbers $\left\{r_k\right\}$ has {\bf (geometric)
growth rate $\gamma$} provided }
$$
\lim_{k \to \infty} {{\ln |r_k|} \over k} = \gamma.
$$
%\end{defn}

\smallskip

%\begin{thm}
\noindent{\bf Theorem 3.2.} {\it
For $k=1,2 \dots$, let $A_k$ be $m \times m$ upper triangular matrices,
and define $S_n = A_n \cdots A_1$.  Assume the magnitudes of the entries
of $A_k$ are bounded independent of $k$, and that the diagonal entries of
$S_n$ have growth rates $\gamma_1, \dots, \gamma_m$ as $n \to \infty$.  
Then there exists vectors $v_1, \dots, v_m$ such that for each $i=1,
\dots, m$, $\left\| S_n v_i \right\|$ has growth rate $\gamma_i$.
}
%\end{thm}

\PROOF
The proof is by induction on $m$.  Notice that when multiplying two upper
triangular matrices, the lower $i$ rows of the product are independent of
the entries of the two matrices above the lower $i$ rows.  So, the
induction on $m$ will start at the lower right and work upward.

Case $m=1$ is clear; the vector $v_1 = (1)$ works.  For the general case,
assume that the theorem holds for all such sequences of $(m-1) \times
(m-1)$ matrices.  We use the partitioned notation for $A_k$ and $S_n$ from
Lemma \ref{TS1}.  By the induction hypothesis, we will assume that by
denoting by $\hat v_2, \dots, \hat v_m$ the columns of the matrix
$$
V_{m-1} = \begin{pmatrix}
1 & v_{23} & \dots  & v_{2m} \\
  &    1   & \dots  & v_{3m} \\
  &        & \ddots & \vdots \\
  &        &        &   1    
\end{pmatrix},$$
the sequence $\left\|T_n \hat v_i\right\|$ has growth rate $\gamma_i$ for
$i=2, \dots, m$.  (Recall from Lemma \ref{TS1} that $T_n$ is the lower
right $(m-1) \times (m-1)$ submatrix of $S_n$.)  In particular, for each
$\epsilon>0$ there exists a constant $K$ such that
\begin{equation}\label{EQNB.1}
{1 \over K} e^{(\gamma_i-\epsilon)n} \leq \left\|T_n \hat v_i\right\| \leq
K e^{(\gamma_i+\epsilon)n} 
\end{equation}
for $i=1, \dots, m$, and furthermore, by assumption
\begin{equation}\label{EQNB.2}
{1 \over K} e^{(\gamma_1-\epsilon)n} \leq s_n = a_n \cdots a_1 \leq K
e^{(\gamma_1+\epsilon)n}.
\end{equation}
We will add a top row to $V_{m-1}$ to get 
$$
V_m = \begin{pmatrix}
1 & v_{12} & \dots  & v_{1m} \\
  &    1   & \dots  & v_{2m} \\
  &        & \ddots & \vdots \\
  &        &        &   1   
\end{pmatrix}
$$
such that using the new columns $v_1, \dots, v_m$ of $V_m$, the sequence
$\left\|S_nv_i\right\|$ has growth rate $\gamma_i$ for $i=1, \dots, m$.  
In fact, this is already clear for $i=1$; we just have to make sure none
of the other growth rates were changed by adding the top row of $V_m$.

Now we give the definitions of the new entries in the top row, $v_{12},
\dots, v_{1m}$.  If $\gamma_1 \leq \gamma_i$, set $v_{1i} = 0$.  If
$\gamma_1 \geq \gamma_i$, define
\begin{equation}\label{EQNB.3}
v_{1i} = -\sum_{k=1}^\infty {{b_k T_{k-1} \hat v_i} \over {a_k \cdots
a_1}}.
\end{equation}
This series converges by comparison to a geometric series, because for
each $\epsilon>0$, the magnitude of the $k$th term is bounded above by a
constant (independent of $k$) times ${{e^{(\gamma_i+\epsilon)k}} \over
{e^{(\gamma_1-\epsilon)k}}} = e^{(\gamma_i-\gamma_1+2\epsilon)k}$, using
equations (\ref{EQNB.1}) and (\ref{EQNB.2}).  Here we used the assumption
that the magnitude of the entries of $b_k$ from $A_k$ obey a bound
independent of $k$.

To finish, we need to show that for each $i=2, \dots, m$, the growth rate
of $\left\|S_n v_i\right\|$ is $\gamma_i$.  There are two cases.  First,
we consider $i$ such that $\gamma_1 \leq \gamma_i$.
\begin{equation}\label{EQNB.4}
S_n v_i = 
\begin{pmatrix} t_n \hat v_i \\ T_n \hat v_i \end{pmatrix} = 
\begin{pmatrix} 
a_n \cdots a_1 \sum_{k=1}^n {{b_k T_{k-1} \hat v_i} \over
{a_k \cdots a_1}} \\
T_n \hat v_i 
\end{pmatrix}
\end{equation}
where we have used Lemma \ref{TS1} to rewrite $t_n$.  Equations
(\ref{EQNB.1}) and (\ref{EQNB.2}) imply that for each $\epsilon>0$ there
exist constants independent of $n$ such that
\begin{equation*}
\begin{split}
\left\|a_n \cdots a_1 \sum_{k=1}^n {{b_k T_{k-1} \hat v_i} \over {a_k
\cdots a_1}}\right\|
&\leq constant \cdot e^{(\gamma_1+\epsilon)n} \sum_{k=1}^n
{{e^{(\gamma_i+\epsilon)k}} \over {e^{(\gamma_1-\epsilon)k}}} \\
&\leq constant \cdot e^{(\gamma_1+\epsilon)n}
{{e^{(\gamma_i-\gamma_1+2\epsilon)(n+1)} - 1} \over
{e^{\gamma_i-\gamma_1+2\epsilon} - 1}} \\
&\leq constant \cdot e^{(\gamma_1+\epsilon)n}
{{e^{(\gamma_i-\gamma_1+2\epsilon)n} \ e^{(\gamma_i-\gamma_1+2\epsilon)}}
\over {e^{\gamma_i-\gamma_1+2\epsilon} - 1}} \\
&\leq constant \cdot e^{(\gamma_i + 3\epsilon)n}
\end{split}
\end{equation*}
Since both entries of $S_n v_i$ have $\gamma_i$ as an upper bound for the
lim sup of the growth factor as $n \to \infty$ (using the above inequality
and equation (\ref{EQNB.1})), and since the lower entry has $\gamma_i$ as
a lower bound for the lim inf of the growth factor (again from
(\ref{EQNB.1})), the growth factor limit exists and is $\gamma_i$.

Second, we treat the case where $\gamma_1 > \gamma_i$.
\begin{equation*}
\begin{split}
S_n v_i
&= \begin{pmatrix} 
s_n v_{1i} + t_n \hat v_i \\ T_n \hat v_i 
\end{pmatrix} \\
&= \begin{pmatrix}
-a_n \cdots a_1 \sum_{k=1}^\infty {{b_k T_{k-1} \hat v_i}
\over {a_k \cdots a_1}} + a_n \cdots a_1 \sum_{k=1}^n {{b_k T_{k-1} \hat
v_i} \over {a_k \cdots a_1}} \\ T_n \hat v_i 
\end{pmatrix} \\
&= \begin{pmatrix}
-a_n \cdots a_1 \sum_{k=n+1}^\infty {{b_k T_{k-1} \hat v_i}
\over {a_k \cdots a_1}} \\  T_n \hat v_i
\end{pmatrix}.
\end{split}
\end{equation*}
We can bound the top entry as follows.  For each $\epsilon >0$ there exist
constants independent of $n$ such that
\begin{equation*}
\begin{split}
\left\|a_n \cdots a_1 \sum_{k=n+1}^\infty {{b_k T_{k-1} \hat v_i} \over
{a_k \cdots a_1}}\right\|
&\leq constant \cdot e^{(\gamma_1+\epsilon)n} \sum_{k=n+1}^\infty
{{e^{(\gamma_i+\epsilon)k}} \over {e^{(\gamma_1-\epsilon)k}}} \\
&= constant \cdot e^{(\gamma_1+\epsilon)n}
{{e^{(\gamma_i-\gamma_1+2\epsilon)(n+1)}} \over {1 -
e^{\gamma_i-\gamma_1+2\epsilon}}} \\
&\leq constant \cdot e^{(\gamma_i + 3\epsilon)n}.
\end{split}
\end{equation*}
Just as in the first case, both entries of $S_n v_i$ have $\gamma_i$ as an
upper bound for the growth factor, and the lower entry has $\gamma_i$ as a
lower bound for the growth factor.  Thus, the growth factor is $\gamma_i$.
\QED

% References
\nocite{*}

%The file is refer.bib and it has been included 
%  via mythesis.tex. 

\addcontentsline{toc}{chapter}{Bibliography}
% Enter the name of bibtex  bibliography database, if you wish
\bibliographystyle{unsrt}
\bibliography{refer}
\end{document}